\documentclass[a4paper,11pt]{article}

\usepackage{graphicx, amstext, amsmath, amssymb,color}
\usepackage{siunitx}
\usepackage[colorlinks=true,linkcolor=blue]{hyperref}
\usepackage{subcaption}
\usepackage[a4paper, left=2.5cm, right=2.5cm, top=1.8cm, bottom=1.8cm, includehead, includefoot, head=30pt]{geometry}

\usepackage{pdflscape}

\title{Bias Reduced Peaks over Threshold Tail Estimation}
\author{ J. Beirlant $^{a,b}$\footnote{Corresponding author: Jan Beirlant, KU Leuven, Dept of Mathematics, Celestijnenlaan 200B, 3001 Heverlee, Belgium; Email: jan.beirlant@kuleuven.be }, G. Maribe $^{b}$,  Ph. Naveau$^{c}$, A. Verster$^{b}$  \\
{\fontsize{8pt}{11pt} \selectfont $^a$ Dept. of Mathematics, LStat and LRisk, KU Leuven}
\\
{\fontsize{8pt}{11pt} \selectfont $^b$ Dept. of Mathematical Statistics and Actuarial Science, Free State University }
\\
{\fontsize{8pt}{11pt} \selectfont $^c$ Laboratoire des Sciences du Climat et de l'Environnement, CNRS, Universit\'e Paris-Saclay}
}
\begin{document}

\maketitle
\begin{abstract}
{\noindent In recent years several attempts have been made to extend tail modelling towards the modal part of the data. 
Frigessi et al. (2002) introduced dynamic mixtures of two components with a weight function $\pi=\pi (x)$ smoothly connecting the bulk and the tail of the distribution. 
Recently,  Naveau et al. (2016) reviewed this topic, and, continuing on the work by Papastathopoulos and Tawn (2013), proposed a statistical model which is in compliance with extreme value theory and allows for a smooth transition  between the modal and tail part.
Incorporating second order rates of convergence for distributions of peaks over thresholds (POT), Beirlant et al. (2002, 2009) constructed  models that can be viewed as special cases from both approaches discussed above. When fitting such second order models it turns out that the bias of the resulting extreme value estimators is significantly reduced compared to the classical tail fits using only the first order tail component based on the Pareto or generalized Pareto fits to peaks over threshold distributions. 

 In this paper we provide novel bias reduced tail fitting techniques, improving upon the classical generalized Pareto (GP) approximation for POTs using the flexible semiparametric GP modelling introduced in Tencaliec et al. (2018). We also revisit and extend the second-order refined POT approach started in Beirlant et al. (2009) to all max-domains of attraction using flexible semiparametric  modelling of the second order component. In this way we relax the classical second order regular variation assumptions. 
 }
\end{abstract}




\section{Introduction} 
\label{Sec1}              

 Extreme value (EV) methodology starts from the assumption that the distribution of the available sample $X_1, X_2,\ldots,X_n$ belongs to the domain of attraction of a generalized extreme value distribution, i.e.  there exists sequences $(b_n)_n$ and $(a_n>0)_n$ such that as $n \to \infty$
\begin{equation}
{\max (X_1, X_2,\ldots,X_n)-b_n \over a_n} \to _d Y_{\xi},
\label{maxd}
\end{equation}
where $\mathbb{P} (Y_\xi >y) = \exp (-(1+\xi y)^{-1/\xi})$, for some $\xi \in \mathbb{R}$ with $1+\xi y>0$. The parameter $\xi$ is termed the extreme value index (EVI). It is well-known (see e.g. Beirlant et al., 2004, and de Haan and Ferreira, 2006)) that \eqref{maxd} is equivalent to the existence of a positive function $t \mapsto \sigma_t$, such that 
\begin{equation}
\mathbb{P}\left({X-t \over \sigma_t} >y| X>t \right)
= {\bar{F}(t+y\sigma_t) \over \bar{F}(t)} \to_{t \to x_+}
\bar{H}^{GP}_{\xi}(y)= (1+\xi y)^{-1/\xi},  
\label{POT}
\end{equation}
where $\bar{F}(x)=\mathbb{P}(X>x)$   and $x_+$ denotes the endpoint of the distribution of $X$. The conditional distribution of $X-t$ given $X>t$ is called the peaks over threshold (POT) distribution, while $\bar{H}_{\xi}^{GP}$ is the survival function of the generalized Pareto distribution (GPD).\\
In case $\xi >0$, the limit in \eqref{maxd} holds if and only if $F$ is of Pareto-type, i.e.
\begin{equation}
\bar{F}(x) = x^{-1/\xi}\ell (x),
\label{Patype}
\end{equation} 
for some slowly varying function $\ell$, i.e. satisfying $\frac{\ell(yt)}{\ell (t)} \to 1 $ as $t \to \infty$, for every $y>1$. Pareto-type distributions satisfy a simpler POT limit result: as $t \to \infty$
\begin{equation}
\mathbb{P} \left( {X \over t} >y | X>t \right) \to \bar{H}^P_{\xi}(y) := y^{-1/\xi}, y>1. 
\label{POTPa}
\end{equation}

Estimation of $\xi$ and tail quantities such as return periods is then based on fitting a GPD to the observed excesses $X-t$ given $X>t$, respectively a simple Pareto distribution with survival function $y^{-1/\xi}$ to $X/t$ given $X>t$ in case $\xi >0$.
The main difficulty in such an EV application is the choice of the threshold $t$. Most often, the threshold $t$ is chosen as one of the top data points $X_{n-k,n}$ for some $k \in \{1,2, \ldots,n \}$ where $X_{1,n} \leq X_{2,n} \leq \ldots \leq X_{n,n}$ denotes the ordered sample. The limit results in \eqref{POT} and \eqref{POTPa} require $t$ to be chosen as large as possible (or, equivalently, $k$ as small as possible) for the bias in the estimation of $\xi$ and other tail parameters to be limited. However, in order to limit the estimation variance, $t$ should be as small as possible, i.e. the number of data points $k$  used in the estimation should be as large as possible. Several adaptive procedures for choosing $t$ or $k$ have been proposed, but mainly in the Pareto-type case with $\xi >0$ under further second-order specifications of \eqref{Patype} or \eqref{POTPa}, see for instance Chapter 3 in Beirlant et al. (2004), or Matthys and Beirlant (2000). \\
In case of a real-valued EVI, the selection of an appropriate threshold is even more difficult and only a few methods are available. Dupuis (1999) suggested a robust model validation mechanism  to guide the threshold selection, assigning weights between 0 and 1 to each data point where a high weight means that the point should be retained since a GPD model is fitting it well. However, thresholding is required at the level of the weights and hence the method cannot be used in an unsupervised manner.

Another approach consists of proposing penultimate limit distributions in \eqref{POT} and \eqref{POTPa}. In case $\xi >0$, under the mathematical theory of second-order slow variation, i.e. assuming that 
\begin{equation}
\frac{\ell(yt)}{\ell (t)} - 1 =\delta_t 
\left( y^{-\beta}-1 \right),
\label{SO}
\end{equation}
where $\delta_t=\delta (t)= t^{-\beta}\tilde{\ell}(t)$, with $\beta >0$ and $\tilde\ell$ slowly varying at infinity (see section 2.3 in de Haan and Ferreira, 2006), the left hand side of \eqref{POTPa} equals
\[
\frac{\bar{F}(yt)}{\bar{F}(t)} = 
y^{-1/\xi} \frac{\ell(yt)}{\ell (t)} =
y^{-1/\xi} \left( 1 + \delta_t (y^{-\beta}-1)\right), \; y>1.
\]
This then leads to the extension of the Pareto distribution (EPD) to approximate the distribution of $X/t$ given $X>t$ as $t \to \infty$:
\begin{equation}
\bar{H}_{\xi,\delta}^{EP}(y) := y^{-1/\xi}\left( 1+ \delta_t \left( (y^{-1/\xi})^{\beta\xi}-1\right)\right), \; y>1,
\label{EP}
\end{equation}
with  $\delta_t$ satisfying $\delta_t \downarrow 0$ as $t \to \infty$.
  {\it In cases where the second order model \eqref{SO} holds}, such a mixture model  $\bar{H}_{\xi,\delta}^{EP}$ will improve the approximation of  $\left( {X \over t} >u | X>t \right)$ for values of $t$ which are smaller than the appropriate $t$-values when modelling the POTs using $\bar{H}_{\xi}^{P}$. So the extension can work when modelling large and moderate extremes. As a byproduct however, at instances, it may even work for the full sample.
\\
In Beirlant et al. (2009), using an external estimator  of $\rho=-\beta\xi$, the parameters $(\xi, \delta)$ are estimated fitting the  EPD (slightly adapted, with survival function $ \left\{y (1+ \tilde{\delta}_t -\tilde{\delta}_t y^{-\beta}) \right\}^{-1/\xi}$ and $\tilde\delta _t = \delta_t \xi$) by maximum likelihood on excesses over a random threshold $X_{n-k,n}$, $k=1,2,\ldots,n$. 
The result of this procedure is two-fold:
\begin{itemize}
\item First,  the estimates $\hat{\xi}_k^{EP}$ of $\xi$ are more stable as a function of $k$ compared to the original ML estimator derived by Hill (1975)
$$
H_{k,n} = {1 \over k} \sum_{j=1}^k \log {X_{n-j+1,n}\over X_{n-k,n}}
$$
 which is obtained by fitting the Pareto distribution $\bar{H}_{\xi}^P$ to the excesses 
 $\{ {X_{n-j+1,n}\over X_{n-k,n}}, j=1,\ldots,k\}$ following \eqref{POTPa}. Indeed, the bias in the simple POT model \eqref{POTPa} is estimated when fitting $\bar{H}_{\xi,\delta}^{EP}$ and it is shown that, under the assumption that the EP model for the excesses $X/t$ is correct and that $\beta$ is estimated consistently,  the asymptotic bias of  $\hat{\xi}_k^{EP}$ is 0 as long as  $k (k/n)^{2\beta\xi} \to \lambda \geq 0$ as $k,n \to \infty$, while the asymptotic bias of $H_{k,n}$ is only 0 when $k (k/n)^{2\beta\xi} \to 0$.
 \item On the other hand,  the asymptotic variance of $\hat{\xi}_k^{EP}$ equals $\left({1-\rho\over \rho}\right)^2 {\xi^2 \over k}$, where ${\xi^2 \over k}$ is the asymptotic variance of $H_{k,n}$. 
\end{itemize}
As an example Figure 1 shows both the Hill estimates $H_{k,n}$ and the bias reduced estimates $\hat{\xi}_k^{EP}$, obtained from maximum likelihood fitting of  \eqref{EP}  using $\rho=-\xi\beta=-0.25, -0.5$ and $-1$, as a function of $k$ for a dataset of Belgian ultimate car insurance claims from  1995 and 2010 discussed in more detail in Albrecher et al. (2017). Note that the bias reduced estimates helps to interpret the original Hill "horror" plot. Here from the bias reduced estimator a $\xi$ level around 0.5 becomes apparent for $k \geq 200$ and a lower value between 0.3 and 0.4 for smaller values of $k$. In fact in insurance claim data mixtures in the ultimate tail do appear quite often. Moreover the EPD fit appears to extend quite well down to the lower threshold value, i.e. with $k$ up to 600 (but not when using almost all data, $k>600$).
In this sense, classical first order extreme value modelling can in some cases be extended using mixture modelling in order to capture the characteristics of the bulk of the data. 
\begin{figure}[!ht]
  \centering
\includegraphics[width=0.70\textwidth]{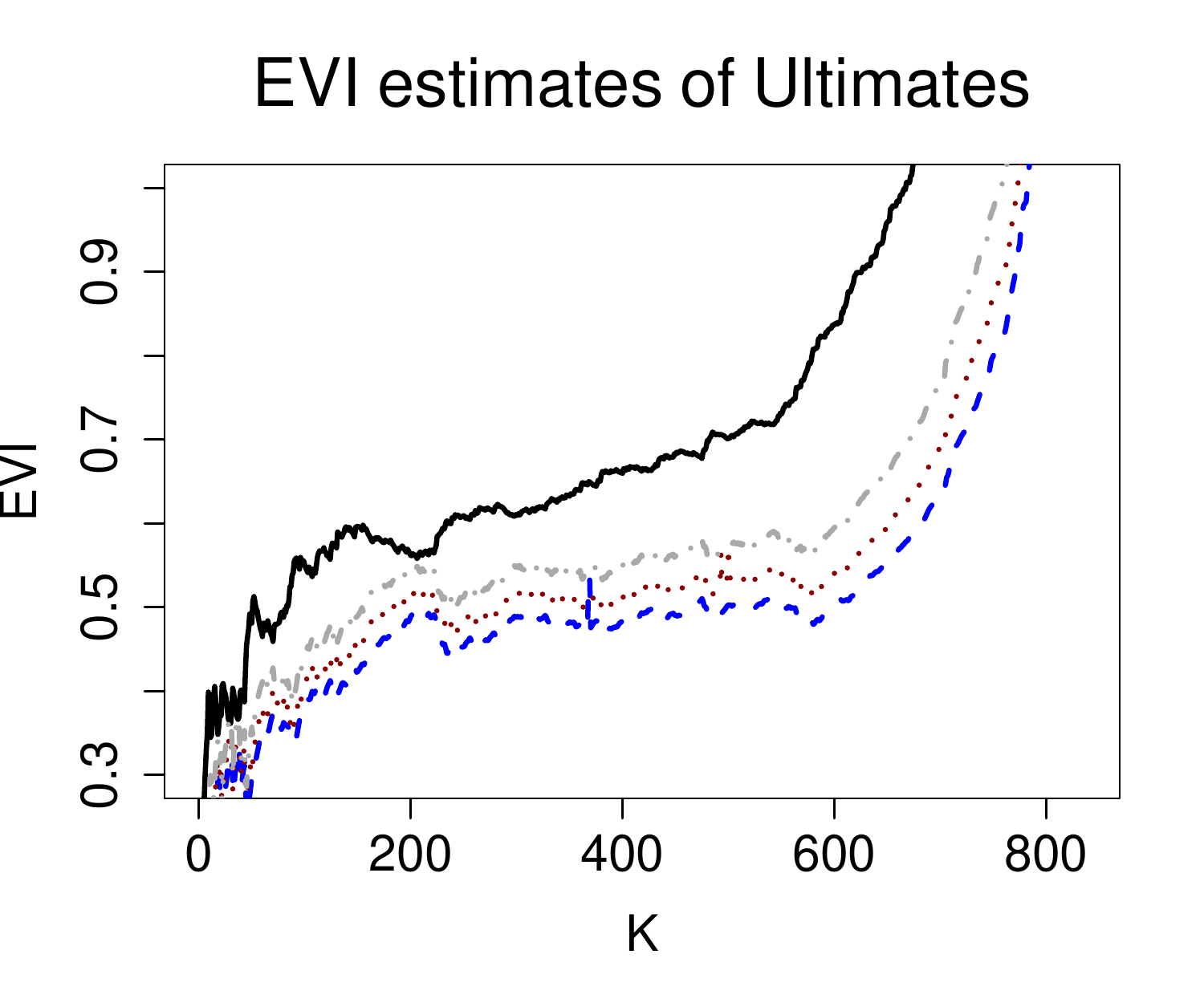} 
\caption{ Ultimates of Belgian car insurance claims: bias reduction of Hill estimator (full line) using $\bar{H}_{\xi,\delta}^{EP}$ with $\rho=-0.25$ (dashed line), $\rho=-0.5$ (dotted line) and $\rho=-1$ (dash-dotted line).}
\end{figure}

Other bias reduction techniques in the Pareto-type case $\xi >0$ have been proposed among others in Feuerverger and Hall (1999), Gomes et al. (2000), Beirlant et al. (1999, 2002) and Gomes and Martins (2002). In Caeiro and Gomes (2011) methods are proposed to limit the variance of bias-reduced estimators to the level of the variance of the Hill estimator $H_{k,n}$. The price to pay is then to assume a third-order slow variation model specifying \eqref{SO} even further. These methods focus on the distribution of the $\log$-spacings of high order statistics.  Other  construction methods for asymptotically unbiased estimators of $\xi >0$ were introduced in Peng (1998) and Drees (1996).
\\

In this paper we concentrate on bias reduction when using the GPD approximation to the distribution of POTs $X-t|X>t$, on which the literature is quite limited.  This allows to extend bias reduction to  the general case $\xi > -1/2$.   We apply the flexible semiparametric GP modelling introduced in Tencaliec et al. (2018) to the POT distributions. We also extend the second-order refined POT approach using $\bar{H}^{EP}_{\xi,\delta}$ from \eqref{EP}  to all max-domains of attraction.
Here the corresponding basic second order regular variation theory can be found in 
Theorem 2.3.8 in de Haan and Ferreira (2006) stating that
 \begin{equation}
 \lim_{t \to x_+}{ \mathbb{P}(X-t >y\sigma_t|X>t)- (1+\xi y)^{-1/\xi} \over \delta (t)} = (1+\xi y)^{-1-1/\xi}\Psi_{\xi,\tilde{\rho}}((1+\xi y)^{1/\xi}),
 \label{secondorder}
 \end{equation}
 with $\delta(t) \to 0$ as $t \to x_+$ and $\Psi_{\xi,\tilde{\rho}}(x)={1 \over \tilde\rho}\left({x^{\xi +\tilde{\rho}}-1 \over \xi +\tilde\rho}- {x^{\xi}-1 \over \xi}\right)$ which for the cases $\xi=0$ and $\tilde\rho=0$ is understood to be equal to the limit as $\xi \to 0$ and $\tilde\rho \to 0$.
We further allow more flexible second-order models than the ones arising from second-order regular variation theory such as in \eqref{secondorder} using non-parametric  modelling of the second-order component. These new methods are also  applied to the specific case of Pareto-type distributions. 
\\

 In the next section we propose our transformed and extended GPD models, and detail the estimation methods. Some basic asymptotic results are provided in section 3. In the final  section we discuss simulation results of the proposed methods and some practical case studies. We then also discuss the evaluation of the overall goodness-of-fit behaviour of the fitted models.    

\section{Transformed and extended GPD models}

Recently, Naveau et al. (2016), generalizing Papastathopoulos and Tawn (2013),  proposed to use full models for rainfall intensity data that are able to capture low, moderate and heavy rainfall intensities without a threshold selection procedure. These authors, considering only applications with a positive EVI however, propose to model all data jointly using transformation models with survival function 
\begin{equation}
\bar{F}(x) =  1-\bar{G}_0 \left( H^{GP}_{\xi} ({x \over \sigma})\right)=: G_0\left( \bar{H}^{GP}_{\xi} ({x \over \sigma})\right),
\label{naveau}
\end{equation}
with $\bar{G}_0$ and $G_0$  distribution functions on $[0,1]$ linked by $G_0(u)=1-\bar{G}_0 (1-u)$ ($0<u<1$), and satisfying constraints to preserve the classical tail GPD fit and a power behaviour for small rainfall intensities:
\begin{itemize}
\item $\lim_{u \downarrow 0} \frac{G_0(u)}{u}=a$, for some $a>0$,
\item  $\lim_{u \downarrow 0} \frac{\bar{G}_0(u)}{u^\kappa}=c$, for some  $c>0$ and $\kappa >0$.
\end{itemize}
In Naveau et al. (2016) the authors propose  parametric examples for $G_0$, such as  
$G_0(u) = {1+D \over D} u (1-\frac{u^D}{1+D})$, $v \in (0,1)$ with $D>0$. 
In Tencaliec et al. (2018) a non-parametric approach is taken using Bernstein polynomials of degree $m$ to approximate $G_0$, i.e. using $G_{0}^{(m)} (u) = \sum_{j=0}^m G({j \over m})b_{j,m}(u)$ with beta densities
$$b_{j,m}(u)=\left( \begin{array}{c} m \\ j \end{array}\right)u^j (1-u)^{m-j},\; u \in (0,1).
$$  
\\
In Naveau et al. (2016) and Tencaliec et al. (2018) the primary goal is the search for a model fitting the whole outcome set, 
while the fit of the proposed model to POT values $X-t |X>t$ for extrapolation purposes in order to estimate extreme quantiles and tail probabilities is imposed using the condition $\lim_{u \downarrow 0} \frac{G_0(u)}{u}=a$. However the bias and MSE properties of the estimators of $\xi$ and $\sigma$ are still to be analyzed.\\

To encompass the above mentioned methods from Beirlant et al. (2009), Naveau et al. (2016) and Tencaliec et al. (2018) we propose  to approximate $\mathbb{P}\left(X-t  >y| X>t \right)$ with a transformation model with right tail  function of the type
\[\hspace{-3.5cm} ({\cal{T}}): \hspace{0.5cm}
\bar{F}^T_t (y) = G_t \left(\bar{H}^{GP}_{\xi}({y \over \sigma})\right)
\]
where $G_t(u)/u \to 1$ for all $u \in (0,1)$ as $t \to x_+$. Note here that for $u \in (0,1)$ and \\ $Y=_d X-t|X>t$, 
\begin{equation}
G_t (u) = \mathbb{P} \left(\bar{H}^{GP}_{\xi}({Y \over \sigma}) \leq u\right).
\label{Gt}
\end{equation} 

\noindent
We  also consider a submodel of $({\cal{T}})$,  approximating the POT distribution with an extended GPD model
\[
({\cal{E}}) : \hspace{0.5cm} \bar{F}^E_t(y)= \bar{H}^{GP}_\xi ({y \over \sigma})\left\{1 +\delta_t B_{\eta} \left( \bar{H}^{GP}_\xi ( {y \over \sigma}) \right) \right\}, 
\]
where 
\begin{itemize}
\item $\delta_t =\delta (t) \to 0$ as $t \to x_+$,
\item $B_{\eta} (1)=0$ and $\lim_{u \to 0} u^{1-\epsilon}B_{\eta}(u)=0$ for every $1>\epsilon >0$,
\item $B_{\eta}$ is twice continously differentiable. 
\end{itemize}
Here the parameter $\eta$ represents a second order (nuisance) parameter. 
For negative $\delta$-values one needs $\delta_t > \{\min_u (1-{d \over du}\, (uB_{\eta}(u))\}^{-1}$ to obtain a valid distribution.  
  At $t=0$ the function $u \mapsto u(1+\delta_0 B_{\eta}(u))$ then corresponds to $u \mapsto G_0(u)$ in \eqref{naveau}, while  ${G_t (u) \over u} \to 1$ as $t \to \infty$ leads to the GPD survival function $\bar{H}^{GP}_\xi (x/\sigma)$ at large thresholds.
 \\
  Note that model ($\cal{E}$) is a direct generalization of the EPD model \eqref{EP} replacing the Pareto distribution $y^{-1/\xi}$ by the GPD $\bar{H}^{GP}_\xi$ and considering a general function  $B_\eta (u)$ rather than $ u^{\beta\xi}-1 = u^{-\rho}-1$. 
\\
  
\noindent  Now several possibilities for bias reduction appear:
  \begin{enumerate}
  \item[(1)] {\bf Estimation under the transformed model (${\cal T}$).} Modelling the distribution  of $Y=X-t|X>t$
  with model ($\cal{T}$) 
and estimating $G_t$ and $(\xi,\sigma)$ for every $t$, we propose to use the algorithm from  Tencaliec et al. (2018) for every $t$ or $k=1,\ldots,n$. This approach is further denoted with ($T\bar{p}$). \\
 Here we apply Bernstein approximation and estimation of $G_t$ which is the distribution function of $\bar{H}_{\xi}^{GP}(Y/\sigma)$. 
The Bernstein approximation of order $m$ of a continuous distribution function $G$ on $[0,1]$
is given by 
\[
G^{(m)}(u) = \sum_{j=0}^m G\left( {j \over m}\right)\left( \begin{array}{c} m \\ j \end{array}\right)u^j (1-u)^{m-j},
\; u \in [0,1].
\]
As in Babu et al. (2002) one then replaces the unknown distribution function $G$ itself with the empirical distribution function $\hat{G}_n$ of the available data in order to obtain a smooth estimator of $G$:
\[
\hat{G}_n^{(m)}(u) = \sum_{j=0}^m \hat{G}_n\left( {j \over m}\right)\left( \begin{array}{c} m \\ j \end{array}\right)u^j (1-u)^{m-j}. 
\] 
In the present application, data from $G_t$ are only available after imputing a value for $(\xi,\sigma)$. This then leads to the iterative algorithm from Tencaliec et al. (2018), which is applied to every threshold $t$, or every number of top $k$ data. We here detail the algorithm for excesses $Y_{j,k}=X_{n-j+1,n}-X_{n-k,n}$ $(j=1,\ldots,k)$, using the reparametrization $(\xi,\tau)$ with $\tau=\xi/\sigma$:

\vspace{0.3cm}\noindent
{\it Algorithm} ($A_{\cal T}$) 
\begin{enumerate}
\item[(i)] Set starting values ($\hat{\xi}_k^{(0)},\hat{\tau}_k^{(0)}$). Here one can use ($\hat{\xi}_k^{ML},\hat{\tau}_k^{ML}$) from using $G_t (u)=u$.
\item[(ii)]
Iterate for $r=0,1,\ldots$ until the difference in loglikelihood taken in ($\hat{\xi}_k^{(r)},\hat{\tau}_k^{(r)}$) and ($\hat{\xi}_k^{(r+1)},\hat{\tau}_k^{(r+1)}$) is smaller than a prescribed value
\begin{enumerate}
\item Given ($\hat{\xi}_k^{(r)},\hat{\tau}_k^{(r)}$)
construct rv's 
$ \hat{Z}_{j,k}= \left( 1+ \hat{\tau}_k^{(r)}Y_{j,k}\right)^{-1/\hat{\xi}_k^{(r)}} 
$
\item Construct Bernstein approximation based on $\hat{Z}_{j,k}$ ($1\leq j \leq k$)
\[
\hat{G}_k^{(m)}(u) = \sum_{j=0}^m \hat{G}_k \left( {j \over m}\right)\left( \begin{array}{c} m \\ j \end{array}\right)u^j (1-u)^{m-j}
\]
with $\hat{G}_k$ the empirical distribution function  of $\hat{Z}_{j,k}$
\item Obtain new estimates ($\hat{\xi}_k^{(r+1)},\hat{\tau}_k^{(r+1)}$) with ML:
\begin{eqnarray*}
(\hat{\xi}_k^{(r+1)},\hat{\tau}_k^{(r+1)})&=&
\mbox{argmax} \left\{
\sum_{j=1}^k \log \{ \hat{g}^{(m)}_k ((1+\tau \hat{Z}_{j,k})^{-1/\xi})\} \right.\\
&& \hspace{1.5cm} \left. + \sum_{j=1}^k
\log \{{\tau \over \xi}(1+\tau \hat{Z}_{j,k})^{-1-1/\xi} \} \right\}
\end{eqnarray*}
with $\hat{g}^{(m)}_k $ denoting the derivative of $\hat{G}_k^{(m)}$.
\end{enumerate}
\end{enumerate}

\vspace{0.3cm}\noindent
The final estimates of $(\xi,\tau)$ and $G_t$ are denoted here by $(\hat{\xi}_k^T,\hat{\tau}_k^T)$ and $\hat{G}_k^T$. 
As noted in Tencaliec et al. (2018) the theoretical study of these estimates is difficult. In the simulation study the finite sample characteristics of these estimators $\hat{\xi}_k^T$ are given using $m=k^a$ with  a fixed value of $a$ using $\hat{a} = argmin \sum_{k=2}^n (\hat{\xi}_k^T - \bar{\hat{\xi}}^T_n)^2$ in order to stabilize the plots of the estimates of $\xi$ as much as possible.  Note that this estimation procedure is computationally demanding. 

\vspace{0.2cm}\noindent
Estimates of small tail probabilities $\mathbb{P}(X>c)$ can be obtained through 
\[
\hat{\mathbb{P}}_k^T(X>c) = {k \over n} \hat{G}^T_k 
\left( \bar{H}_{\hat{\xi}^T_k} ({\hat{\tau}^T_k \over \hat{\xi}^T_k}(c-X_{n-k,n}) )\right).
\]
Finally, bias reduced  estimators of extreme $1-p$ quantiles for small $p$ are obtained by setting the above formulas equal to $p$ and solving for $c$.  

 \item[(2)] {\bf Estimation under the extended model (${\cal E}$).} Modelling the distribution of the exceedances $Y$ with model ($\cal{E}$) leads to maximum likelihood estimators based on the excesses $Y_{j,k}=X_{n-j+1,n}-X_{n-k,n}$ $(j=1,\ldots,k)$:  
\begin{eqnarray}
(\hat{\xi}^E_k,\hat{\tau}^E_k, \hat{\delta}_k)&=&
\mbox{argmax} \left\{
\sum_{j=1}^k \log\left( 
1+\delta_k b_{\eta}((1+\tau Y_{j,k})^{-1/\xi}) \right)
 \right. \nonumber \\
&& \hspace{1.5cm} \left. + \sum_{j=1}^k
\log \{{\tau \over \xi}(1+\tau Y_{j,k})^{-1-1/\xi} \}
 \right\}
 \label{MLE}
\end{eqnarray}
with $b_{\eta}(u) ={d \over du} (uB_{\eta}(u))$ for a given choice of $B_\eta$.\\
\noindent
Estimates of small tail probabilities $\mathbb{P}(X>c)$ are then  obtained through 
\[
\hat{\mathbb{P}}_k^E(X>c) = {k \over n} \bar{H}^{GP}_{\hat{\xi}^E_k}\left({\hat{\tau}^E_k \over \hat{\xi}^E_k}(c-X_{n-k,n}) \right) 
\left( 1+ \hat{\delta}_k^E \hat{B}_{\eta}\left(\bar{H}^{GP}_{\hat{\xi}^E_k} ({\hat{\tau}^E_k \over \hat{\xi}^E_k}(c-X_{n-k,n})\right)\right).
\]
 
 \noindent
  As in Naveau et al. (2016), respectively Tencaliec et al. (2018), two approaches can be taken towards the bias function $B_{\eta}$: a parametric approach, respectively a non-parametric approach. 
\begin{itemize}
\item[(a)] {\it In the parametric approach}, denoted ($Ep$), the second-order result \eqref{secondorder} leads to the parametric choice $B_{\xi,\tilde{\rho}}(u)= {u^{\xi} \over \tilde\rho}\left({u^{-\xi -\tilde{\rho}}-1 \over \xi +\tilde\rho}- {u^{-\xi}-1 \over \xi}\right)$ in case  $\xi+\tilde\rho \neq 0$ and $\xi \neq 0$. \\
  Model (${\cal{E}}$)  allows for bias reduced estimation of  $(\xi,\tau)$ under the assumption that the corresponding second-order model \eqref{secondorder} is correct for the POTs  $X-t|X>t$. Note  that  ($Ep$) generalizes the approach taken in Beirlant et al. (2009) to all max-domains of attraction. When model (${\cal E}$)  is used as a model for all observations, i.e. taking $t=0$, this model directly encompasses the models from Frigessi et al. (2002) 	and Naveau et al. (2016).\\
  Here
\[b_\eta (u)= u^{-\tilde\rho}\left( {1-\tilde\rho\over \tilde\rho (\xi +\tilde\rho)}\right)
+ u^\xi \left( {1-\xi \over \xi (\xi +\tilde\rho)}\right)
- {1 \over \xi\tilde\rho},
\]
in which the classical estimator of $\xi$ (with $\delta_k=0$), or an appropriate value $\xi_0$, is used to substitute $\xi$, next to an appropriate value of $\tilde\rho$. One can also choose a value of ($\xi_0,\tilde\rho$)  minimizing the variance in the plot of the resulting estimates of $\xi$ as a function of $k$.
 
 \item[(b)] Alternatively, {\it  a non-parametric approach} (denoted $E{\bar{p}}$) can be  performed using the Bernstein polynomial  algorithm from Tencaliec et al. (2018). In fact in practice a particular distribution probably follows laws of nature, environment or business and does not have to follow the second-order regular variation assumptions as in  \eqref{secondorder}. Moreover in the case of a real-valued EVI, the function $B_{\eta}$ can take different mathematical forms depending on ($\xi,\tilde\rho$)  and $\xi+\tilde\rho$ being close to 0 or not.\\
 Here a Bernstein type approximation is obtained for $u\mapsto uB_\eta (u)$ from
$\hat{G}_{k_*}^{(m)}(u) -u$ obtained through algorithm ($A_{\cal T}$), and reparametrizing $\delta_k$ by $\delta_k/\delta_{k_*}$ with $k_*$ an appropriate value of the number of top data used.
The function $b_\eta (u)$ is then substituted by
$-1 + {d \over du}\hat{G}_{k_*}^{(m)}(u)$.
 \end{itemize}
 
\end{enumerate}
 
 \vspace{0.5cm}\noindent
 The methods described above of course can be rewritten for the specific case of Pareto-type distributions where the distribution of POTs $Y={X \over t}|X>t$ are approximated by transformed Pareto distributions. The models are then rephrased as 
 \[\hspace{-2.9cm} ({\cal{T}^+}): \hspace{0.5cm}
\bar{F}^E_t (y) = G_t \left(\bar{H}^{P}_{\xi}(y)\right),
\]
where for $u \in (0,1)$  
\begin{equation*}
G_t (u) = \mathbb{P} \left(\bar{H}^{P}_{\xi}(Y) \leq u\right),
\end{equation*} 
and
\[
({\cal{E}^+}) : \hspace{0.5cm} \bar{F}^E_t(y)= \bar{H}^{P}_\xi (y)\left\{1 +\delta_t B_{\eta} \left( \bar{H}^{P}_\xi (y) \right) \right\}. 
\]
The above algorithms, now based on the exceedances $Y_{j,k}= X_{n-j+1,n}/X_{n-k,n}$ ($j=1,\ldots,k$), are then adapted as follows:\\
$\bullet$ In algorithm ($A_{\cal T}$) step (ii.c) is replaced by 
\[
\hat{\xi}_k^{(r+1)}=
\mbox{argmax} \left\{
\sum_{j=1}^k \log \{ \hat{g}^{(m)}_k ( (\hat{Z}_{j,k})^{-1/\xi})\}  + \sum_{j=1}^k
\log \{{1 \over \xi}(\hat{Z}_{j,k})^{-1-1/\xi} \} \right\},
\]
with $ \hat{Z}_{j,k}= Y_{j,k}^{-1/\hat{\xi}_k^{(r)}} $. 
The resulting estimates are denoted with $\hat{\xi}^{T+}_{k}$ and $\hat{G}^{T+}_k $.
 \\
 $\bullet$ In approach (${\cal E}$) the likelihood solutions are given by
  \begin{equation}
(\hat{\xi}^{E+}_k, \hat{\delta}^{E+}_k)=
\mbox{argmax} \left\{
\sum_{j=1}^k \log\left( 
1+\delta_k b_{\eta}(Y_{j,k}^{-1/\xi}) \right)+
  \sum_{j=1}^k
\log \{{1 \over \xi} (Y_{j,k})^{-1-1/\xi} \}
 \right\}.
\label{E+}
\end{equation}
Note that the ($Ep^+$) approach using the parametric version $B_\eta (u) = u^{-\rho}-1$ for a particular fixed $\rho <0$  equals the EPD method from Beirlant et al. (2009), while ($E\bar{p}^+$) is new.

\vspace{0.2cm}\noindent
Estimators of tail probabilities are then given  by  
\[
\hat{\mathbb{P}}_k^{T+}(X>c) = {k \over n} \hat{G}^{T+}_k 
\left( \bar{H}_{\hat{\theta}^{T+}_k} \; (c/X_{n-k,n})\right),
\]
respectively
\[
\hat{\mathbb{P}}_k^{E+}(X>c) = {k \over n} \bar{H}^{P}_{\hat{\xi}^{E+}_k}\left( c/X_{n-k,n} \right) 
\left( 1+ \hat{\delta}_k^{E+} \hat{B}_{\eta}\left(\bar{H}^{P}_{\hat{\xi}^{E+}_k} (c/X_{n-k,n})\right)\right).
\]

\section{Basic asymptotics under model (${\cal E}$)}

We discuss here in detail the asymptotic properties of the maximum likelihood estimators solving \eqref{MLE} and \eqref{E+}. To this end, as in Beirlant et al. (2009) we develop the likelihood equations up to linear terms in $\delta_k$ since $ \delta_k \to 0$ with decreasing value of $k$. 
Below we set $\bar{H}_{\theta}(y)=(1+\tau y)^{-1/\xi}$ when using extended GPD modelling, while $\bar{H}_{\theta}(y)=y^{-1/\xi}$ when using extended Pareto modelling under $\xi >0$.

\vspace{0.2cm}\noindent
{\it Extended Pareto POT modelling}. The likelihood problem \eqref{E+} was already considered in Beirlant et al. (2009) in case of  parametric modelling for $B_\eta$. We here propose a more general treatment. The limit statements in the derivation can be obtained using the methods from Beirlant et al. (2009). The likelihood equations following from \eqref{E+} are given by
\begin{equation}
\left\{
\begin{array}{lcl}
{\partial \over \partial \xi} \ell &=&
-{k \over \xi }+ {1 \over \xi^2} \sum_{j=1}^k \log Y_{j,k}
 + {\delta_k \over \xi^2} \sum_{j=1}^k \frac{ b'_\eta (\bar{H}_{\theta}(Y_{j,k}))\bar{H}_{\theta}(Y_{j,k})\log Y_{j,k}}{1+\delta_k b_\eta (\bar{H}_{\theta}(Y_{j,k}))} \\
 {\partial \over \partial \delta_k} \ell &=&
 \sum_{j=1}^k b_\eta (\bar{H}_{\theta}(Y_{j,k}))-\delta_k \sum_{j=1}^k b^2_\eta (\bar{H}_{\theta}(Y_{j,k})).
\end{array}
\right.
\label{systemE+}
\end{equation}

\vspace{0.2cm}\noindent
{\it Extended Generalized Pareto POT modelling}. 
The likelihood equations following from \eqref{MLE} up to linear terms in $\delta_k$ are now given by
\[
\left\{
\begin{array}{lcl}
{\partial \over \partial \xi} \ell &=&
-{k \over \xi }+ {1 \over \xi^2} \sum_{j=1}^k \log (1+\tau Y_{j,k})
 + {\delta_k \over \xi^2} \sum_{j=1}^k  b'_\eta (\bar{H}_{\theta}(Y_{j,k}))\bar{H}_{\theta}(Y_{j,k})\log (1+\tau Y_{j,k}) \\
 {\partial \over \partial \tau} \ell &=&
{k \over \xi \tau}
\left\{ -1+ (1+\xi) {1 \over k}\sum_{j=1}^k {1 \over 1+\tau Y_{j,k}} \right. \\
&& \hspace{1cm} \left. 
 -{\delta_k \over k}\sum_{j=1}^k b'_\eta (\bar{H}_{\theta}(Y_{j,k})) (\tau Y_{j,k}) (1+\tau Y_{j,k})^{-1-1/\xi}
   \right\}
 \\
 {\partial \over \partial \delta_k} \ell &=&
 \sum_{j=1}^k b_\eta (\bar{H}_{\theta}(Y_{j,k}))-\delta_k \sum_{j=1}^k b^2_\eta (\bar{H}_{\theta}(Y_{j,k})),
\end{array}
\right.
\]
from which
\begin{equation}
 \left\{
\begin{array}{l}
\hat{\delta}_k = \frac{\sum_{j=1}^k b_\eta (\bar{H}_{\hat{\theta}_k}(Y_{j,k}))}{\sum_{j=1}^k b^2_\eta (\bar{H}_{\hat{\theta}_k}(Y_{j,k}))}, \\
{1 \over k}\sum_{j=1}^k \log (1+\hat{\tau}_k Y_{j,k})
= \hat{\xi}_k- 
{\hat{\delta}_k \over k}\sum_{j=1}^k b'_{\eta}(\bar{H}_{\hat{\theta}_k}(Y_{j,k})) \bar{H}_{\hat{\theta}_k}(Y_{j,k}) \log (1+\hat{\tau}_k Y_{j,k}), \\
{1 \over k}\sum_{j=1}^k {1 \over 1+\hat{\tau}_k Y_{j,k}}
= {1 \over 1+ \hat{\xi}_k}  
+ {\hat{\delta}_k \over 1+ \hat{\xi}_k}
\left\{ 
{1\over k}\sum_{j=1}^k b'_{\eta}(\bar{H}_{\hat{\theta}_k}(Y_{j,k}))\bar{H}_{\hat{\theta}_k}(Y_{j,k}) \right.\\
 \hspace{6cm} \left.
- {1\over k}\sum_{j=1}^k b'_{\eta}(\bar{H}_{\hat{\theta}_k}(Y_{j,k}))\bar{H}_{\hat{\theta}_k}(Y_{j,k}) {1 \over 1+\hat{\tau}_k Y_{j,k}}
 \right\}.
\end{array}
\right.
\label{systemE}
\end{equation}

\vspace{0.3cm}\noindent
Under the extended model we now state the asymptotic distribution of the estimators $\hat{\xi}_k^{E+}$ and $\hat{\xi}_k^{E}$. To this end let $Q$ denote the quantile function of $F$, and let $U(x)=Q(1-x^{-1})$ denote the corresponding tail quantile function. 
Model assumption $({\cal{E}})$ can be rephrased in terms of $U$: 
\[
({\tilde{\cal{E}}}):\;\;
\frac{\frac{U(vx)-U(v)}{\sigma_{U(v)}} -h_{\xi}(x)}{\delta (U(v))} \to _{v \to \infty} x^{\xi} B_{\eta}(1/x),
\]
where $h_{\xi}(x)= (x^{\gamma}-1)/\gamma$ and $\delta (U)$ regularly varying with index $\tilde\rho<0$. 
Moreover in the mathematical derivations one needs the extra condition that for every $\epsilon,\nu>0$, and  $v, vx$ sufficiently large
\[
({\tilde{\cal{E}}}_2):\;\;
\left| \frac{\frac{U(vx)-U(v)}{\sigma_{U(v)}} -h_{\xi}(x)}{\delta (U(v))} - x^{\xi} B_{\eta}(1/x) \right| \leq \epsilon x^{\xi}|B_{\eta}(1/x)| \max\{x^{\nu},x^{-\nu}\}.
\]
Similarly, $({\cal{E}}^+)$ is rewritten as
\[
({\tilde{\cal{E}}}^+):\;\;
\frac{\frac{U(vx)}{U(v)} - x^{\xi}}{\xi \delta(U(v)))} \to_{v \to \infty}  x^{\xi} B_{\eta}(1/x).
\]
The analogue of 
$({\tilde{\cal{E}}}_2)$ in this specific case is given by
\[
({\tilde{\cal{E}}}_2^+):\;\;
\left| \frac{\frac{U(vx)}{U(v)} - x^{\xi}}{\xi\delta(U(v))} -
x^{\xi} B_{\eta}(1/x) \right|
 \leq \epsilon x^{\xi}|B_{\eta}(1/x)| \max\{x^{\nu},x^{-\nu}\},
\]
with $\delta(U)$ regularly varying with index $\rho <0$.\\
Finally, in the expression of the asymptotic variances we use 
\[
Eb^2_{\eta} = \int_0^1 b^2_{\eta} (u)du, \;\;
EB_{\eta} = \int_0^1 B_{\eta} (u)du, \;\;
EC_{\eta} = \int_0^1 u^{\xi}B_{\eta} (u)du.
\]
The proof of the next theorem is outlined in the Appendix. It allows to construct confidence intervals for the estimators of $\xi$ obtained under the extended models.\\
{\bf Theorem 1} {\it Let $k=k_n$ be a sequence such that $k,n \to \infty$ and $k/n \to 0$ such that $\sqrt{k}\delta (U(n/k)) \to \lambda \in \mathbb{R}$. Moreover assume that in \eqref{MLE}
and \eqref{E+}, $B_{\eta}$ is substituted by a consistent estimator as $n\to \infty$. Then
\begin{enumerate}
\item[i.] when $\xi>0$ with $({\tilde{\cal{E}}}_2^+)$ 
$$\sqrt{k}\left( \hat{\xi}_k^{E+} -\xi \right) \to_d \mathcal{N}\left(0,\xi^2 \frac{Eb^2_{\eta}}{Eb^2_{\eta}-(EB_{\eta})^2}\right),$$
\item[ii.] when $\xi > -1/2$ with $({\tilde{\cal{E}}}_2)$ 
$$\left( \sqrt{k}( \hat{\xi}_k^{E} -\xi), \sqrt{k} ({\hat{\tau}^{E}_k\over \tau}-1) \right)\to_d \mathcal{N}_2({\bf 0},\Sigma)$$
\[
\Sigma ={\xi^2 \over D}
\left( 
\begin{array}{ll}
{1 \over (1+\xi)^2(1+2\xi)} - \frac{ (EC_{\eta})^2}{Eb^2_{\eta}} 
 & 
{1 \over \xi(1+\xi)^3 }-\frac{EB_{\eta}EC_{\eta}}{\xi(1+\xi)Eb^2_{\eta}}
\\
{1 \over \xi(1+\xi)^3 }-\frac{EB_{\eta}EC_{\eta}}{\xi(1+\xi)Eb^2_{\eta}}
&
{1 \over \xi^2(1+\xi)^2}\left(1- \frac{(EB_{\eta})^2}{Eb^2_{\eta}} \right)
\end{array}
\right)
\]
where
\[
D=\left( {1 \over (1+\xi)^2(1+2\xi)} - \frac{ (EC_{\eta})^2}{Eb^2_{\eta}} \right)\left(1- \frac{(EB_{\eta})^2}{Eb^2_{\eta}}\right)
- \left( {1 \over (1+\xi)^2 }- \frac{EB_{\eta}EC_{\eta}}{Eb^2_{\eta}}\right)^2,
\]
\end{enumerate} 
}

\vspace{0.3cm}\noindent
{\bf Remark 1.} The asymptotic variance of $\hat{\xi}_k^{E+}$ is larger than the asymptotic variance $\xi^2$ of the Hill estimator $H_{k,n}$. Indeed, 
\begin{eqnarray*}
(EB_\eta)^2 &=& \left( \int_0^1 \log (1/u) b_{\eta}(u) du \right)^2 \\
&=& \left( \int_0^1 (\log (1/u) -1) b_{\eta}(u) du \right)^2 \\
&\leq & \left( \int_0^1 (\log (1/u) -1)^2 du \right)
\left( \int_0^1 b^2_{\eta}(u) du \right)\\
&=& (Eb^2_\eta),
\end{eqnarray*}
where the above inequality follows using the Cauchy-Schwarz inequality.  \\
Similarly, one can show that
\[
(EC_\eta)^2= \xi^{-2} \left(\int_0^1 (u^\xi -{1 \over 1+\xi})b_\eta du \right)^2 \leq {1 \over (1+2\xi)(1+\xi)^2}
(Eb^2_\eta).
\]
The asymptotic variance of  $\hat{\xi}_k^{E}$ equals 
\[
{(1+\xi)^2 \over k} \;
\frac{1- (1+\xi)^2(1+2\xi)(EC_\eta)^2/(Eb_\eta^2)}
{1- {(1+\xi)^4(1+2\xi)\over \xi^2} (Eb_\eta^2)^{-1}
[(EC_\eta)^2 -2 {(EC_\eta)(EB_\eta)\over (1+\xi)^2}+ {(EB_\eta)^2 \over (1+\xi)^{2}(1+2\xi)}]
}
\]
which can be shown to be larger than the asymptotic variance $(1+\xi)^2/k$ of the classical GPD maximum likelihood estimator. In the parametric case with $B_\eta (u) = {u^{\xi} \over \tilde\rho}\left({u^{-\xi -\tilde{\rho}}-1 \over \xi +\tilde\rho}- {u^{-\xi}-1 \over \xi}\right)$, one obtains  $EB_\eta = (1+\xi)^{-1}(1-\tilde\rho)^{-1}$, $EC_\eta = (1+\xi)^{-1}(1+2\xi)^{-1}
(\xi-\tilde\rho+1)^{-1}$ and $Eb^2_\eta = 2 (1+2\xi)^{-1}(1-2\tilde\rho)^{-1}(\xi-\tilde\rho+1)^{-1}$. It then follows that the asymptotic variance of $\hat{\xi}_k^{E}$ equals ${(1+\xi)^2 \over k} \left(\frac{1-\tilde\rho}{\tilde\rho}\right)^2$.\\
In case $\xi >0$  with $B_\eta (u)= u^{-\rho}-1$, the asymptotic variance of  $\hat{\xi}_k^{E+}$ is given by  ${\xi^2 \over k} \left(\frac{1-\rho}{\rho}\right)^2$ as already found in Beirlant et al. (2009).

\vspace{0.3cm}
Since in model (${\cal{E}}$) the $B_{\eta}$ factor is multiplied by $\delta_t$, the asymptotic distribution of tail estimators based on (${\cal{E}}$) will not depend on the asymptotic distribution of the estimator of $B_{\eta}$. As in Beirlant et al. (2009) when using the EPD model in the Pareto-type setting,  one can rely in the parametric approach on consistent estimators of the nuisance parameter $\eta$  using a larger proportion $k_*$ of the data.
 Alternatively, one can also consider different values of $\eta$ in the parametric approach, and of $(k_*,m)$ in the non-parametric setting, and search for values of this nuisance parameter which stabilizes the plots of the EVI estimates as a function of $k$ using the minimum variance principle for the estimates as a function of $k$. Clearly one loses the asymptotic unbiasedness in Theorem 1 if $B_{\eta}$ is not consistently estimated. However as becomes clear from the simulation results in many instances the extreme value index estimators are not very sensitive to such a misspecification, especially in the non-parametric approaches $T\bar{p}$, $T\bar{p}^+$, $E\bar{p}$
and $E\bar{p}^+$, and the proposed estimators can still outperform the classical maximum likelihood estimators based on the first order approximations of the POT distributions.

 \section{Simulations and case studies}
 
 Simulation results and practical cases  are proposed on
 
 \url{https://phdshinygao.shinyapps.io/ExtendedModels/}
 
 \vspace{0.3cm}\noindent
  Under {\it Simulations} one finds simulation results with sample sizes $n= 200$ for different distributions from each max-domain of attraction. The bias and MSE for the different estimators are plotted as a function of the number of  exceedances $k$. 
  Using the notation from the preceding sections one has a choice to apply the technique with $\bar{H}_{\theta}$ equal to the GPD, respectively the simple Pareto distribution (only when $\xi>0$). \\

\noindent 
In case of the transformed models (${\cal T}$) one finds a slider to adapt the degree $m$ of the Bernstein polynoms along $m=k^a$ for different values of $a \in (0,1)$. One can also choose $a$ adaptively per sample so as to minimize  the variance of $\hat{\xi}_k$ over $k=2,\ldots,n$ (in order to have stable plots over $k$).
 \\
 
 \noindent
 In case of the extended models (${\cal E}$) one finds  sliders for the following parameters:
 \begin{itemize}
 \item in case of Pareto modelling: $\rho$ in $Ep^+$, and $(k_*,m)$ in $E{\bar p}^+$ estimation;
 \item in case of GPD modelling: $\tilde\rho$ in $Ep$, and  $(k_*,m)$ in $E{\bar p}$ estimation.
  \end{itemize}
  Again one can indicate to choose these parameters so as to minimize  the variance of $\hat{\xi}_k$ over $k=2,\ldots,n$.  
 The value of $\xi$ in the parametric function $B_{\xi,\tilde\rho}$ in $Ep$ is imputed with the classical GPD-ML estimator at the given value of $k$.
\\

\noindent 
Also bias and RMSE plots of the corresponding tail probability estimates of $p=\mathbb{P}(X>c)$ are given, where $c$ is chosen so that these probabilities equal $p=0.005$ or $p=0.003$. Here the bias, respectively RMSE, are expressed as the average, respectively the average of squared values, of $\log (p/\hat{p})$.  
 \\ One can also change the vertical scale of the plots, smooth the figures by taking moving averages of a certain number of estimates. Finally one can download the figures in pdf. \\

\noindent

  While on the above link, several other distributions are used and sliders are provided for the different parameters $a$, $m$, $\rho$, $\tilde\rho$, and $k_*$, we collect here the resulting figures for estimation of $\xi$ and estimating 0.003 tail probabilities, when using the minimum variance principle for all parameters, in case of the following subset of models: 
\begin{itemize}
\item
	{\it The Burr$\left(\tau,\lambda\right)$ distribution} with  $\bar{F}(x)= \left(1+x^{\tau} \right)^{-\lambda}$ for $x>0$ with $\tau=1$ and $\lambda=2$, so that $\xi= {1 \over \tau\lambda}={1 \over 2}$ and $\rho= -{1 \over \lambda}= -{1 \over 2}$. 
\item
	{\it The Fr\'echet$\,(2)$ distribution} with  $\bar{F}(x)= 1-\exp \left(-x^{-2} \right)$ for $x>0$, so that $\xi = \frac{1}{2}$ and $\rho= \tilde\rho=-1$.
	
	\item 
	{\it The standard normal distribution} with  $\xi=0$ and $\tilde{\rho}=0$.
	\item 	{\it The Exponential distribution} with $\bar{F}(x) = e^{-\lambda x}$ for $x>0$, so that $\xi=0$ and $\tilde{\rho}=0$.
		\item
	{\it The Reversed Burr distribution} with 
	$\bar{F}(x) =\left(1+(1-x)^{-\tau}\right)^{-\lambda}$
	 for $x< 1$ with $\tau=5$ and $\lambda=1$, so that $\xi = -1/(\tau\lambda)=-\frac{1}{5}$ with $\tilde{\rho}=-1/\lambda=-1$.
	 \item
	{\it The extreme value Weibull distribution} with 
	$F(x) =e^{-(1-x)^{\alpha}}$
	 for $x< 1$ with $\alpha =4$, so that $\xi = -\frac{1}{4}$ with $\tilde{\rho}=-1$.
\end{itemize}

In general the minimum variance principle works well, though in some cases some improved results can be obtained by choosing specific values of the parameters $a$, $\rho$, $m$ and $k_*$. This is mainly the case for the Pareto-type models when using $T\bar{p}$ and $E\bar{p}$, such as for the Fr\'echet distribution. Also, in case of tail probability estimation using $Ep$ for cases with $\xi <0$ particular choices of the corresponding parameters lead to improvements over the minimum variance principle.
\\
When using $GPD$ modelling of the exceedances, overall the $Ep$ approach yields the best results, both in estimation of $\xi$ and tail probabilities. The improvement over the classical GPD maximum likelihood approach is smaller for $E\bar{p}$, and in case of situations where the second order parameter $\tilde\rho$ equals 0 then $E\bar{p}$ basically equals the ML estimators. 
\\
In case of simple Pareto modelling for $\xi >0$ cases (see Figures 3 and 5) the $Ep^+$ and $E\bar{p}^+$  approaches yield serious improvements over the Hill estimator, with small bias for $Ep^+$ and $E\bar{p}^+$, while parametric approach $Ep^+$ naturally exhibits the best RMSE.  Note that when $\tilde\rho=0$ the conditions of the main theorem are not met, in which case the GPD and the bias reductions are known to exhibit a large bias. This is typically the case when $\xi=0$. This is also known to be the case using simple Pareto modelling when $\rho=0$.\\
 
\noindent
 Under {\it Applications} the app also offers the analysis of some case studies, some of which are discussed here in more detail. We use the ultimates of  the Belgian ultimate car insurance claims used in Figure 1, in order to illustrate   $T\bar{p}^+$, $E\bar{p}^+$ and $Ep^+$, and the winter rainfall data at Mont-Aigoual station (1976-2015)   already used in Tencaliec et al. (2018) to illustrate $T\bar{p}$, $E\bar{p}$ and $Ep$. We then present estimates of $\xi$, $\sigma$ and tail probabilities $\mathbb{P}(X>x_{n,n})$ with $x_{n,n}$ denoting the largest observation, so that the estimated probability is supposed to be close to $1/n$. An option is provided to construct confidence intervals for $\xi$ on the basis of Theorem 1. \\
 In case $k=n$ the exceedances correspond to the reversely ordered data, i.e. $Y_{j,n}= X_{n-j+1,n}$. The goodness-of-fit for the {\it complete data set} can be analyzed  choosing a specific value  $\theta_0=(\xi_0,\sigma_0)$ using the estimates of ($\xi,\sigma$) which were obtained  as a function of $k$, and by estimating the transformation $G$ using one (ii.b) step from the transformation algorithm ($A_{\mathcal T}$) with starting value ($\xi_0,\sigma_0$) and with a chosen value $m=n^a$ with $a \in (0,1)$ (slider). We then construct  transformed P-P plots
\begin{eqnarray}
 && \hspace{-1cm} \left(  -\log (1-\hat{F}_n (X_{n-j+1,n})); -\log \hat{G}(\bar{H}_{\theta_0}(X_{n-j+1,n})) \right)
 \nonumber  \\
 &=& 
 \left( \log {n+1 \over j};  -\log \hat{G}_n^{(m)}(\bar{H}_{\theta_0}(X_{n-j+1,n}))  \right), \; j=1,\ldots,n,
 \label{gof}
 \end{eqnarray} 
 where $\hat{F}_n$ denotes the empirical distribution function based on $X_{n-j+1,n}$ ($j=1,\ldots,n$). The closer the plot lies to the diagonal, the better the fit of the model defined by the survival function $\hat{G}^{(m)} (\bar{H}_{\theta_0})$.\\
 
 \noindent
 In Figure 10, the estimates of $\xi$ and $\mathbb{P}(X>4 \, 564 \,759)$ using the minimum variance principle are given, next the goodness-of-fit plot as defined in \eqref{gof} with $\xi_0=0.28$ and $m=n^{0.99}$.  The estimates of $\xi$ obtained from $Ep^+$ and $E\bar{p}^+$ are clearly most stable as a function of $k$ indicating the EVI value 0.4. The tail probability above the largest ultimate observation is also most stable for $Ep^+$ and $E\bar{p}^+$ indicating a value close to $1/n$ (indicated by the horizontal line). While the goodness-of-fit plot shows some deviations from the 45 degree line for the fitted transformation model, the presented overall global fit is useful (correlation equals 0.97). \\

 \noindent
 In Figure 11, for the winter rainfall data the results for $E\bar{p}$ with $m=57$, $k_*=43$ indicate two levels for $\xi$ (0.4 and ultimately at small $k$ close to 0), $\sigma$ (10 and 40 for small $k$) and tail probability $\mathbb{P}(X>162.05)$ (0.005 and 0.002 for small $k$, to be compared with $1/n=0.0018$), which indicates a change of statistical tail behaviour near the top. Method $Ep$ with $\rho=-1$ yields almost the same result as the classical GPD-ML method, except for the tail probability where it is quite unstable. Finally the
  transformation approach $T\bar{p}$ yields very stable plots at compromise values 0.2 for $\xi$, 10 for $\sigma$ and 0.003 for the tail probability. While the goodness-of-fit plot with $m=n^{0.99}$, $\sigma_0=10$ and $\xi_0=0.21$ has a correlation 0.997, the transformation approach seems to be unable to catch the deviating tail component 	near the top data with an EVI value close to 0.

\section{Conclusion} 

In this contribution we have constructed bias reduced estimators of tail parameters extending the classical POT method using the generalized Pareto distribution. The bias can be modelled parametrically (for instance based on second order regular variation theory), or non-parametrically using Bernstein polynomial approximations. A basic asymptotic limit theorem is provided for the estimators of the extreme value parameters which allows to compute asymptotic confidence intervals. A shinyapp has been constructed with which  the characteristics and the effectiveness of the proposed methods are illustrated  through simulations and practical case studies. From this it follows that within the proposed methods it is always possible to improve upon the classical POT method both in bias and RMSE.  

\section{Acknowledgments}  
\label{Sec5}
	\noindent This work is based on the research supported wholly/in part by the National Research Foundation of South Africa (Grant Number 102628) and the DST-NRF Centre of Excellence in Mathematical and Statistical Sciences (COE-Mass). The Grantholder acknowledges that opinions, findings and conclusions or recommendations expressed in any publication generated by the NRF supported research is that of the author(s), and that the NRF accepts no liability whatsoever in this regard. 


\section{Appendix}

\noindent
In this section we provide details concerning the proof of Theorem 1. 
\\

\noindent {\it Asymptotic distribution of 
$\hat{\xi}_k^{E+}$}. \\
From \eqref{systemE+} we obtain up to linear terms in $\delta_k$ that (denoting $\hat{\xi}_k$ for $\hat{\xi}^{E+}_k$)
\[
\left\{
\begin{array}{lcl}
\hat{\delta}_k &=& \frac{\sum_{j=1}^k b_\eta (Y_{j,k}^{-1/\hat{\xi}_k})}{\sum_{j=1}^k b^2_\eta (Y_{j,k}^{-1/\hat{\xi}_k})} \\
\hat{\xi}_k &=& H_{k,n} + \hat{\delta}_k B^{(1)}_{k},
\end{array}
\right.
\]
with $B^{(1)}_{k}= {1 \over k}\sum_{j=1}^k b'_\eta (Y_{j,k}^{-1/\hat{\xi}_k})Y_{j,k}^{-1/\hat{\xi}_k}\log Y_{j,k}$. As $k,n \to \infty$ and $k/n \to 0$ we have $B^{(1)}_{k} \to_p -\xi \int_0^1 b'_{\eta}(u) u \log u du = -\xi EB_\eta$.
\\
Using a Taylor expansion on the numerator of the right hand side of the first equation  leads to
\[
{1 \over k}\sum_{j=1}^k b_\eta (Y_{j,k}^{-1/\hat{\xi}_k})
=
{1 \over k}\sum_{j=1}^k b_\eta (Y_{j,k}^{-1/\xi})
- (\hat{\xi}_k - \xi) \xi^{-1} (EB_\eta) \, (1+o_p(1)), 
\]
so that, with ${1 \over k}\sum_{j=1}^k b^2_\eta (Y_{j,k}^{-1/\hat{\xi}_k}) \to_p E b^2_{\eta}$, up to lower order terms
\[
\hat{\delta}_k = {1 \over E b^2_{\eta}} {1 \over k}\sum_{j=1}^k b_\eta (Y_{j,k}^{-1/\xi})
- (\hat{\xi}_k - \xi) \xi^{-1}\frac{EB_\eta }{E b^2_{\eta}} \, (1+o_p(1)).
\]
Hence, inserting this expansion into $\hat{\xi}_k = H_{k,n} + \hat{\delta}_k B^{(1)}_{k}$, 
finally leads to 
\begin{eqnarray*}
\sqrt{k}(\hat{\xi}_k - \xi)(1+o_p(1)) &=&
\frac{E b^2_{\eta}}{E b^2_{\eta}-(EB_\eta)^2} \sqrt{k}\left(H_{k,n}-\xi \right)
-
\frac{\xi EB_\eta }{E b^2_{\eta}-(EB_\eta)^2 }\sqrt{k} \left( {1 \over k}\sum_{j=1}^k b_\eta (Y_{j,k}^{-1/\xi})\right)\\
&=& \frac{E b^2_{\eta}}{E b^2_{\eta}-(EB_\eta)^2} \sqrt{k}\left(H_{k,n}-\xi - \xi \delta_k EB_\eta\right) 
\\ && \;\;
-
\frac{\xi EB_\eta }{E b^2_{\eta}-(EB_\eta)^2 }\sqrt{k} \left( {1 \over k}\sum_{j=1}^k b_\eta (Y_{j,k}^{-1/\xi})
- \delta_k E b^2_{\eta}
\right),
\end{eqnarray*}
with $\delta_k = \delta (U(n/k))$.
We now show  that this final expression is a linear combination of two zero centered statistics (up to the required accuracy)  which is asymptotically  normal with the stated asymptotic variance.  To this end let $Z_{n-k,n} \leq Z_{n-k+1,n} \leq \ldots \leq Z_{n,n}$ denote the top $k+1$ order statistics of a sample of size $n$ from the standard Pareto distribution with distribution function $z \mapsto z^{-1}$, $z>1$. Then from $({\tilde{\cal{E}}}_2^+) $
\begin{eqnarray*}
H_{k,n} &=& {1 \over k}\sum_{j=1}^k
\left(\log U(Z_{n-j+1,n})- \log U(Z_{n-k,n}) \right) \\
&=& {1 \over k}\sum_{j=1}^k \log \left\{
\left({Z_{n-j+1,n}\over Z_{n-k,n}}\right)^\xi 
\left[
1+\xi \delta (U(Z_{n-k,n}))B_\eta \left( {Z_{n-k,n} \over Z_{n-j+1,n}}\right) \right. \right. \\
&& \hspace{4.6cm}\left.\left.
+o_p(1)|\delta (U(Z_{n-k,n}))| |B_\eta \left( {Z_{n-k,n} \over Z_{n-j+1,n}}\right)| \left({Z_{n-j+1,n}\over Z_{n-k,n}}\right)^\epsilon 
\right]
\right\}\\
&=& \xi {1 \over k}\sum_{j=1}^k \log {Z_{n-j+1,n}\over Z_{n-k,n}}
+ \xi \delta (U(Z_{n-k,n}))B_\eta \left( {Z_{n-k,n} \over Z_{n-j+1,n}}\right)\\
&& \hspace{3.5cm}
+o_p(1)|\delta (U(Z_{n-k,n}))| |B_\eta \left( {Z_{n-k,n} \over Z_{n-j+1,n}}\right)| \left({Z_{n-j+1,n}\over Z_{n-k,n}}\right)^\epsilon. 
\end{eqnarray*}
Now $\log Z_{n-j+1,n}-\log Z_{n-k,n} =_d E_{k-j+1,k}$, the $(k-j+1)$th smallest value from a standard exponential sample $E_1,\ldots,E_k$ of size $k$, so that ${1 \over k}\sum_{j=1}^k \log {Z_{n-j+1,n}\over Z_{n-k,n}}=_d {1 \over k}\sum_{j=1}^k E_j$ and ${1 \over k}\sum_{j=1}^k B_\eta \left( {Z_{n-k,n} \over Z_{n-j+1,n}}\right) =_d {1 \over k}\sum_{j=1}^k B_\eta (e^{-E_j})=_d {1 \over k}\sum_{j=1}^k B_\eta (U_j)$ where $U_1,\ldots,U_k$ is a uniform (0,1) sample. Hence, since $\delta (U(Z_{n-k,n})) / \delta (U(n/k)) \to_p 1$ and  ${1 \over k}\sum_{j=1}^k B_\eta (U_j) \to_p EB_{\eta}$, we have that $H_{k,n} -\xi - \xi \delta_k EB_\eta$ is asymptotically equivalent to $ {1 \over k}\sum_{j=1}^k \xi (E_j-1)$ as $\sqrt{k} \delta_k \to \lambda$. \\
Similarly 
\begin{eqnarray*}
{1 \over k}\sum_{j=1}^k b_\eta (Y_{j,k}^{-1/\xi}) &=&
{1 \over k}\sum_{j=1}^k b_\eta \left(\left[ {U \left({Z_{n-j+1,n}\over Z_{n-k,n}} Z_{n-k,n} \right)\over U(Z_{n-k,n}) }\right]^{-1/\xi} \right)\\
&=& {1 \over k}\sum_{j=1}^k b_\eta
\left(
\left({Z_{n-j+1,n}\over Z_{n-k,n}}\right)^{-1}
\left[
1+\xi \delta (U(Z_{n-k,n}))B_\eta \left( {Z_{n-k,n} \over Z_{n-j+1,n}}\right) \right. \right. \\
&& \hspace{2.5cm}\left.\left.
+o_p(1)|\delta (U(Z_{n-k,n}))| |B_\eta \left( {Z_{n-k,n} \over Z_{n-j+1,n}}\right)| \left({Z_{n-j+1,n}\over Z_{n-k,n}}\right)^\epsilon 
\right]^{-1/\xi}
\right)\\
&=&{1 \over k}\sum_{j=1}^k b_\eta
\left(
\left({Z_{n-j+1,n}\over Z_{n-k,n}}\right)^{-1}
\left[
1-\delta (U(Z_{n-k,n}))B_\eta \left( {Z_{n-k,n} \over Z_{n-j+1,n}}\right) \right. \right. \\
&& \hspace{2.5cm}\left.\left.
+o_p(1)|\delta (U(Z_{n-k,n}))| |B_\eta \left( {Z_{n-k,n} \over Z_{n-j+1,n}}\right)| \left({Z_{n-j+1,n}\over Z_{n-k,n}}\right)^\epsilon 
\right]
\right)
\\
&=& {1 \over k}\sum_{j=1}^k b_\eta (e^{-E_j})\\
&& \hspace{0.3cm}
-\delta (U(Z_{n-k,n})){1 \over k}\sum_{j=1}^k 
b'_\eta \left( {Z_{n-k,n} \over Z_{n-j+1,n}}\right)
B_\eta \left( {Z_{n-k,n} \over Z_{n-j+1,n}}\right) \left( {Z_{n-k,n} \over Z_{n-j+1,n}}\right)(1+o_p(1)).
\end{eqnarray*}
Since $\delta (U(Z_{n-k,n}))/\delta_k \to_p 1$
and ${1 \over k}\sum_{j=1}^k 
b'_\eta \left( {Z_{n-k,n} \over Z_{n-j+1,n}}\right)
B_\eta \left( {Z_{n-k,n} \over Z_{n-j+1,n}}\right) \left( {Z_{n-k,n} \over Z_{n-j+1,n}}\right) \to_p - Eb^2_\eta$
it follows that 
$
{1 \over k}\sum_{j=1}^k b_\eta (Y_{j,k}^{-1/\xi})
- \delta_k E b^2_{\eta}$
is asymptotically equivalent to ${1 \over k}\sum_{j=1}^k b_\eta (e^{-E_j})=_d {1 \over k}\sum_{j=1}^k b_\eta (U_j)$ as $\sqrt{k} \delta_k \to \lambda$, which is centered at 0 since $E(b_\eta (U))=0$.

\vspace{0.3cm}\noindent
{\it Asymptotic distribution of 
$\hat{\xi}_k^{E}$}. \\ This derivation follows similar lines starting from \eqref{systemE}: 
 \[
\left\{
\begin{array}{l}
{1 \over k}\sum_{j=1}^k b'_{\eta}(\bar{H}_{\hat{\theta}_k}(Y_{j,k})) \bar{H}_{\hat{\theta}_k}(Y_{j,k}) \log (1+\hat{\tau}_k Y_{j,k}) \to_p -\xi EB_\eta, \\
{1 \over k}\sum_{j=1}^k b^2_\eta (\bar{H}_{\hat{\theta}_k}(Y_{j,k})) \to_p Eb^2_{\eta}, \\
{1\over k}\sum_{j=1}^k b'_{\eta}(\bar{H}_{\hat{\theta}_k}(Y_{j,k}))\bar{H}_{\hat{\theta}_k}(Y_{j,k}) \to_p b_\eta (1), \\
{1\over k}\sum_{j=1}^k b'_{\eta}(\bar{H}_{\hat{\theta}_k}(Y_{j,k}))\bar{H}_{\hat{\theta}_k}(Y_{j,k}) {1 \over 1+\hat{\tau}_k Y_{j,k}}
\to_p  \xi(1+\xi) EC_{\eta} + b_\eta (1),
\end{array}
\right.
\]
as $k,n\to \infty$ and $k/n\to \infty$,
so that the system of equations is asymptotically equivalent to
\[
\left\{
\begin{array}{l}
\hat{\delta}_k  = \frac{{1 \over k}\sum_{j=1}^k b_\eta (\bar{H}_{\hat{\theta}_k}(Y_{j,k}))}{Eb^2_{\eta}}, \\
{1 \over k}\sum_{j=1}^k \log (1+\hat{\tau}_k Y_{j,k}) = 
\hat{\xi}_k + \hat{\xi}_k\hat{\delta}_kEB_{\eta}
\\
{1 \over k}\sum_{j=1}^k {1 \over 1+\hat{\tau}_k Y_{j,k}} =
{1 \over 1+ \hat{\xi}_k}- \hat{\xi}_k\hat{\delta}_k EC_{\eta}.
\end{array}
\right.
\]
Using a Taylor expansion on the numerator of the right hand side of the first equation leads to
\[
\hat{\delta}_k Eb^2_{\eta}=
{1 \over k } \sum_{j=1}^kb_{\eta}(\bar{H}_{\theta}(Y_{j,k}))
- {EB_{\eta} \over \xi}(\hat{\xi}_k -\xi) 
+ (1+\xi) EC_{\eta} \left({\hat{\tau}_k \over \tau}-1 \right).
\]
Imputing this in the second and third equation in $\xi$ and $\tau$, and expanding these equations linearly around the correct values ($\xi,\tau$), while using, as $k,n \to \infty$ and $k/n \to 0$ 
\[
{1 \over k}\sum_{j=1}^k {\tau Y_{j,k} \over 1+\tau Y_{j,k}}
\to_p {\xi \over 1+\xi} \mbox{ and }{1 \over k}\sum_{j=1}^k {\tau Y_{j,k} \over (1+\tau Y_{j,k})^2} \to_p {\xi \over (1+\xi)(1+2\xi)}, \] 
leads to the linearized equations 
\begin{equation}
\left\{
\begin{array}{l}
\left(\hat{\xi}_k - \xi \right)\left(-1+ \frac{(EB_{\eta})^2}{Eb^2_{\eta}}\right)
+
\left({\hat{\tau}_k \over \tau}- 1 \right)\left({\xi \over 1+\xi} -\xi(1+\xi) \frac{EB_{\eta}\, EC_{\eta}}{Eb^2_{\eta}}\right) \\
\hspace{3cm} = - \left( {1 \over k}\sum_{j=1}^k \log (1+\tau Y_{j,k})-\xi \right)
+ {\xi EB_{\eta}\over Eb^2_{\eta}} {1 \over k}\sum_{j=1}^k
b_{\eta}(\bar{H}_{\theta}(Y_{j,k})), \\
\\
\left(\hat{\xi}_k - \xi \right)\left({1 \over (1+\xi)^2 }- \frac{EB_{\eta}EC_{\eta}}{Eb^2_{\eta}}\right)
+
\left({\hat{\tau}_k \over \tau}- 1 \right)\left(-{\xi \over (1+\xi)(1+2\xi)} +\xi(1+\xi) \frac{ (EC_{\eta})^2}{Eb^2_{\eta}}\right) \\
\hspace{3cm} = - \left( {1 \over k}\sum_{j=1}^k {1 \over 1+\tau Y_{j,k}} -{1 \over 1+\xi} \right)
- {\xi EC_{\eta}\over Eb^2_{\eta}} {1 \over k}\sum_{j=1}^k
b_{\eta}(\bar{H}_{\theta}(Y_{j,k})).
\end{array}
\right.
\label{lineqs}
\end{equation}
It follows that the right hand sides in \eqref{lineqs} can be rewritten as  linear combination of two zero centered  statistics from which the asymptotic normality of $\left( \sqrt{k}(\hat{\xi}^{E}_k-\xi
), \sqrt{k}({\hat{\tau}^{E}_k \over \tau}-1) \right)$ can be obtained, as stated in Theorem 1: 
\begin{equation*}
\left\{
\begin{array}{l}
\left(\hat{\xi}_k - \xi \right)\left(-1+ \frac{(EB_{\eta})^2}{Eb^2_{\eta}}\right)
+
\left({\hat{\tau}_k \over \tau}- 1 \right)\left({\xi \over 1+\xi} -\xi(1+\xi) \frac{EB_{\eta}\, EC_{\eta}}{Eb^2_{\eta}}\right) \\
\hspace{1.5cm} = - \left( {1 \over k}\sum_{j=1}^k \log (1+\tau Y_{j,k})-\xi - \xi\delta_k EB_{\eta} \right)
+ {\xi EB_{\eta}\over Eb^2_{\eta}} \left({1 \over k}\sum_{j=1}^k
b_{\eta}(\bar{H}_{\theta}(Y_{j,k}))-\delta_k Eb^2_{\eta} \right), \\
\\
\left(\hat{\xi}_k - \xi \right)\left({1 \over (1+\xi)^2 }- \frac{EB_{\eta}EC_{\eta}}{Eb^2_{\eta}}\right)
+
\left({\hat{\tau}_k \over \tau}- 1 \right)\left(-{\xi \over (1+\xi)(1+2\xi)} +\xi(1+\xi) \frac{ (EC_{\eta})^2}{Eb^2_{\eta}}\right) \\
\hspace{1.5cm} = - \left( {1 \over k}\sum_{j=1}^k {1 \over 1+\tau Y_{j,k}} -{1 \over 1+\xi} +  \xi\delta_k EC_{\eta}\right)
- {\xi EC_{\eta}\over Eb^2_{\eta}} \left({1 \over k}\sum_{j=1}^k
b_{\eta}(\bar{H}_{\theta}(Y_{j,k}))-\delta_k Eb^2_{\eta}\right).
\end{array}
\right.
\end{equation*}
This is done using similar derivations as in the case $\hat{\xi}_k^{E+}$.

\newpage

 \begin{figure}[!ht]
  \centering
\includegraphics[width=0.45\textwidth]{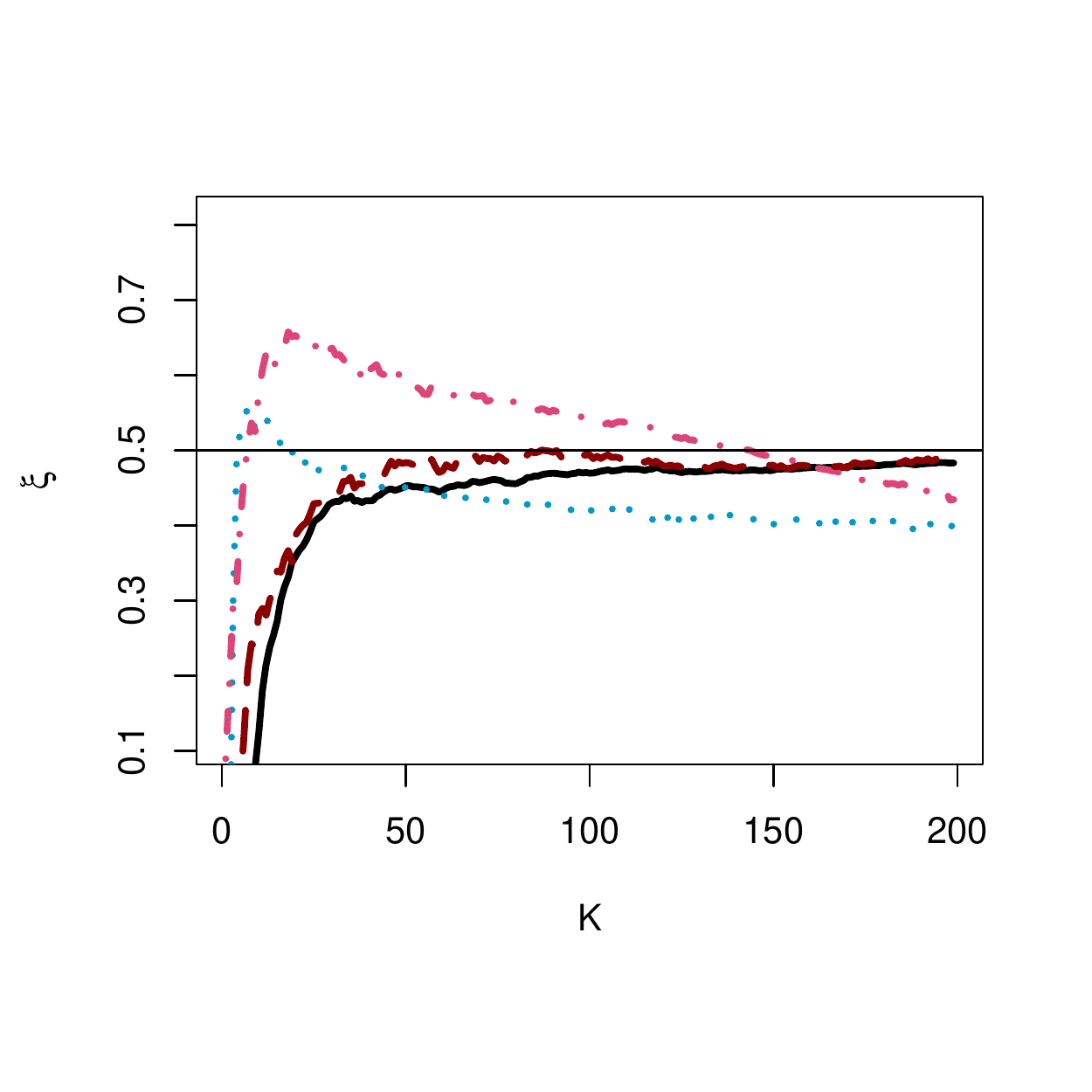} 
\includegraphics[width=0.45\textwidth]{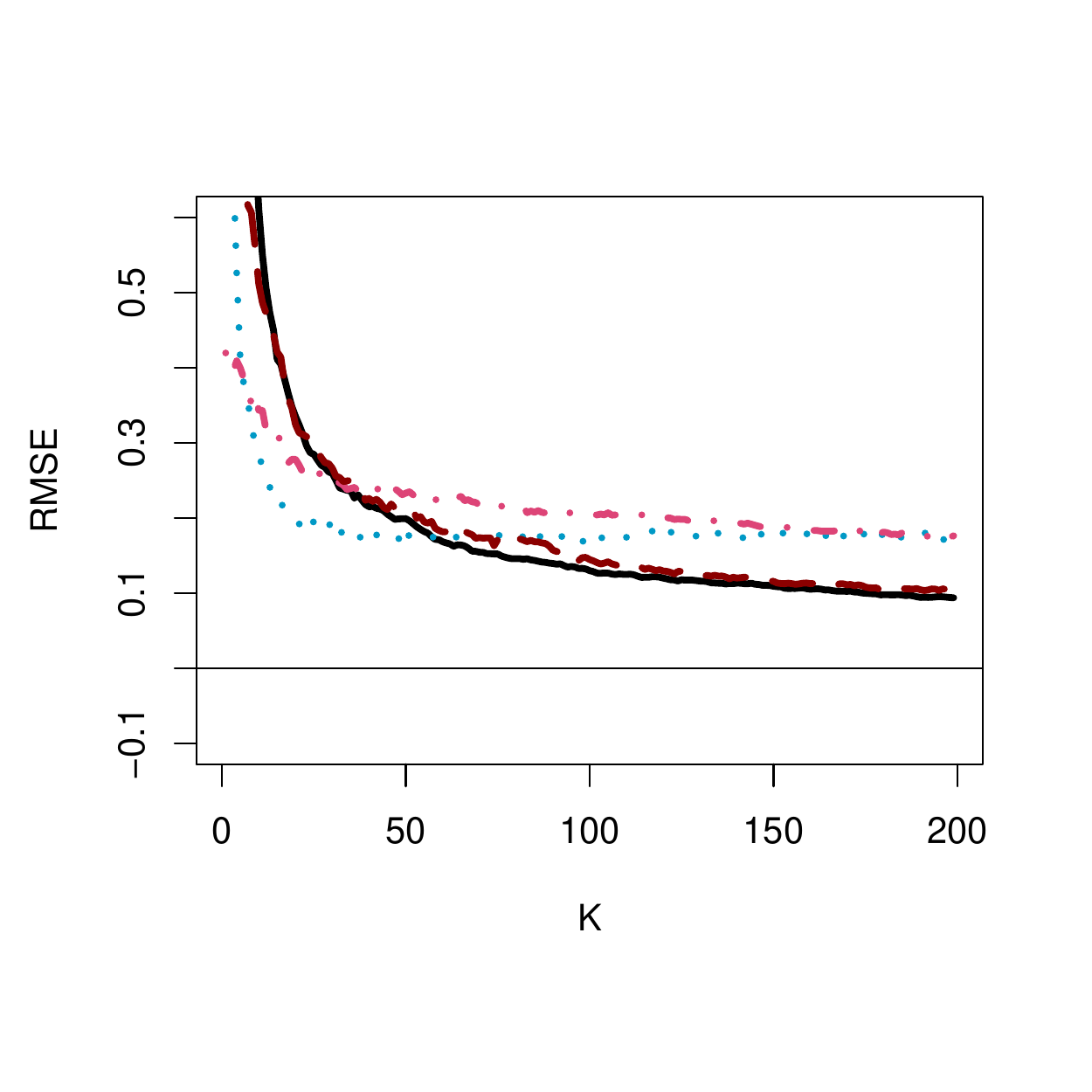} \\
\includegraphics[width=0.45\textwidth]{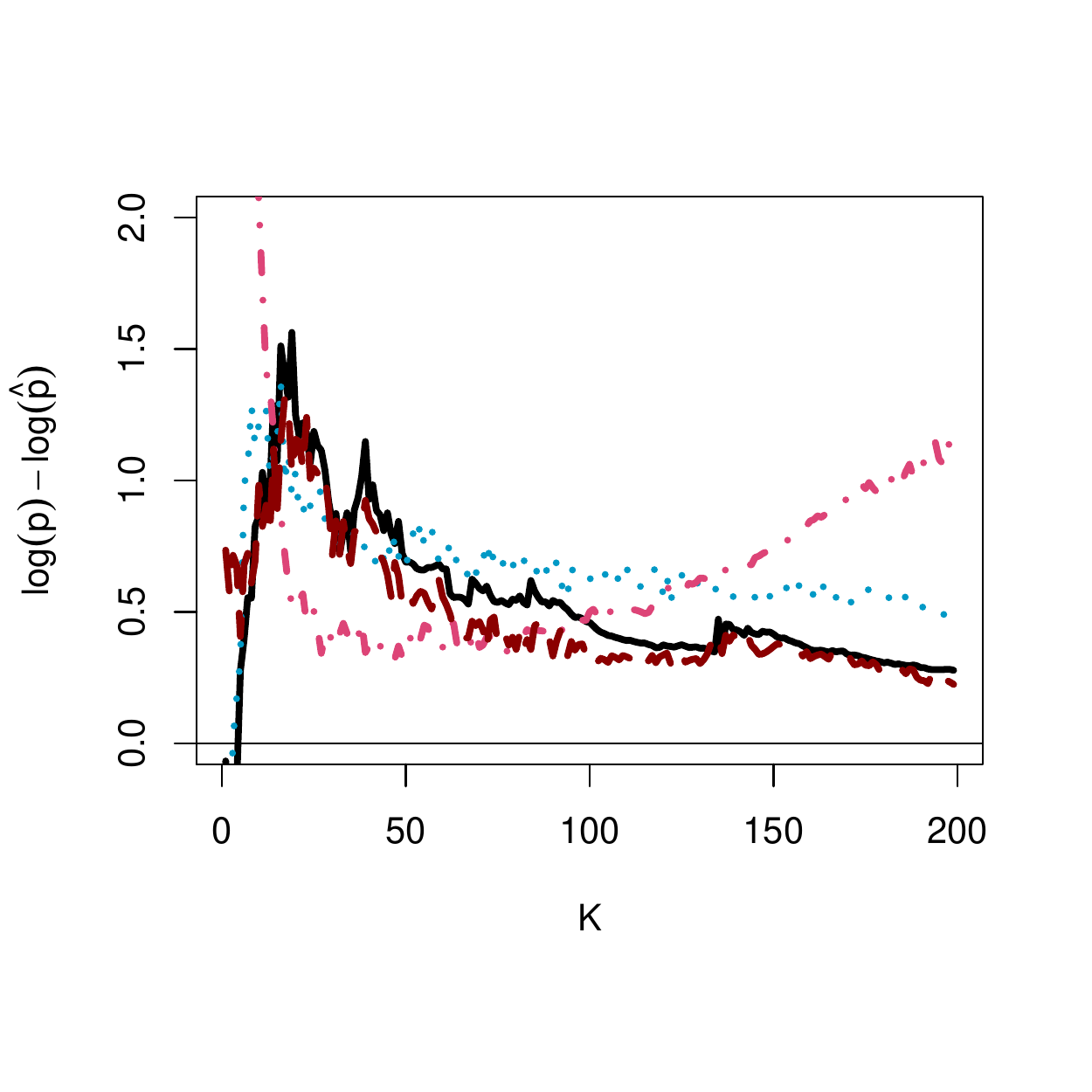}
\includegraphics[width=0.45\textwidth]{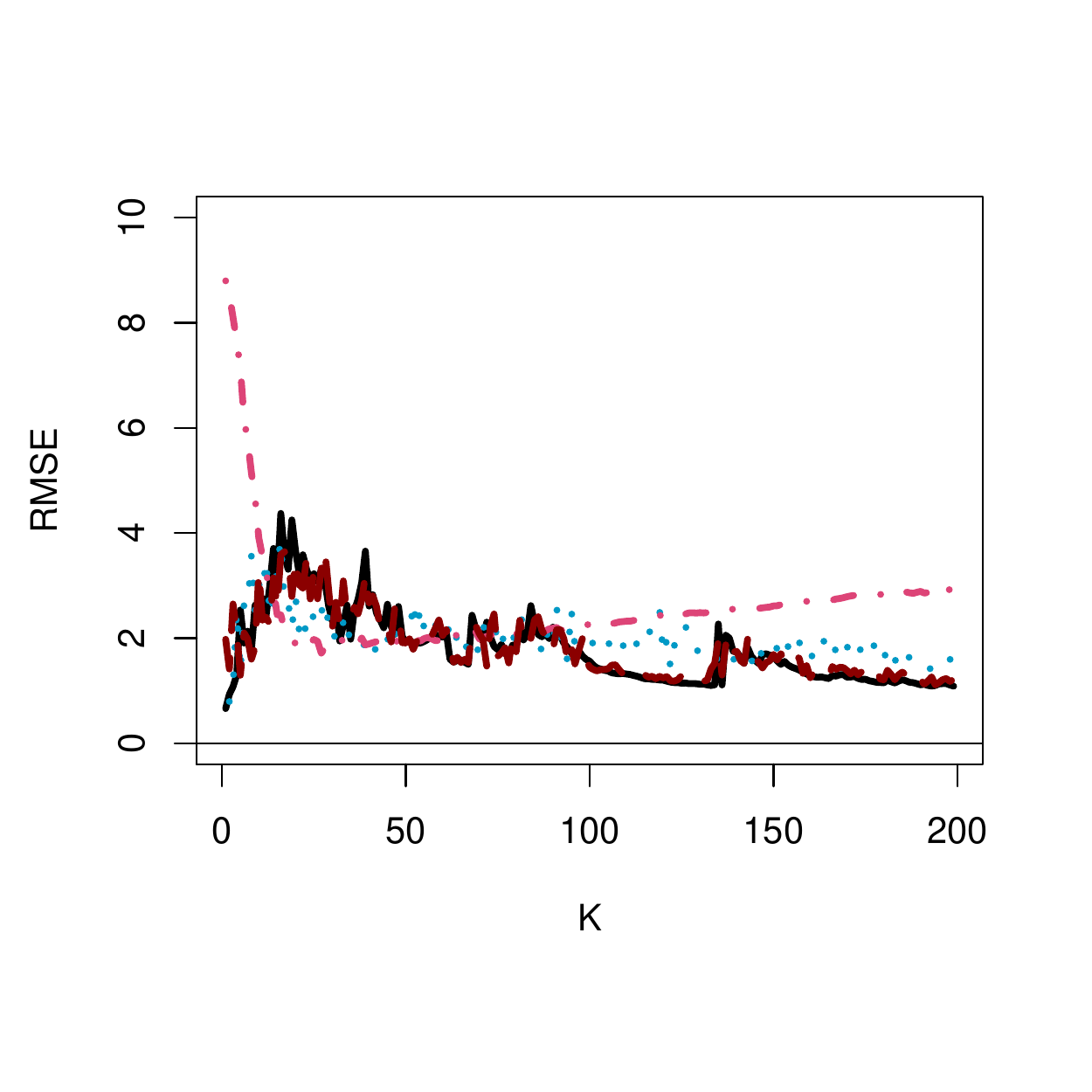}  
 \caption{ Burr distribution with $\xi=0.5$ and $\rho=-0.5$. Estimation of $\xi$ (top) and tail probability (bottom) using minimum variance principle, bias (left), RMSE (right): GPD-ML (full line), $T\bar{p}$ (dotted), $Ep$ (dash-dotted) and $E\bar{p}$ (dashed). }
\end{figure}
 \begin{figure}[!ht]
  \centering
  \includegraphics[width=0.45\textwidth]{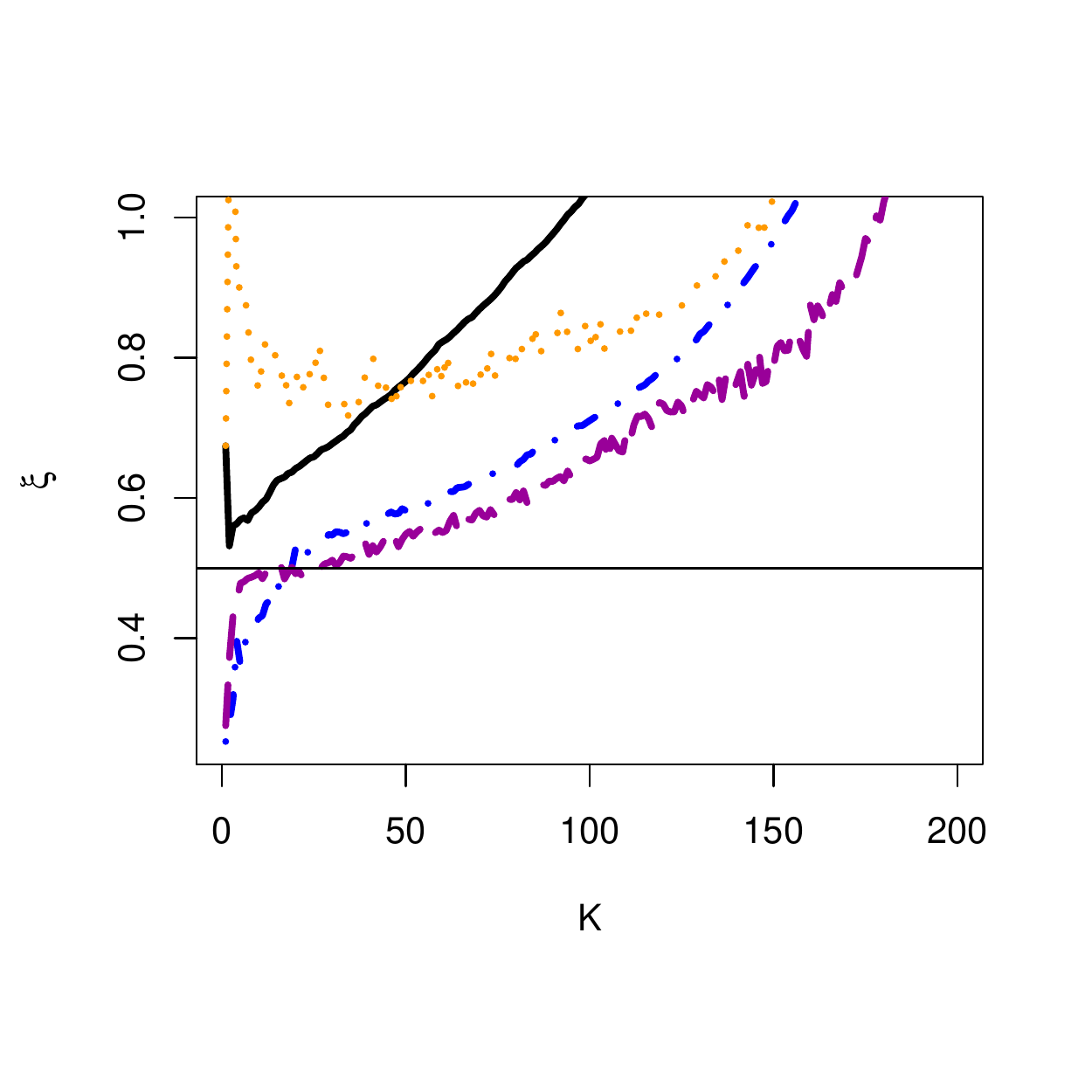} 
\includegraphics[width=0.45\textwidth]{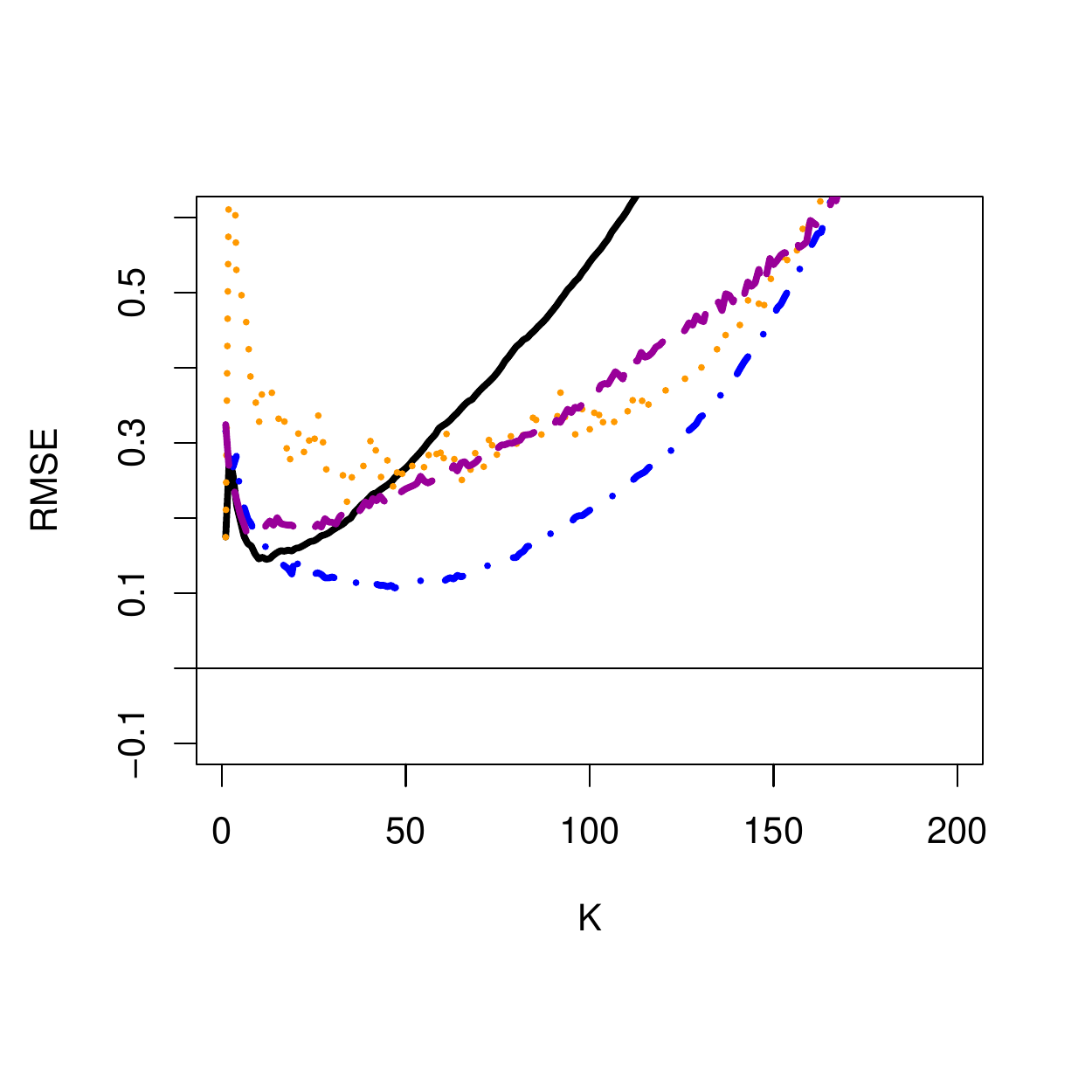} \\
\includegraphics[width=0.45\textwidth]{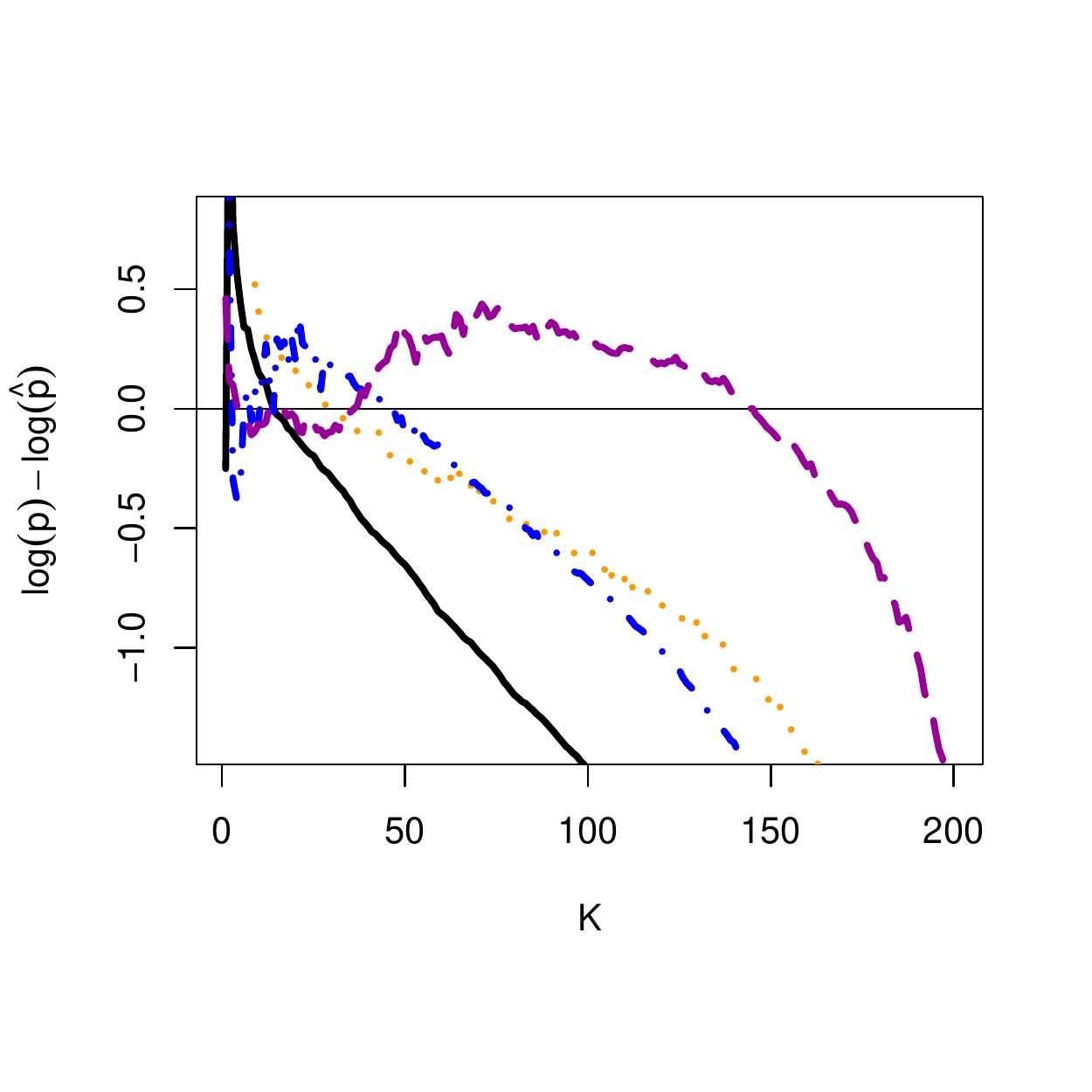}
\includegraphics[width=0.45\textwidth]{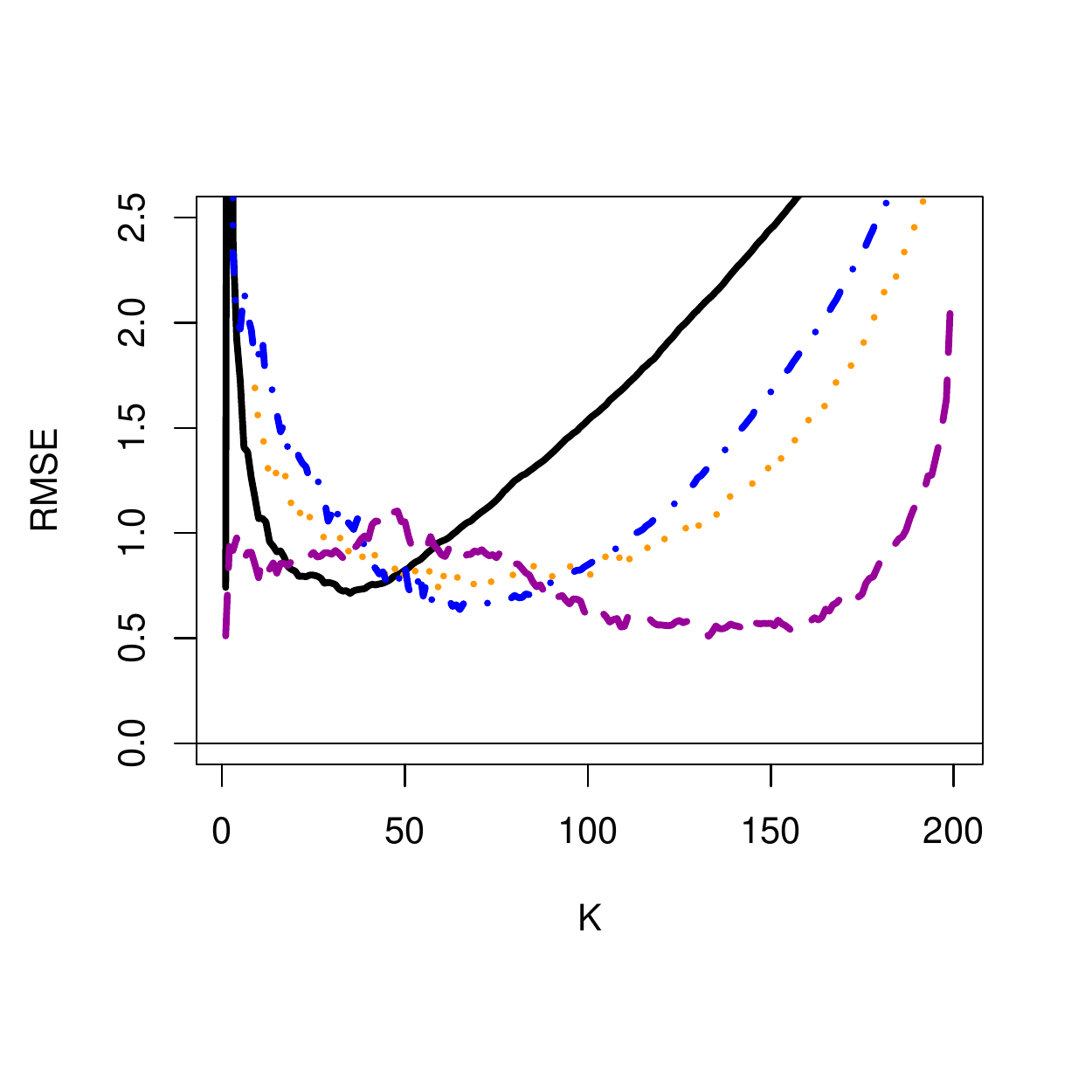} 
\caption{Burr distribution with $\xi=0.5$ and $\rho=-0.5$. Estimation of $\xi$ (top) and tail probability (bottom) using minimum variance principle, bias (left), RMSE (right): Pareto-ML (full line), $T\bar{p}^+$ (dotted), $Ep^+$ (dash-dotted) and $E\bar{p}^+$ (dashed).}
  \end{figure}
  
  \begin{figure}[!ht]
  \centering
\includegraphics[width=0.45\textwidth]{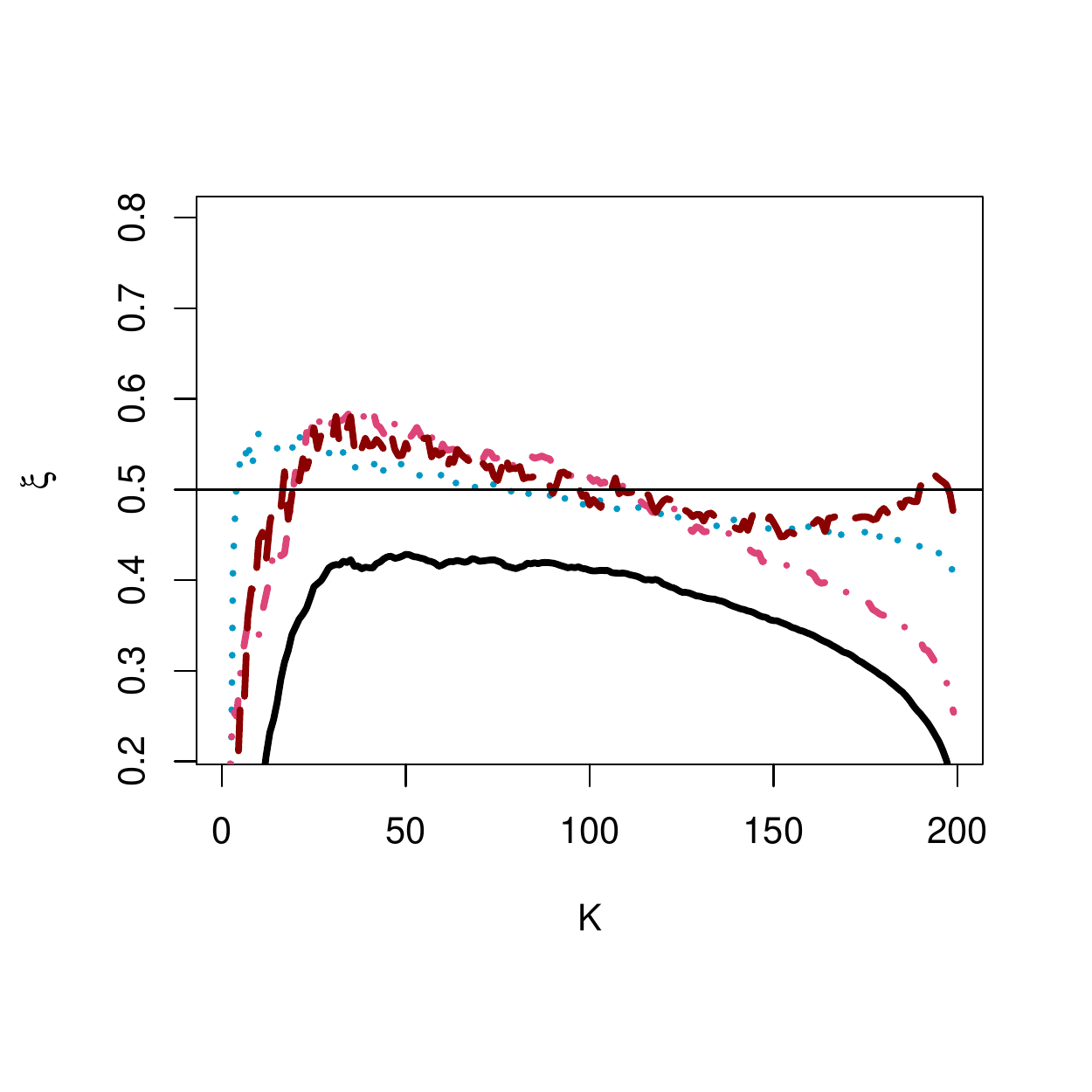} 
\includegraphics[width=0.45\textwidth]{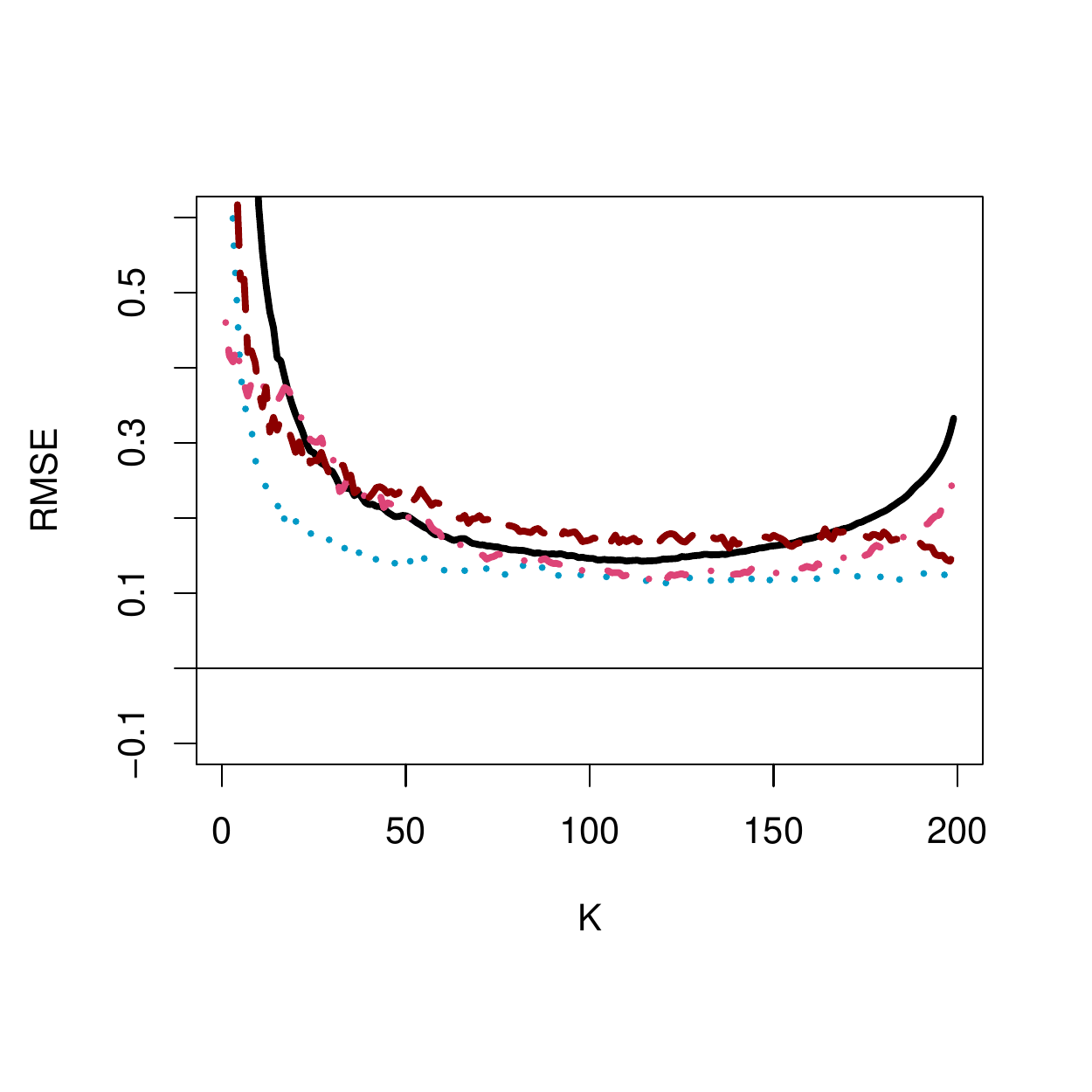} \\
\includegraphics[width=0.45\textwidth]{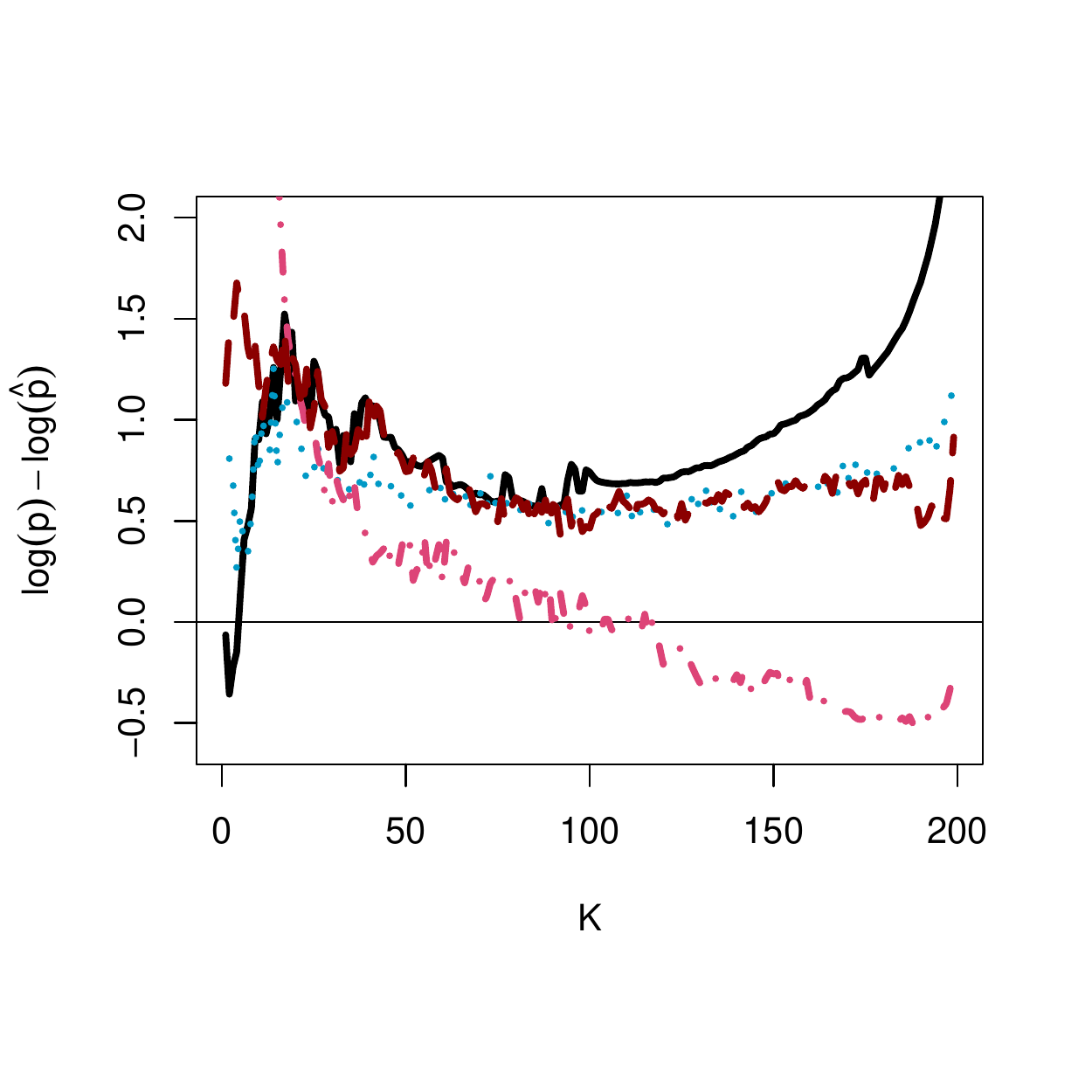}
\includegraphics[width=0.45\textwidth]{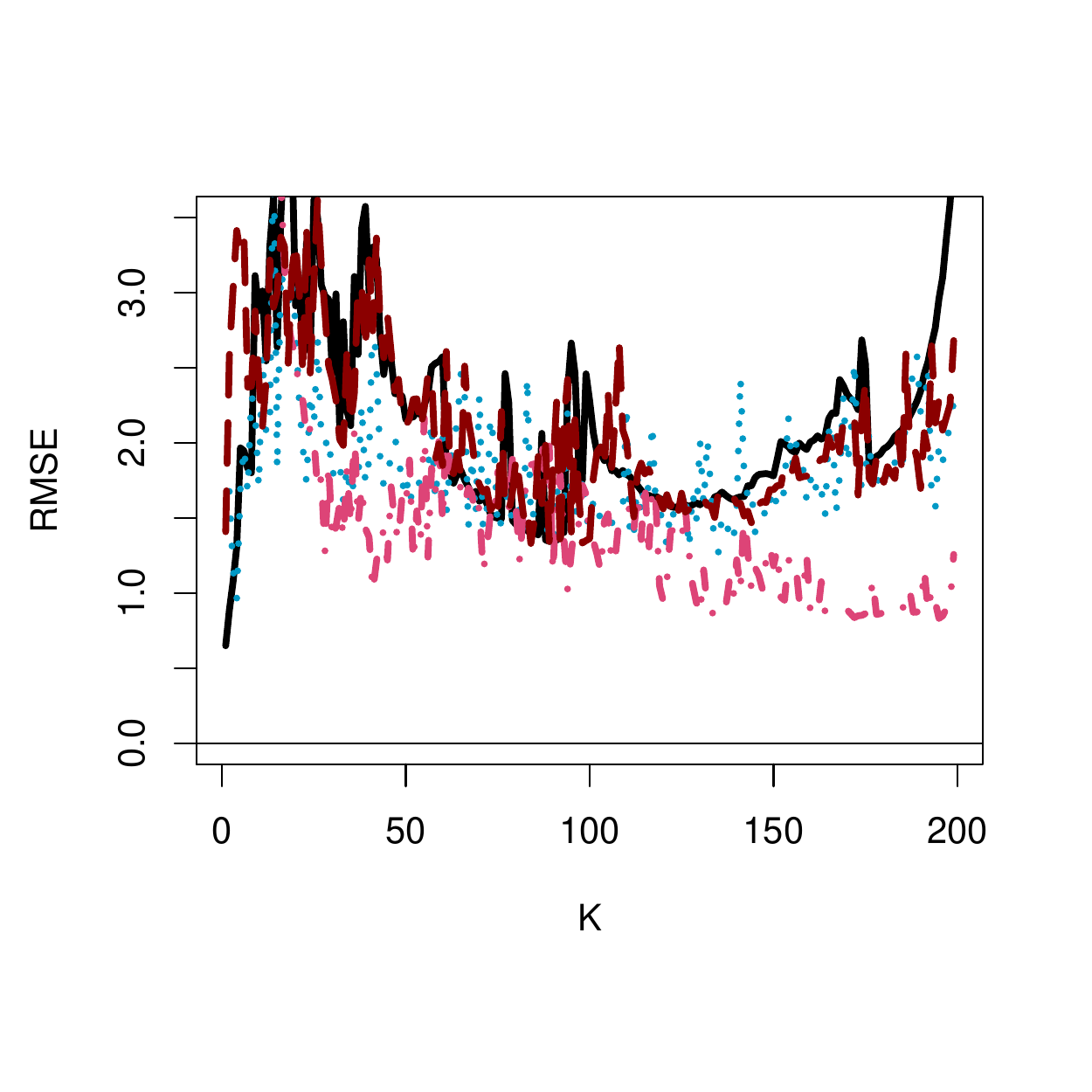}  
 \caption{ Fr\'echet distribution with $\xi=0.5$. Estimation of $\xi$ (top) and tail probability (bottom), bias (left), RMSE (right): GPD-ML (full line), $T\bar{p}$ (dotted, using minimum variance principle), $Ep$ with $\rho=-2$ (dash-dotted) and $E\bar{p}$ with $(k_*,m)=(190,150)$ (dashed). }
\end{figure}
 \begin{figure}[!ht]
  \centering
  \includegraphics[width=0.45\textwidth]{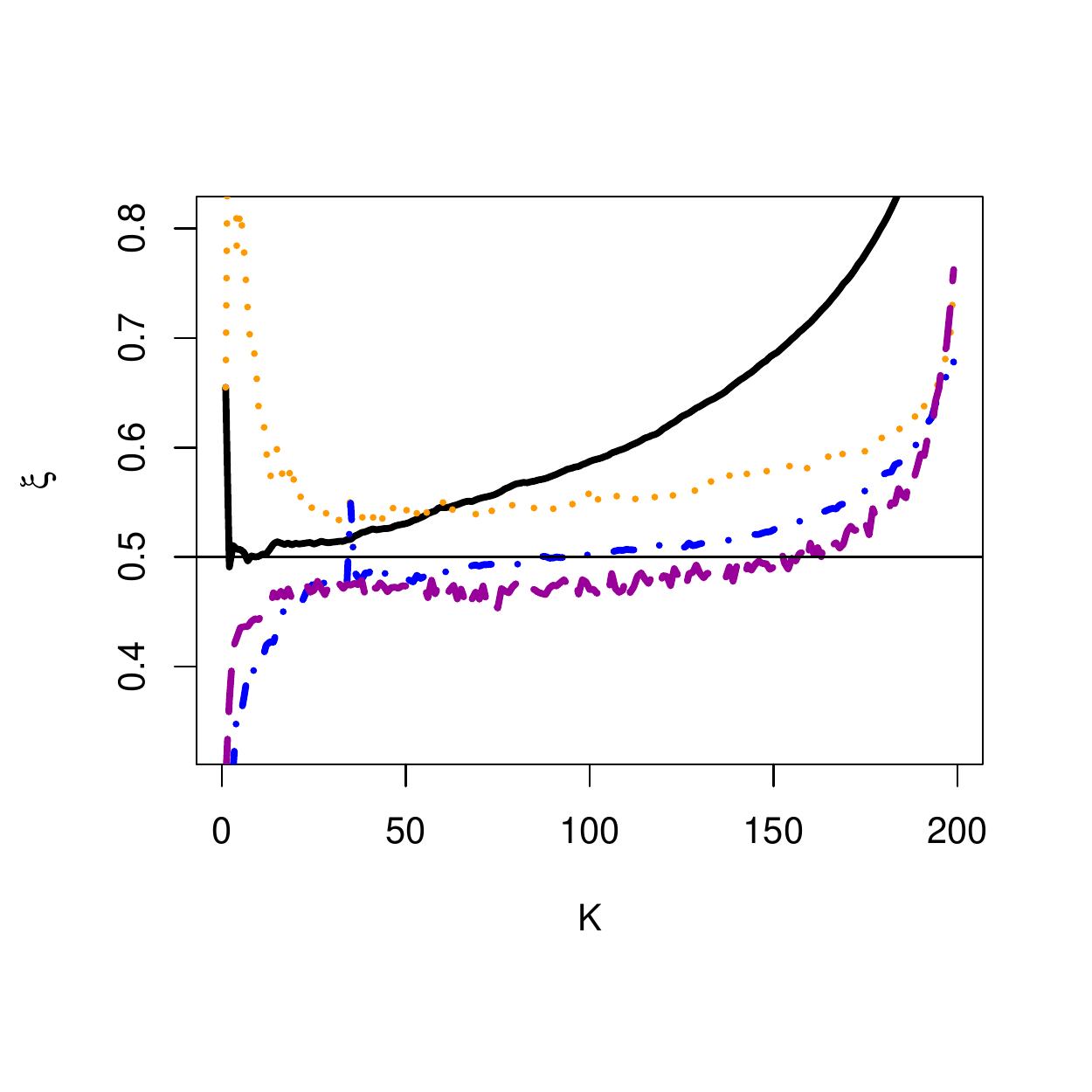} 
\includegraphics[width=0.45\textwidth]{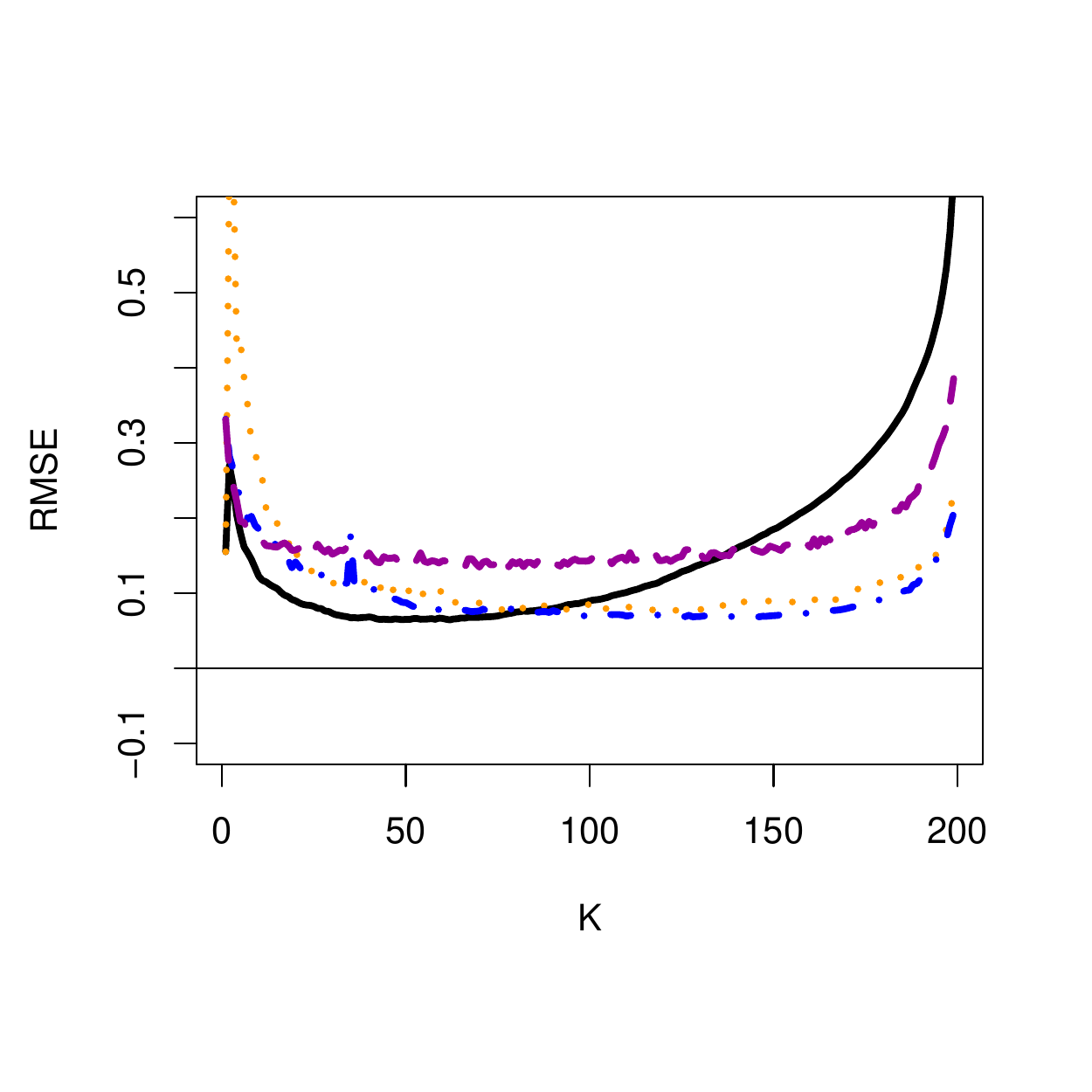} \\
\includegraphics[width=0.45\textwidth]{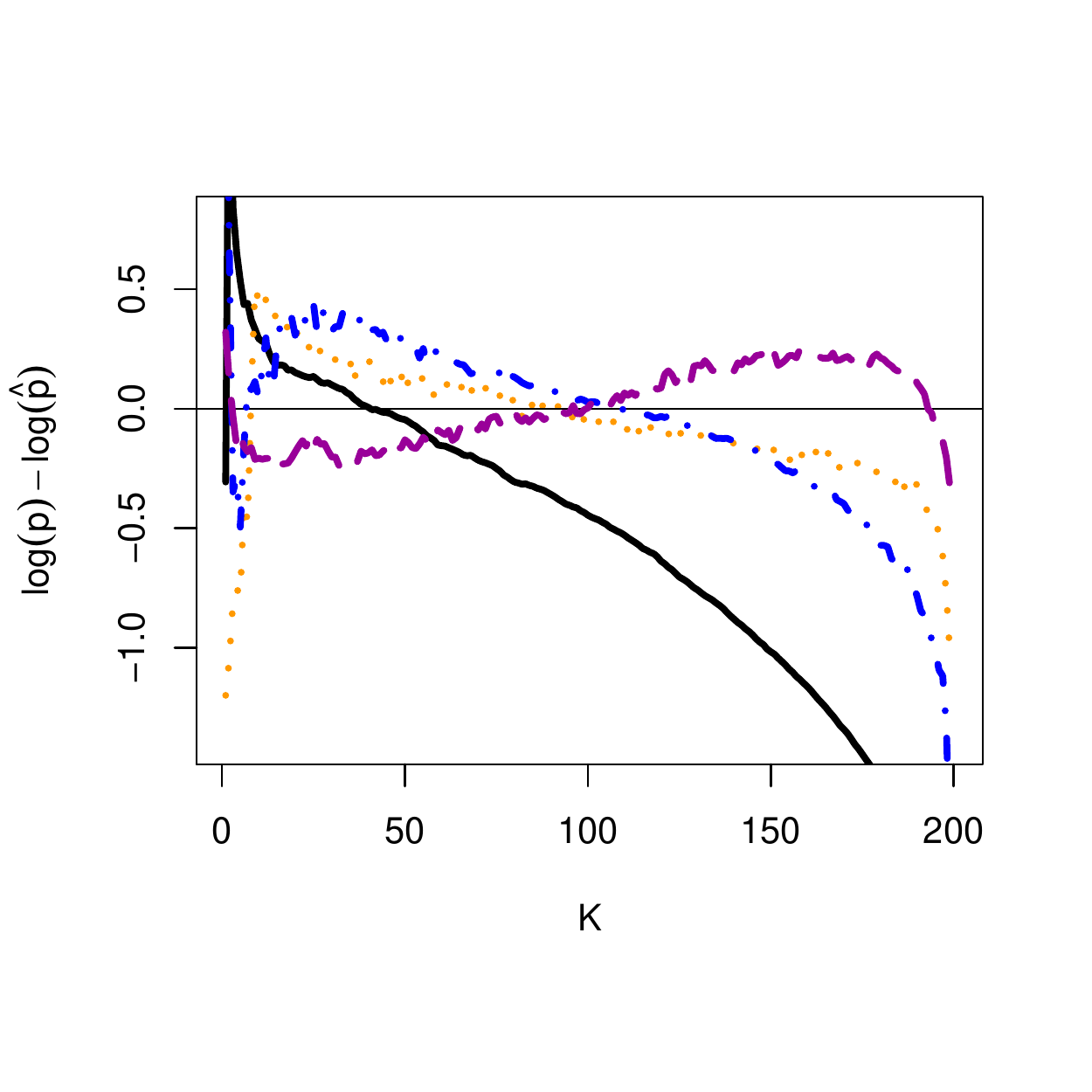}
\includegraphics[width=0.45\textwidth]{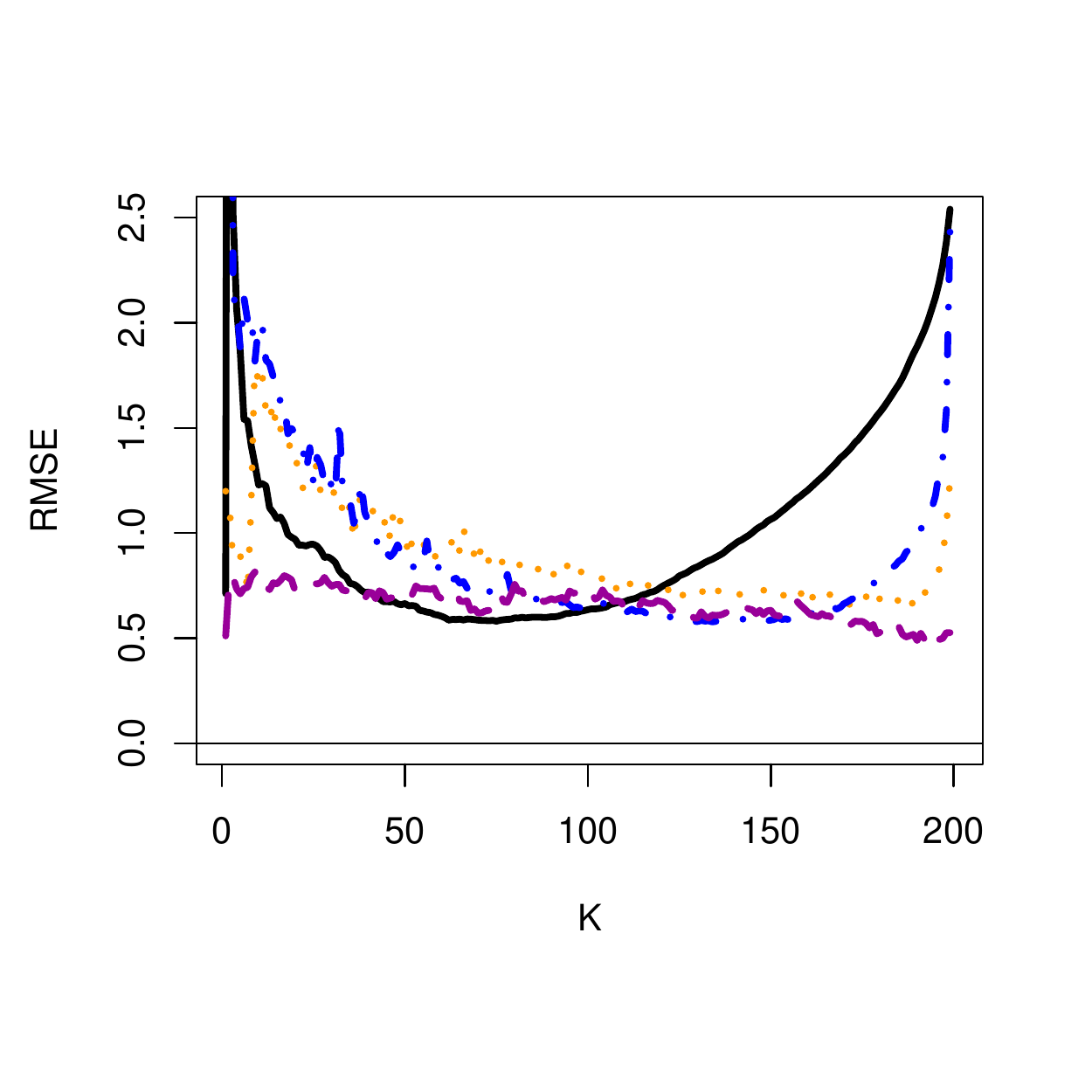} 
\caption{Fr\'echet distribution with $\xi=0.5$. Estimation of $\xi$ (top) and tail probability (bottom) using minimum variance principle, bias (left), RMSE (right): Pareto-ML  (full line), $T\bar{p}^+$ (dotted), $Ep^+$ (dash-dotted) and $E\bar{p}^+$ (dashed).}
  \end{figure}
  \begin{figure}[!ht]
  \centering
\includegraphics[width=0.45\textwidth]{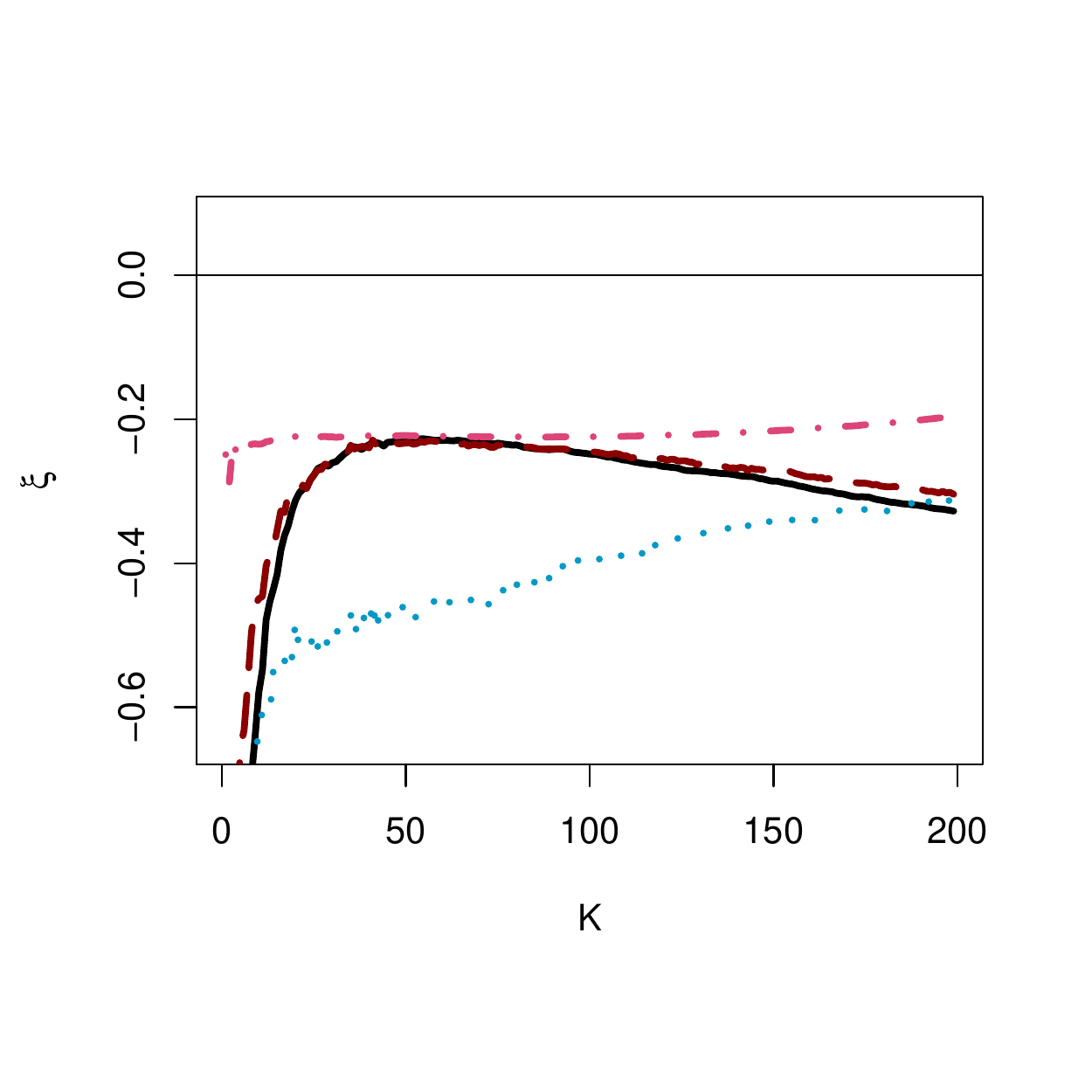} 
\includegraphics[width=0.45\textwidth]{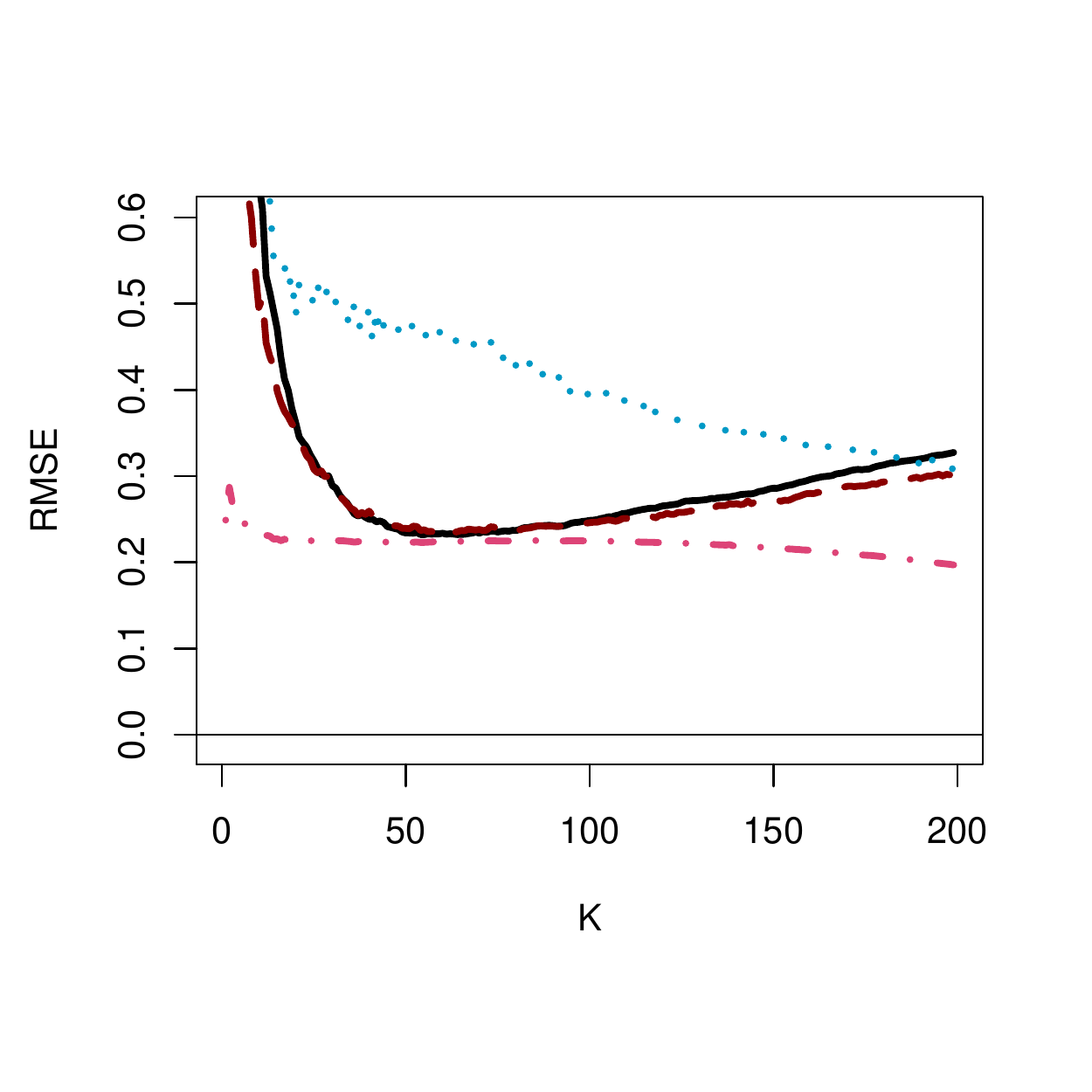} \\
\includegraphics[width=0.45\textwidth]{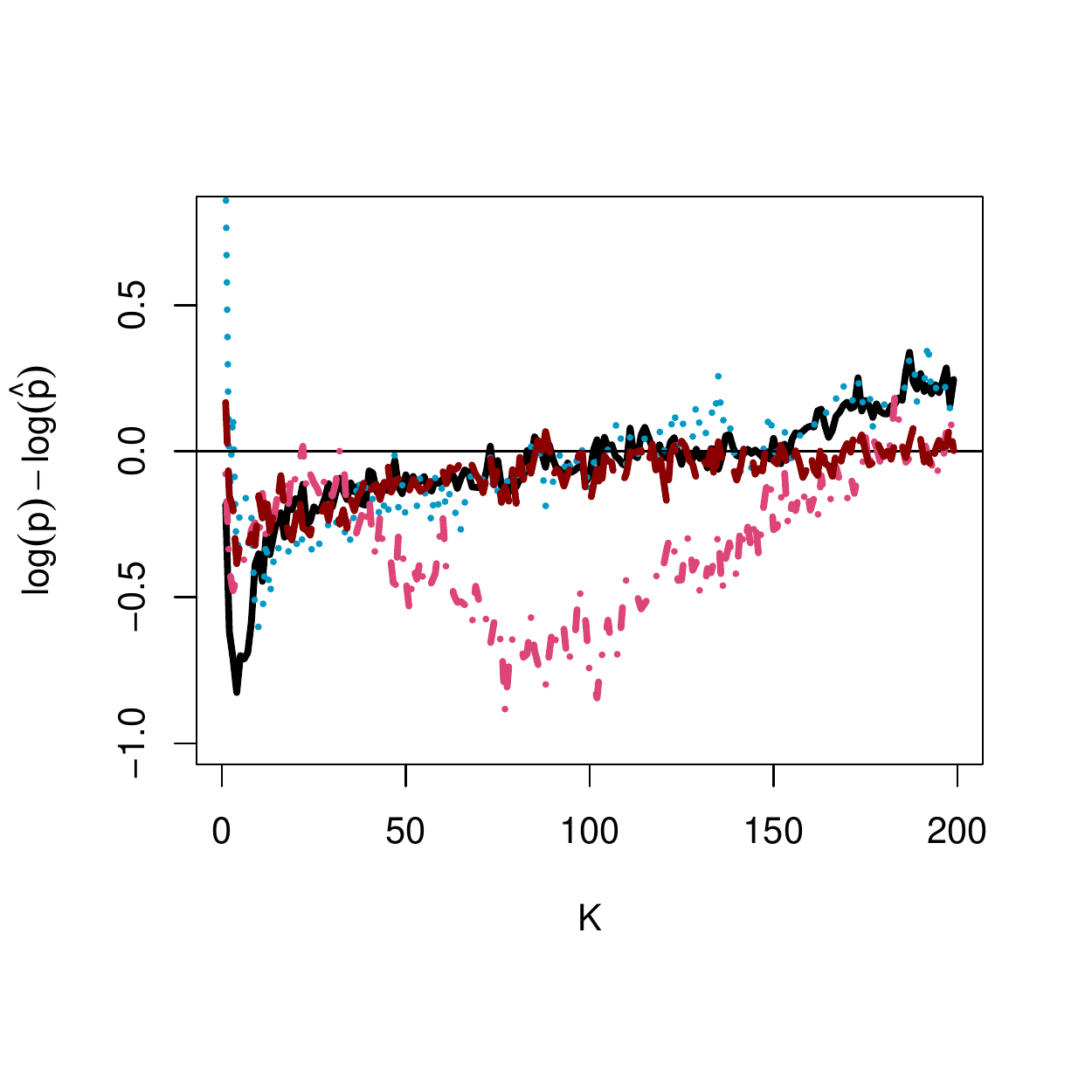}
\includegraphics[width=0.45\textwidth]{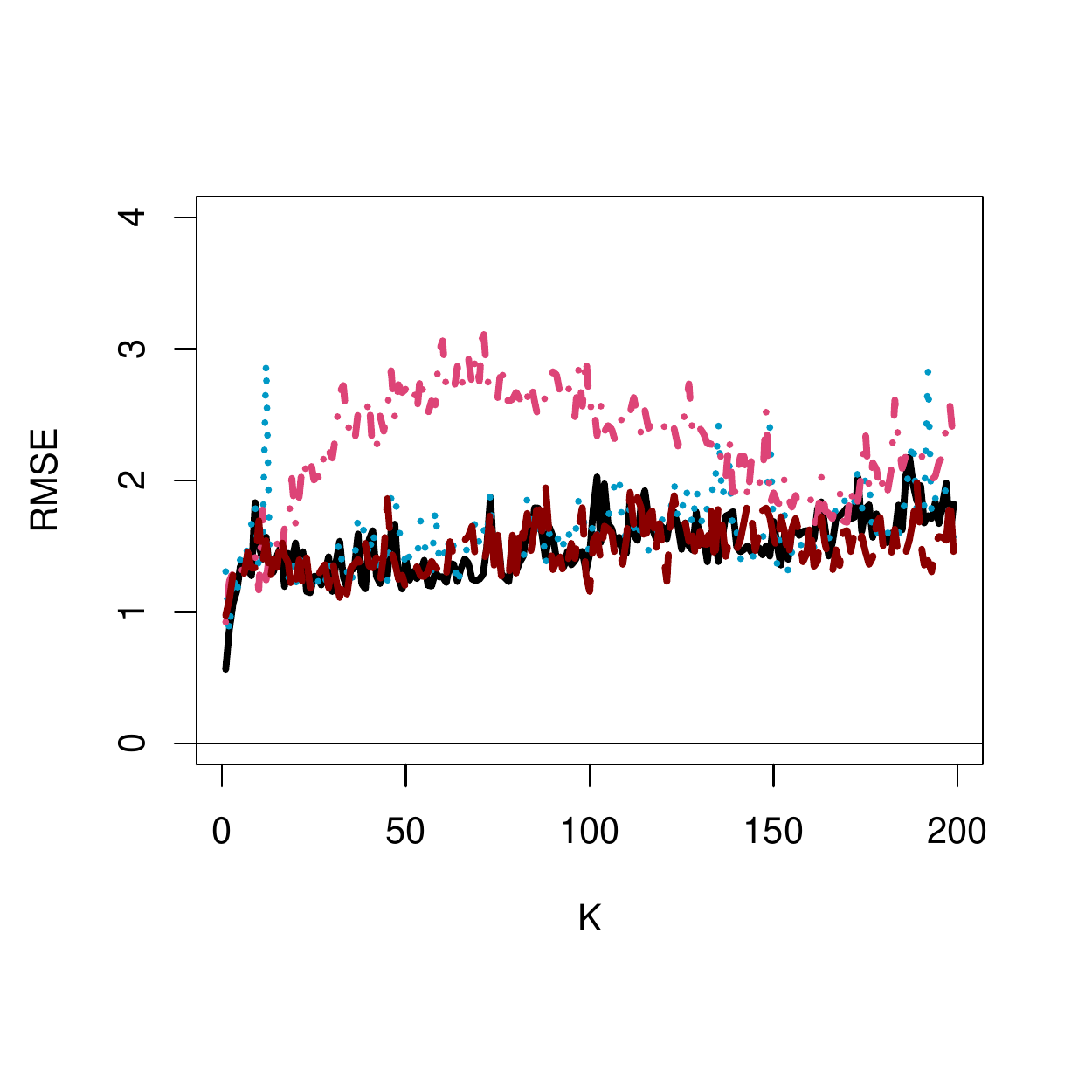}  
 \caption{ Standard normal distribution ($\xi=0$ and $\tilde\rho=0$). Estimation of $\xi$ (top) and tail probability (bottom) using minimum variance principle, bias (left), RMSE (right): GPD-ML (full line), $T\bar{p}$ (dotted), $Ep$ (dash-dotted) and $E\bar{p}$ (dashed).}
\end{figure}

 \begin{figure}[!ht]
  \centering
\includegraphics[width=0.45\textwidth]{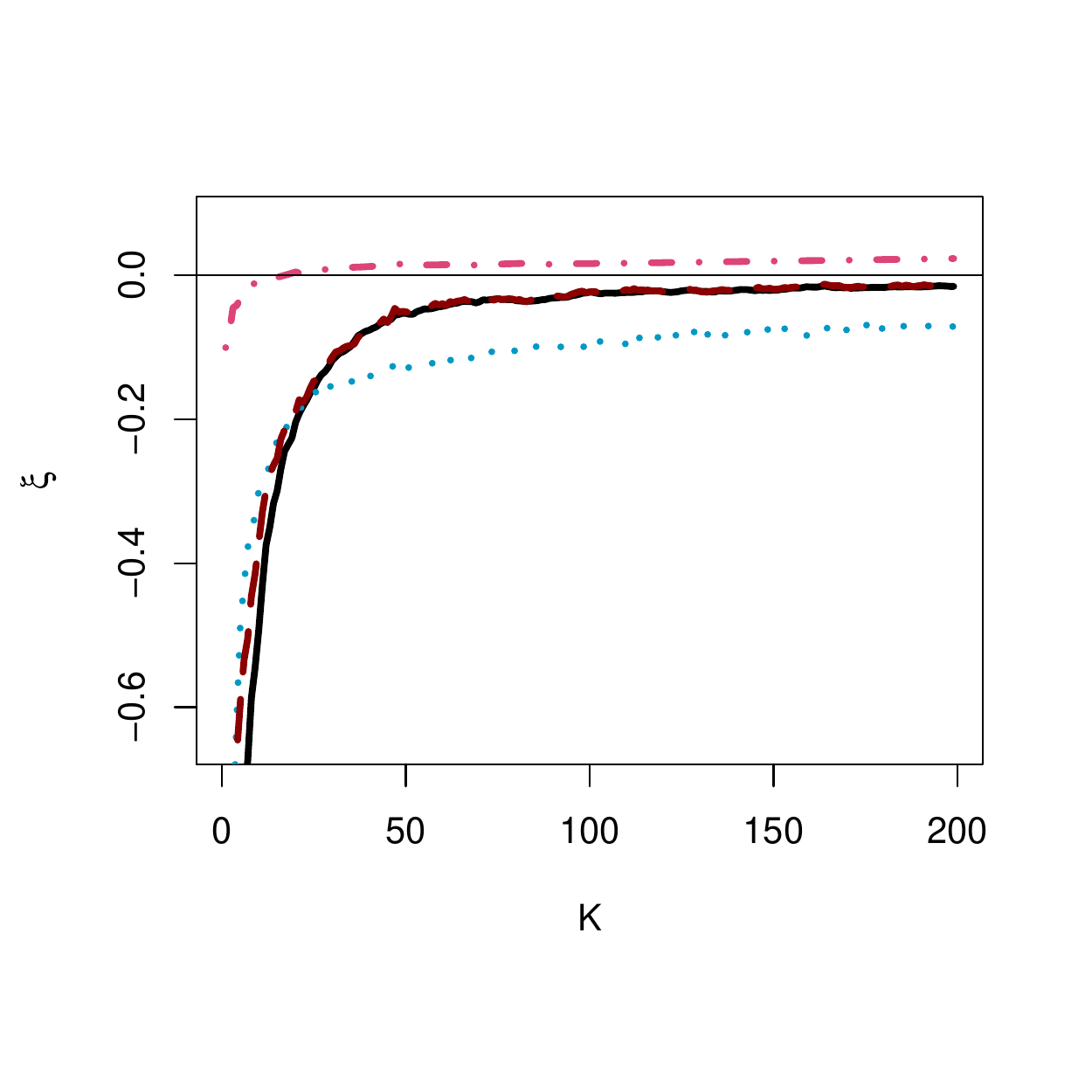} 
\includegraphics[width=0.45\textwidth]{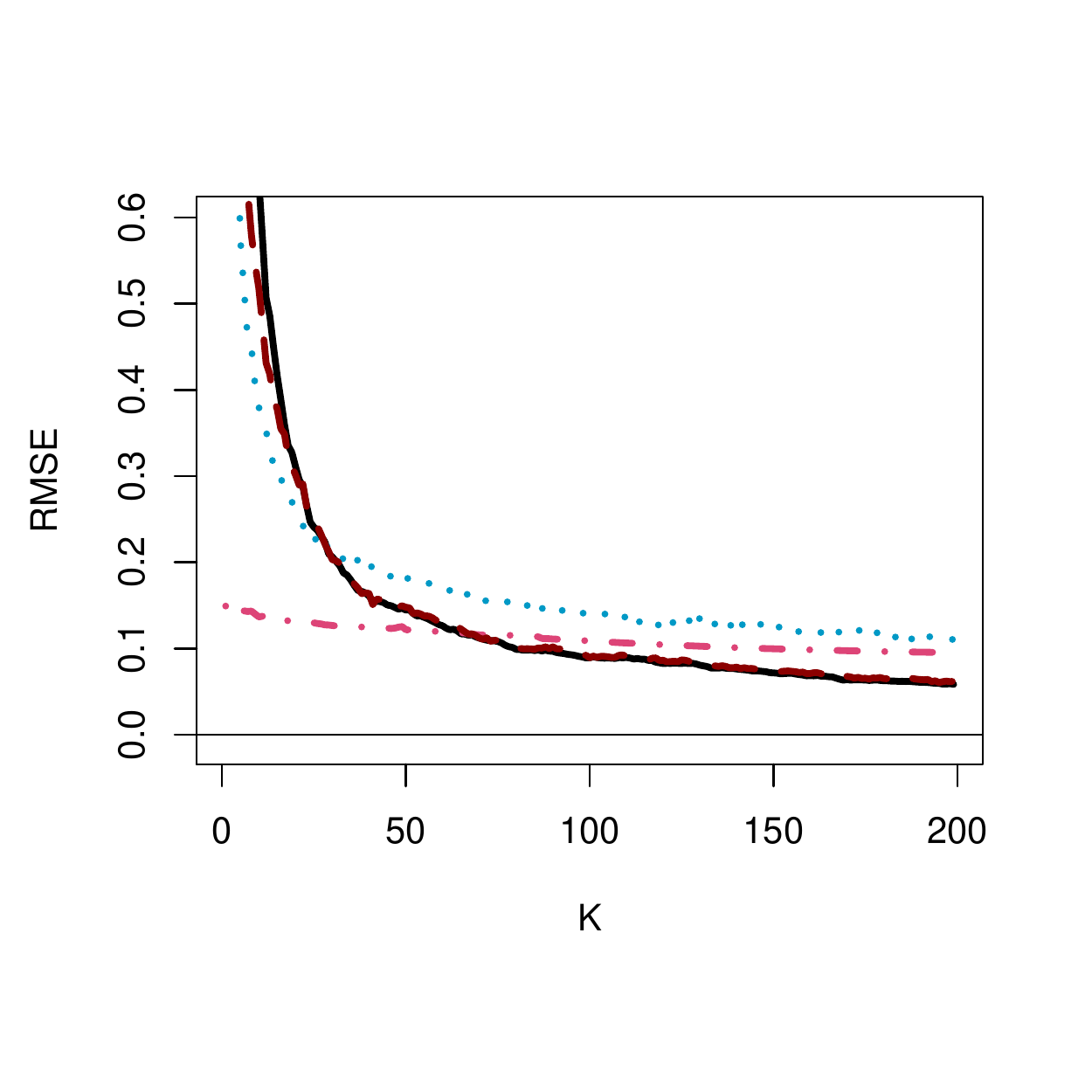} \\
\includegraphics[width=0.45\textwidth]{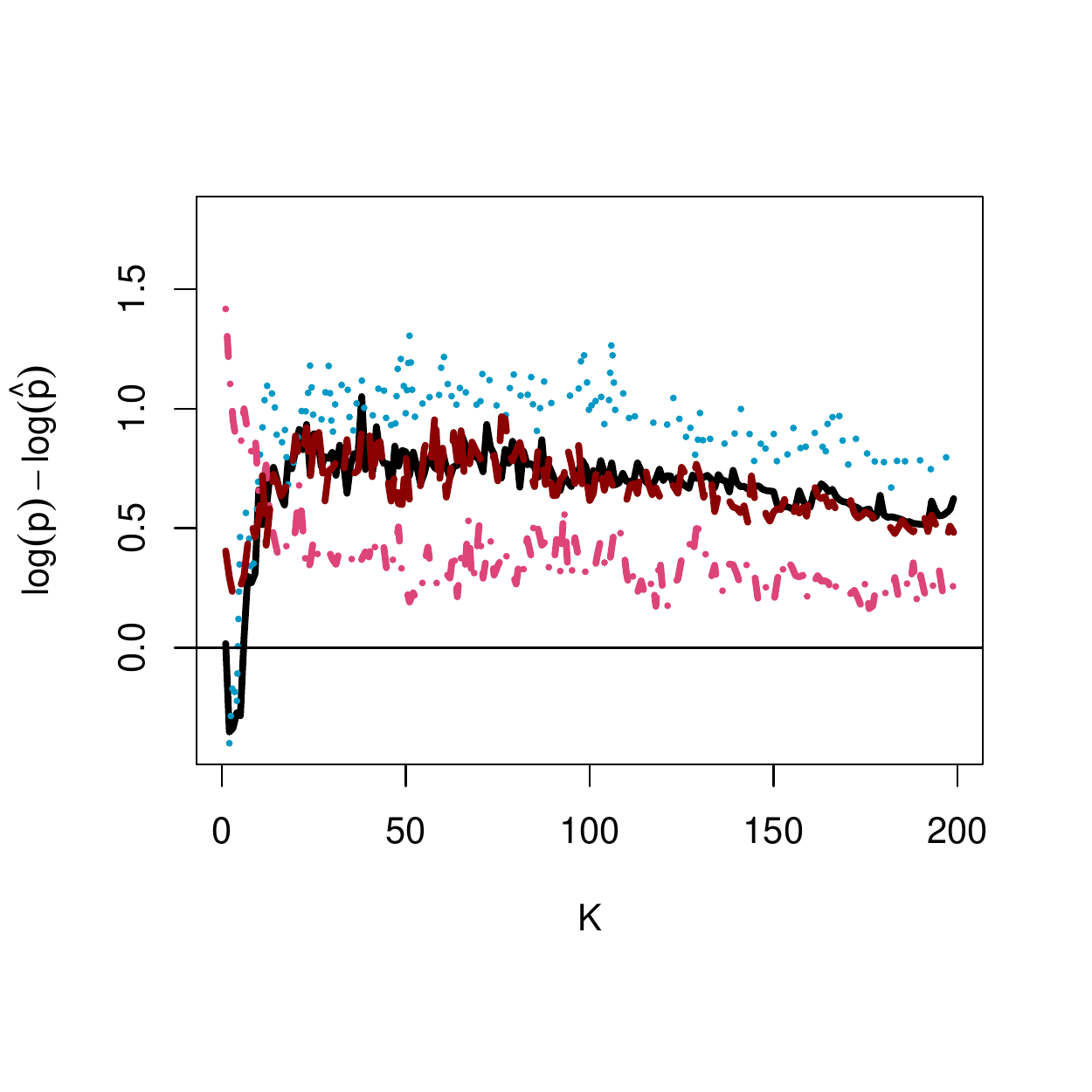}
\includegraphics[width=0.45\textwidth]{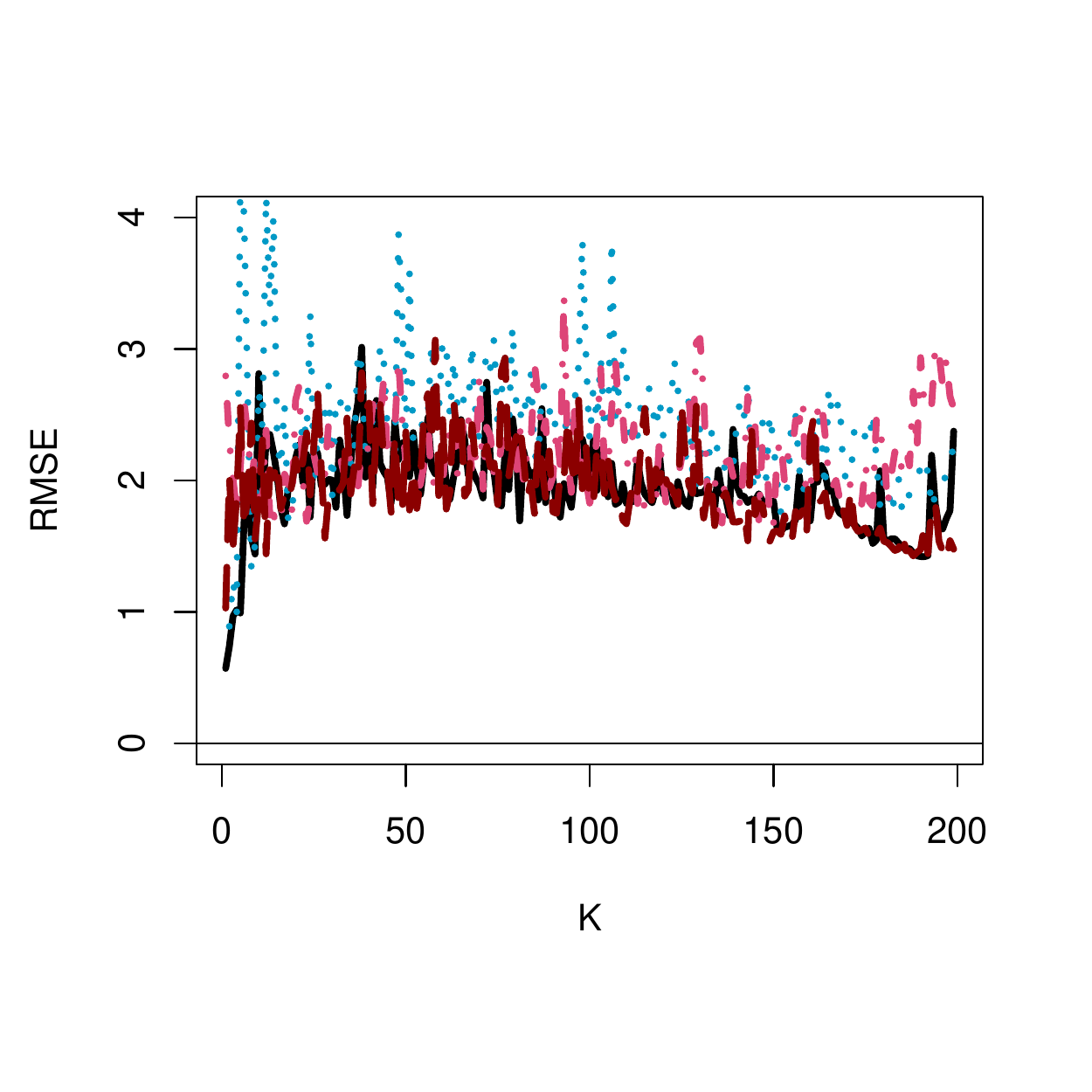}  
 \caption{ The exponential distribution ($\xi=0$ and $\tilde\rho=0$). Estimation of $\xi$ (top) and tail probability (bottom) using minimum variance principle, bias (left), RMSE (right): GPD-ML (full line), $T\bar{p}$ (dotted), $Ep$ (dash-dotted) and $E\bar{p}$ (dashed).}
\end{figure}

 \begin{figure}[!ht]
  \centering
\includegraphics[width=0.45\textwidth]{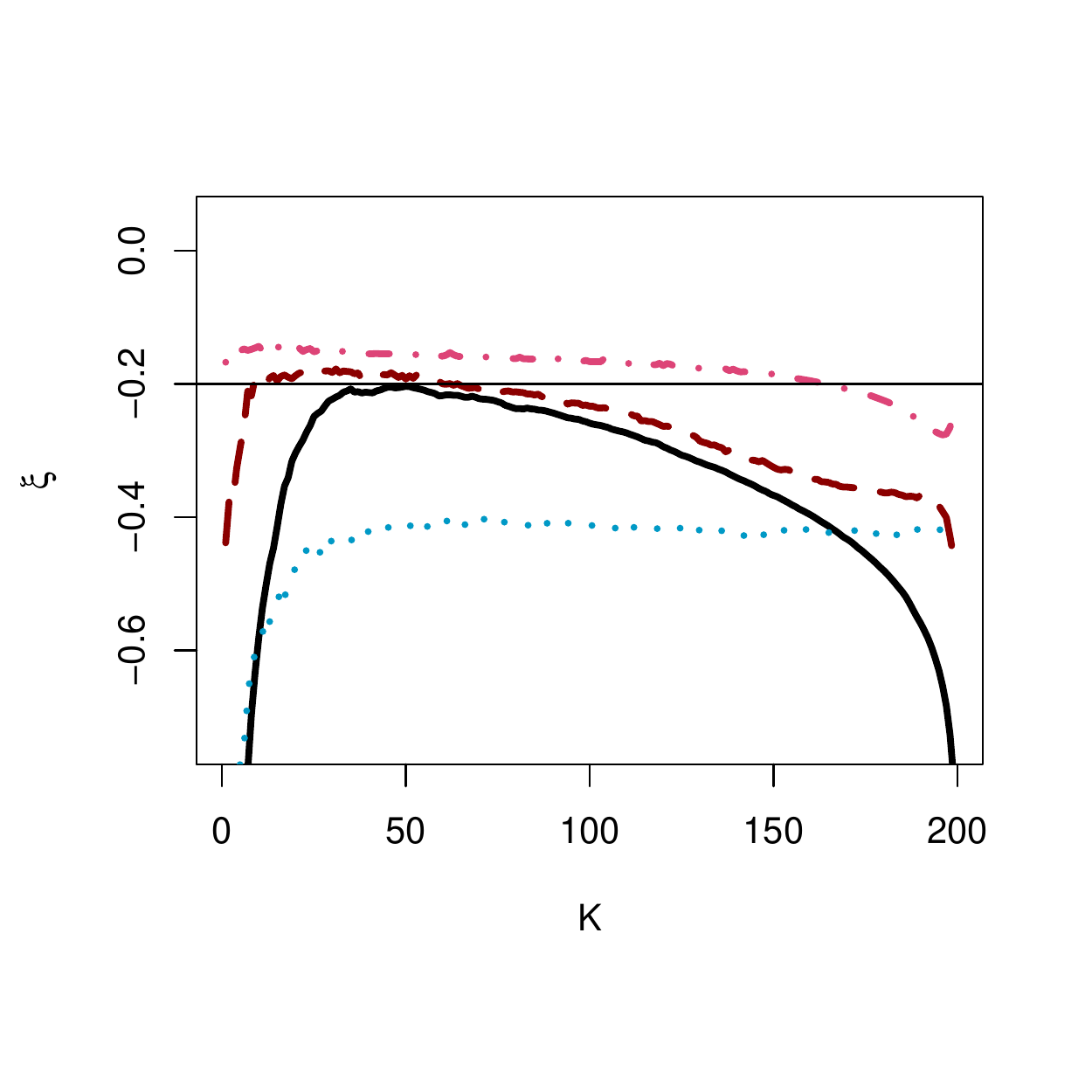} 
\includegraphics[width=0.45\textwidth]{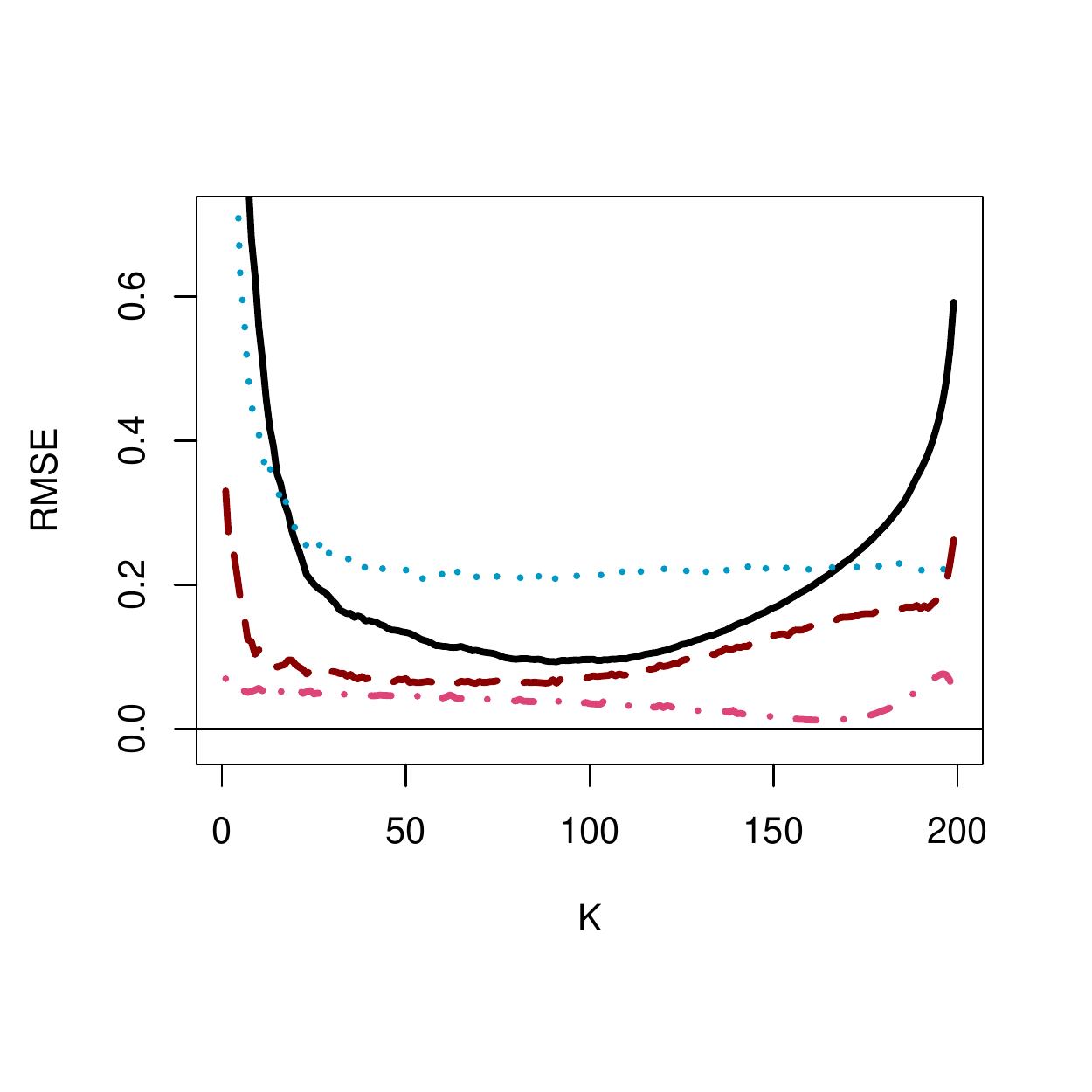} \\
\includegraphics[width=0.45\textwidth]{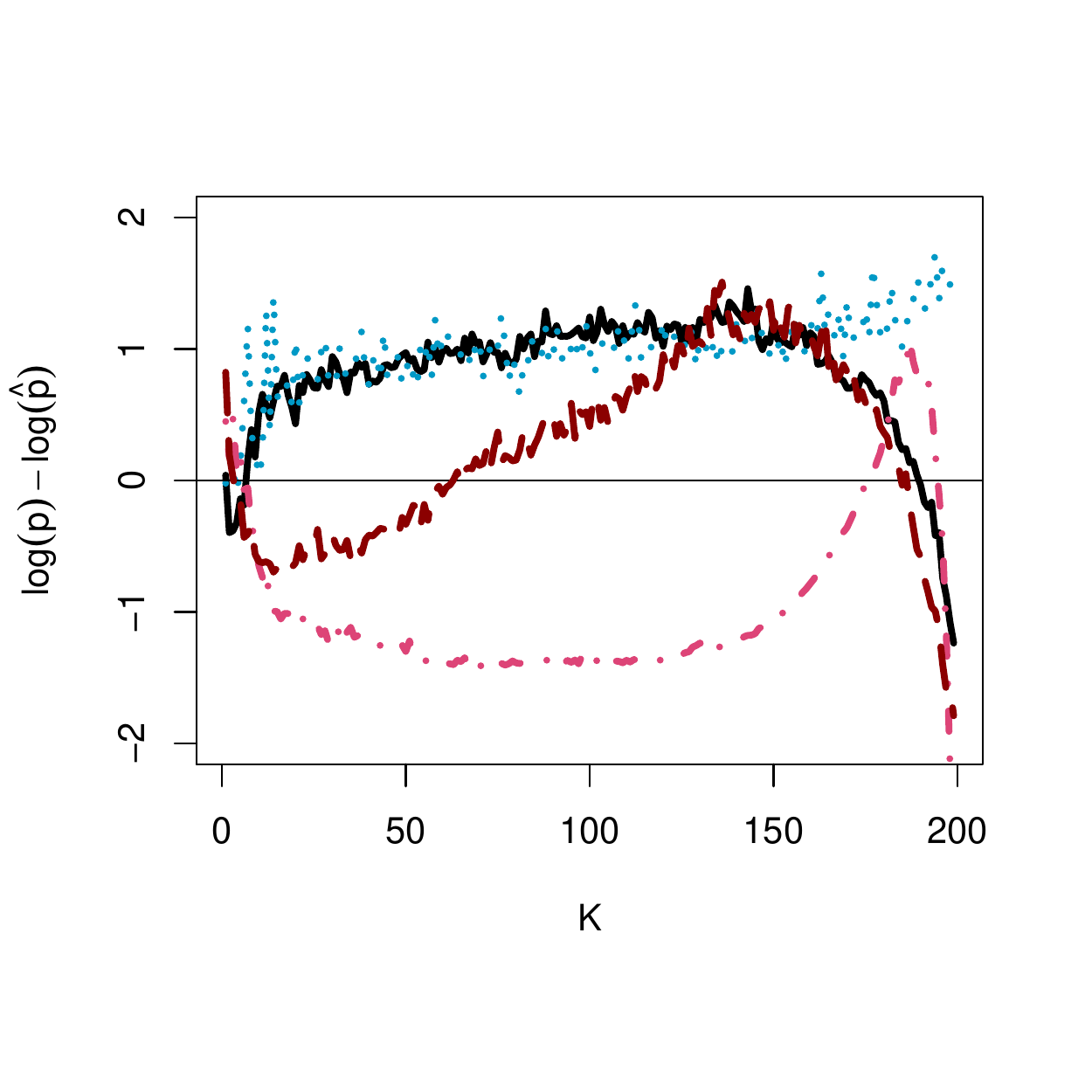}
\includegraphics[width=0.45\textwidth]{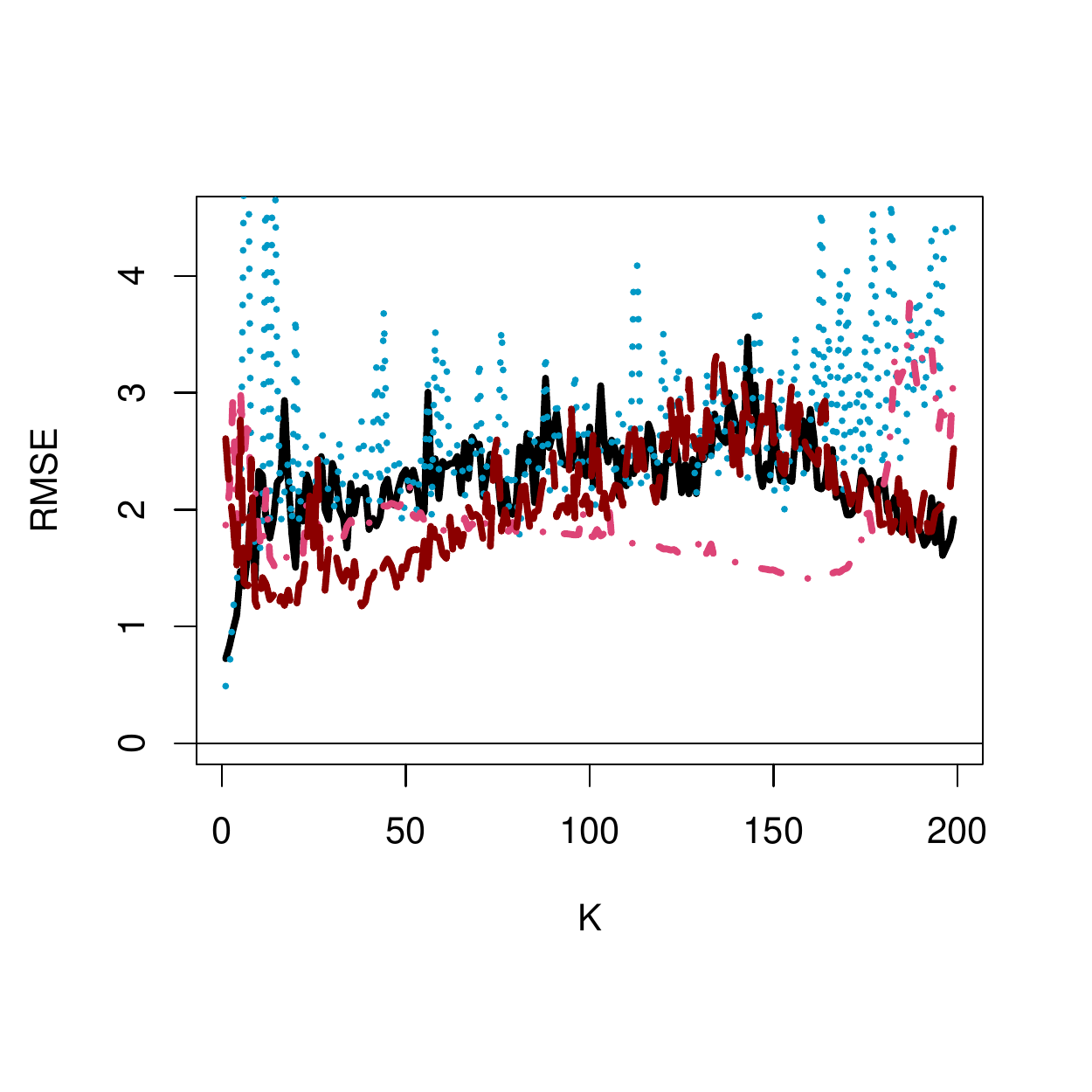}  
 \caption{ Reversed Burr distribution ($\xi=-0.2$ and $\tilde\rho=-1$). Estimation of $\xi$ (top) and tail probability (bottom) using minimum variance principle, bias (left), RMSE (right): GPD-ML (full line), $T\bar{p}$ (dotted), $Ep$ (dash-dotted) and $E\bar{p}$ (dashed).}
\end{figure}

\begin{figure}[!ht]
  \centering
\includegraphics[width=0.45\textwidth]{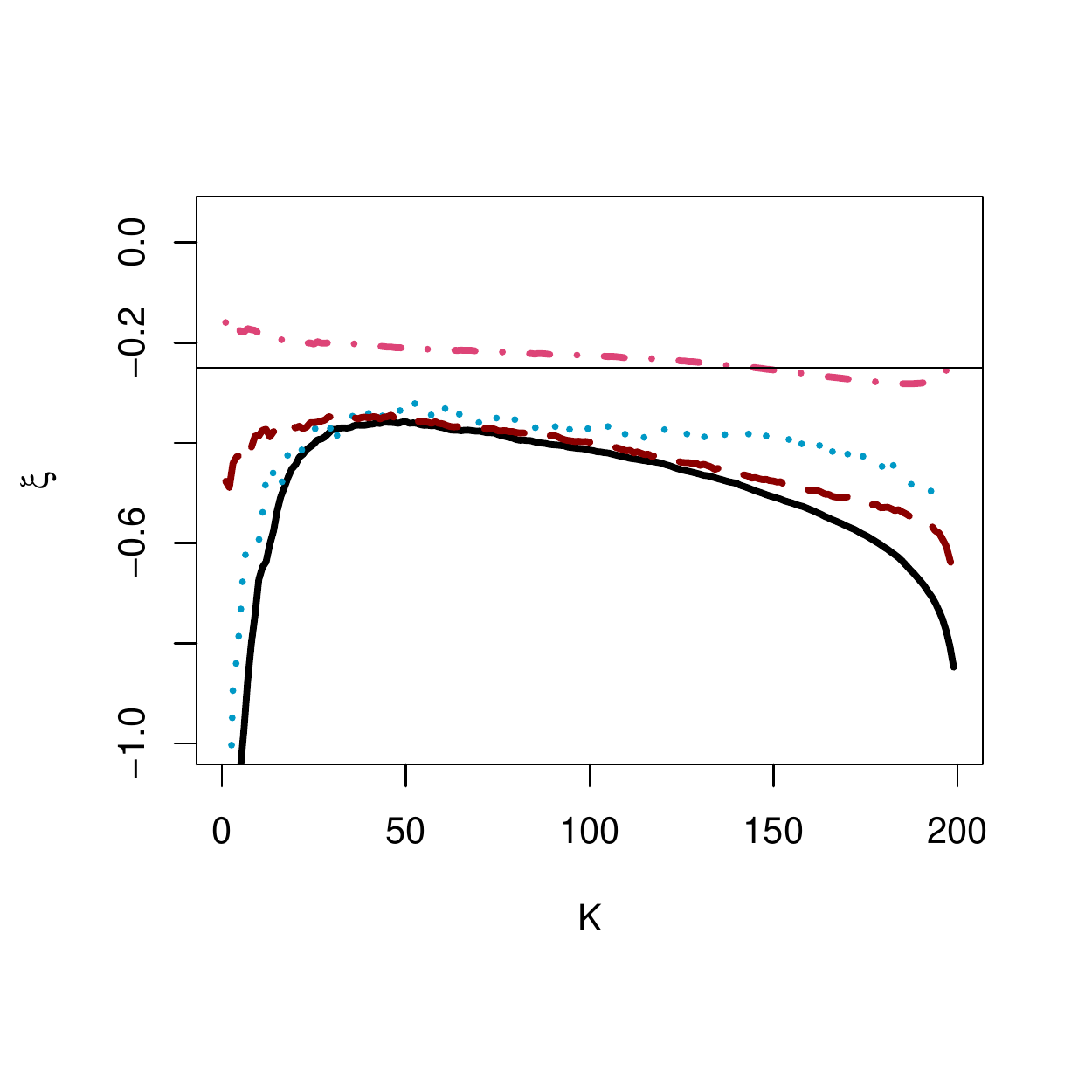} 
\includegraphics[width=0.45\textwidth]{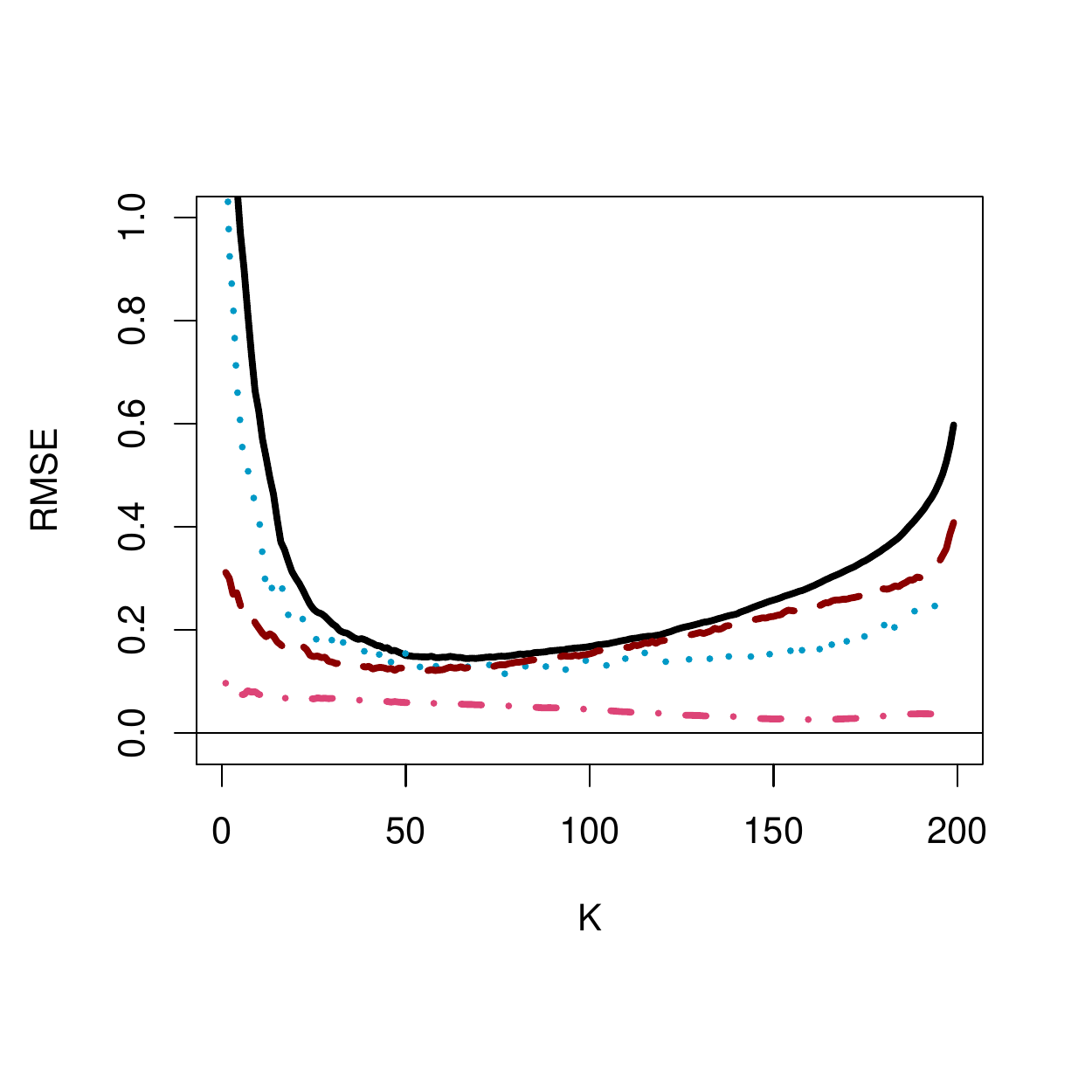} \\
\includegraphics[width=0.45\textwidth]{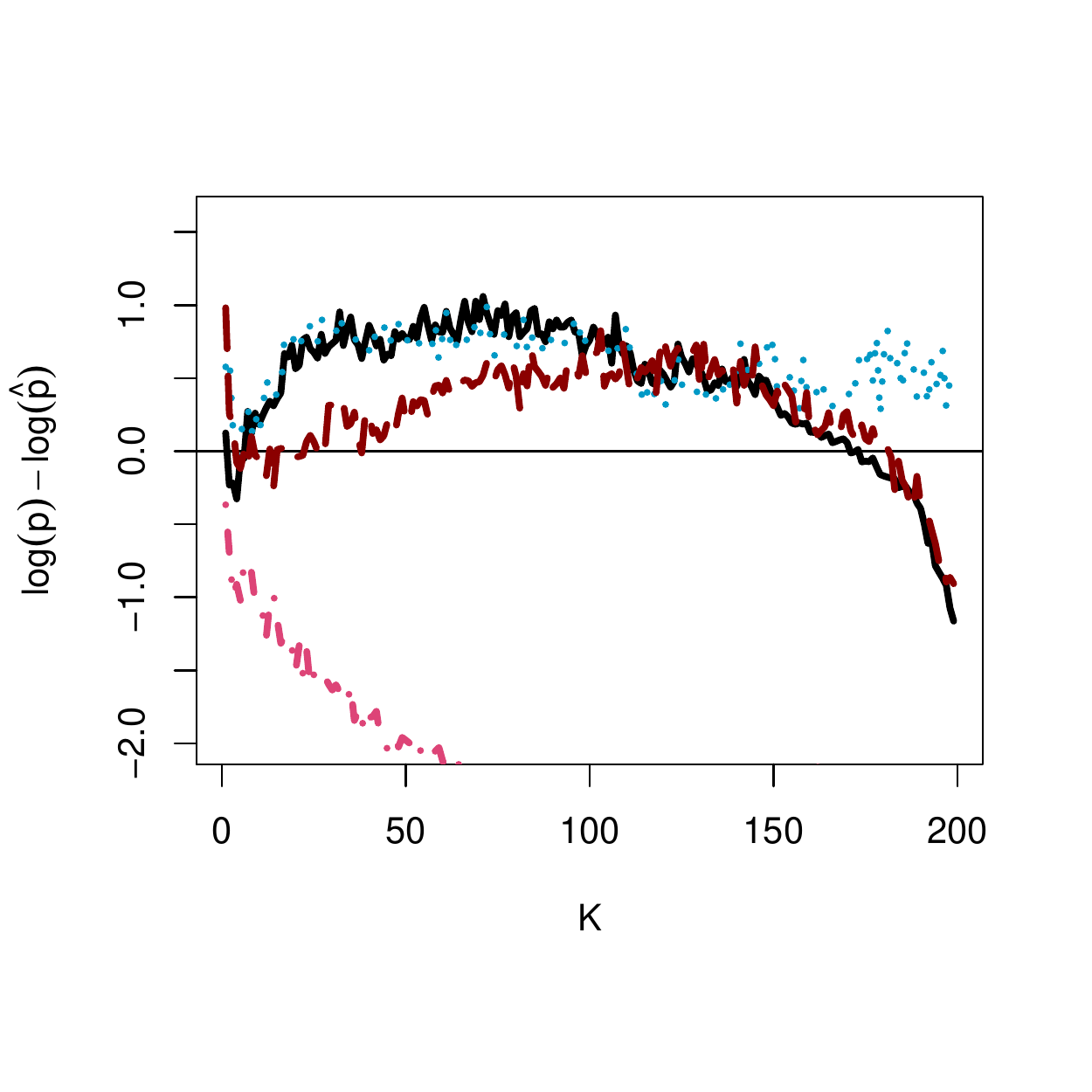}
\includegraphics[width=0.45\textwidth]{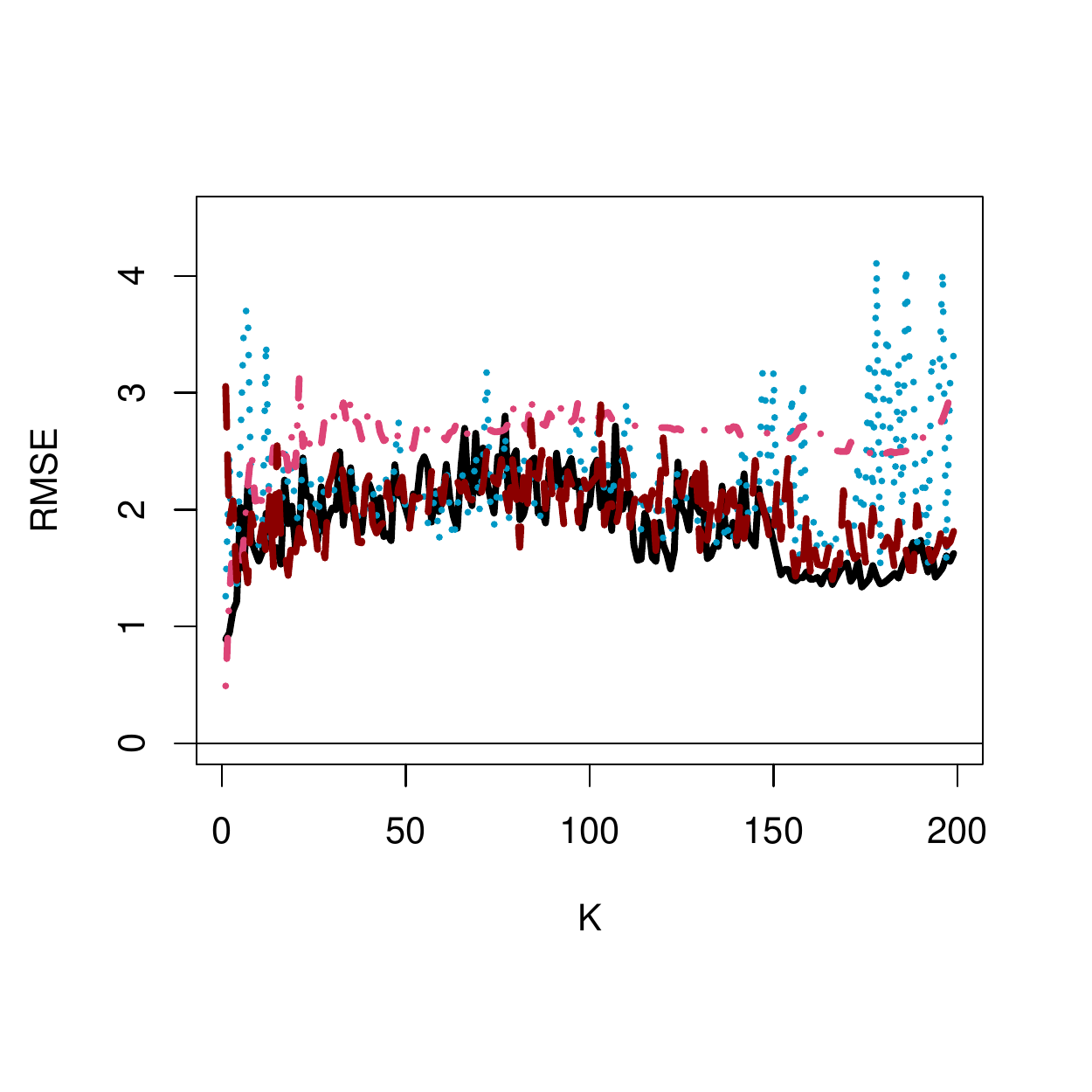}  
 \caption{Extreme value Weibull distribution ($\xi=-0.25$ and $\tilde\rho=-1$). Estimation of $\xi$ (top) and tail probability (bottom) using minimum variance principle, bias (left), RMSE (right): GPD-ML (full line), $T\bar{p}$ (dotted), $Ep$ (dash-dotted) and $E\bar{p}$ (dashed).}
\end{figure}

\begin{figure}[!ht]
  \centering
\includegraphics[width=0.45\textwidth]{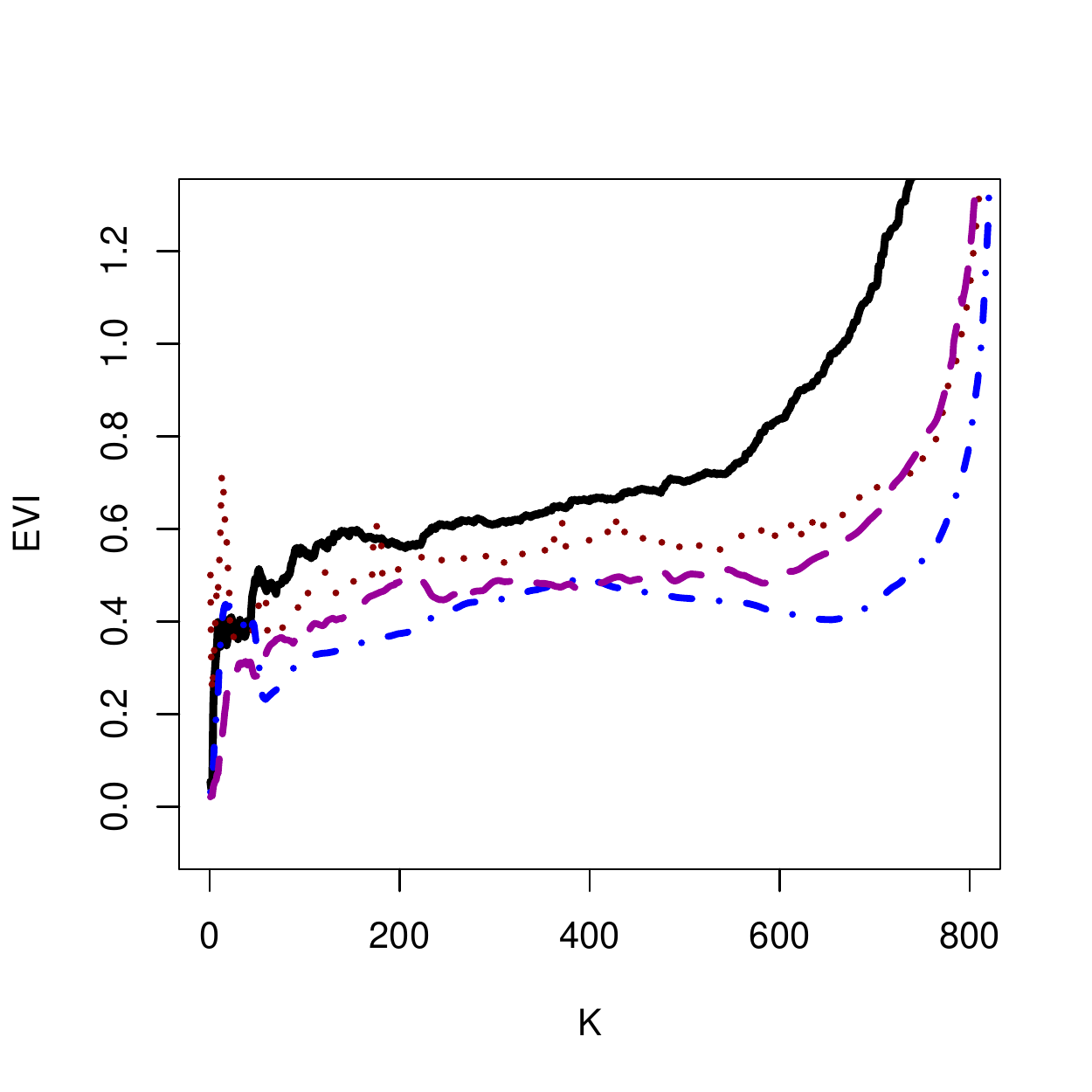} 
\includegraphics[width=0.45\textwidth]{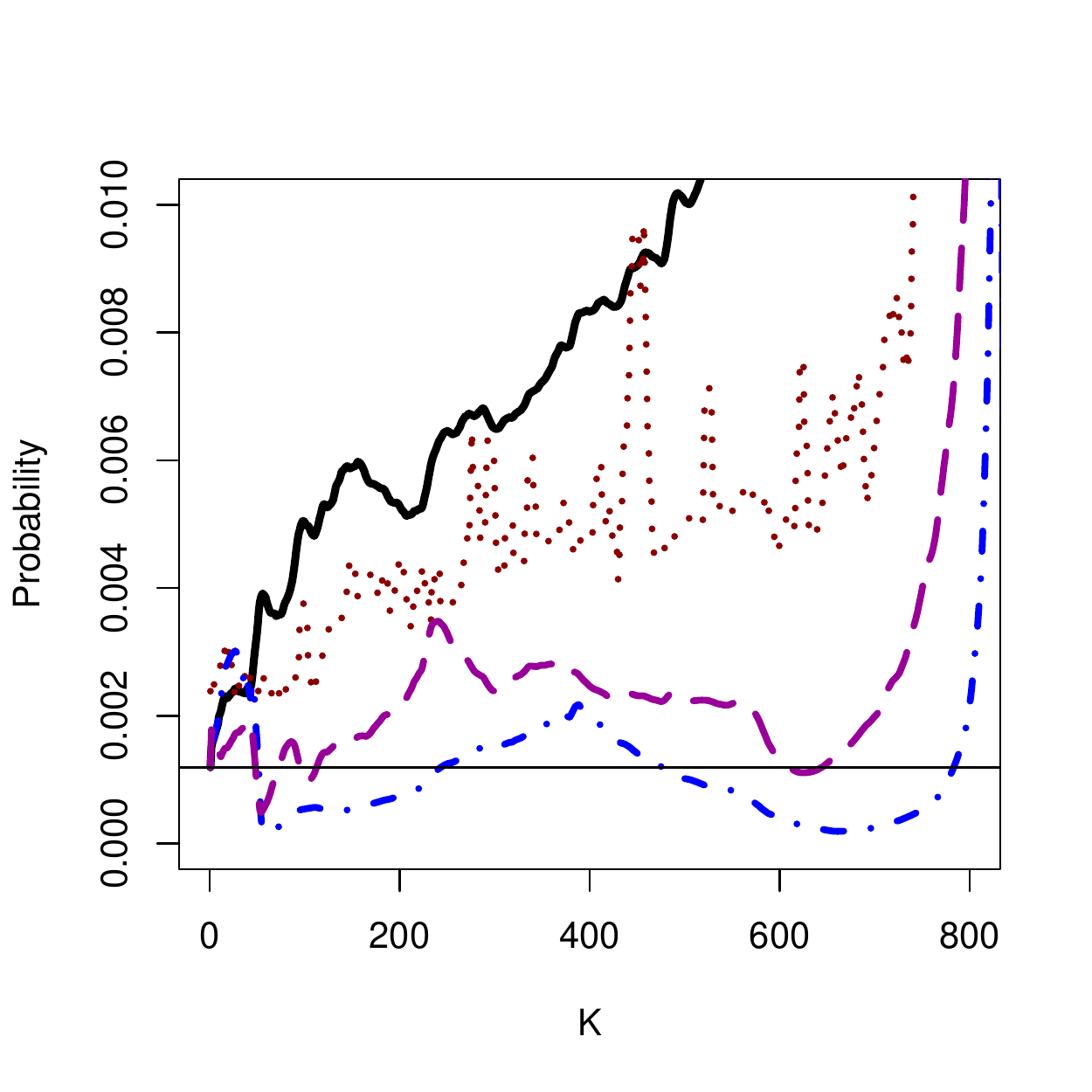} \\
\includegraphics[width=0.45\textwidth]{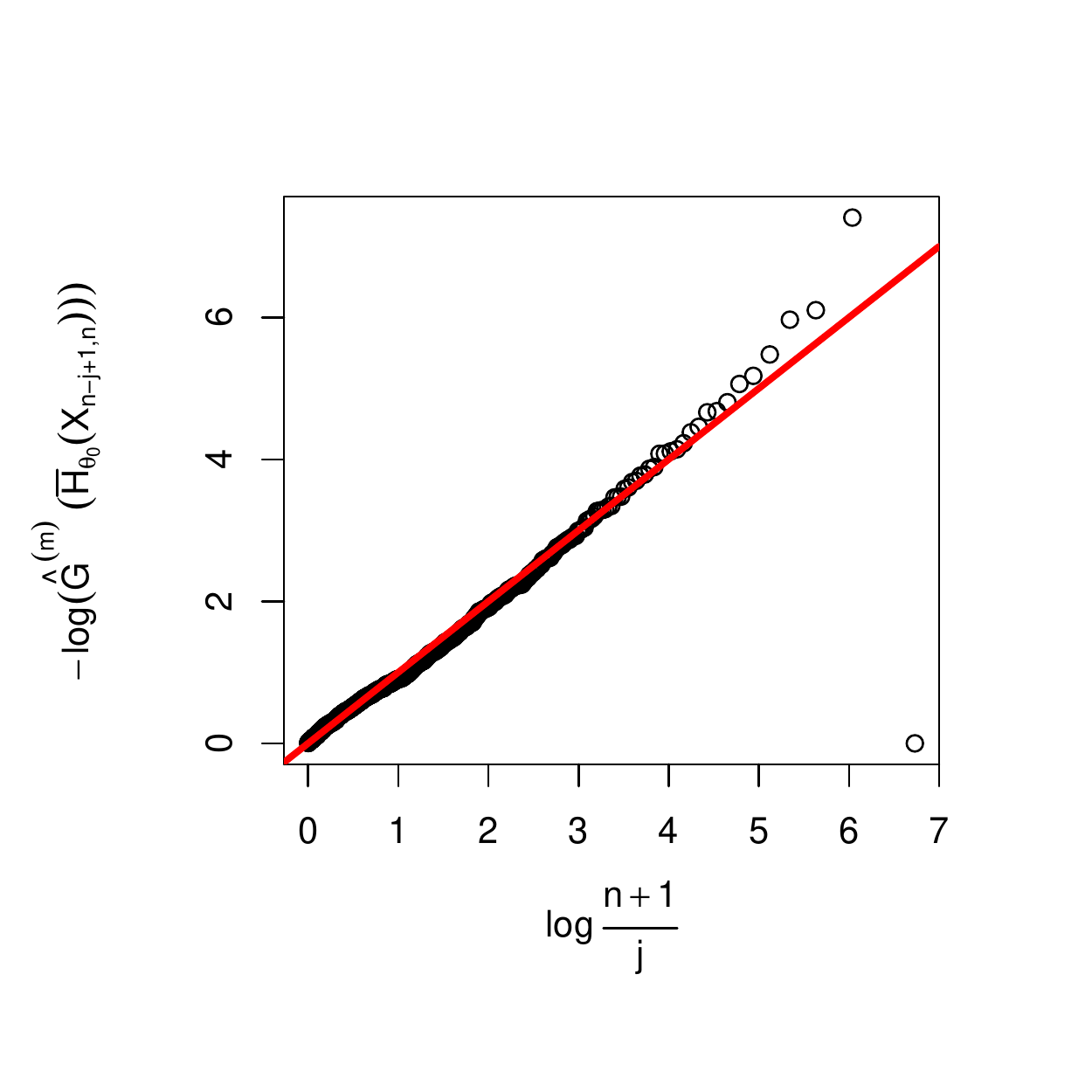}
 
 \caption{Ultimates of Belgian car insurance claims: estimation of $\xi$ (top left), tail probability at maximum observation (top right): Pareto-ML (full line), $T\bar{p}^+$ (dotted), $Ep^+$ (dashed) and $E\bar{p}^+$ (dash-dotted). Goodness-of-fit plot (bottom).}
\end{figure} 

\begin{figure}[!ht]
  \centering
\includegraphics[width=0.45\textwidth]{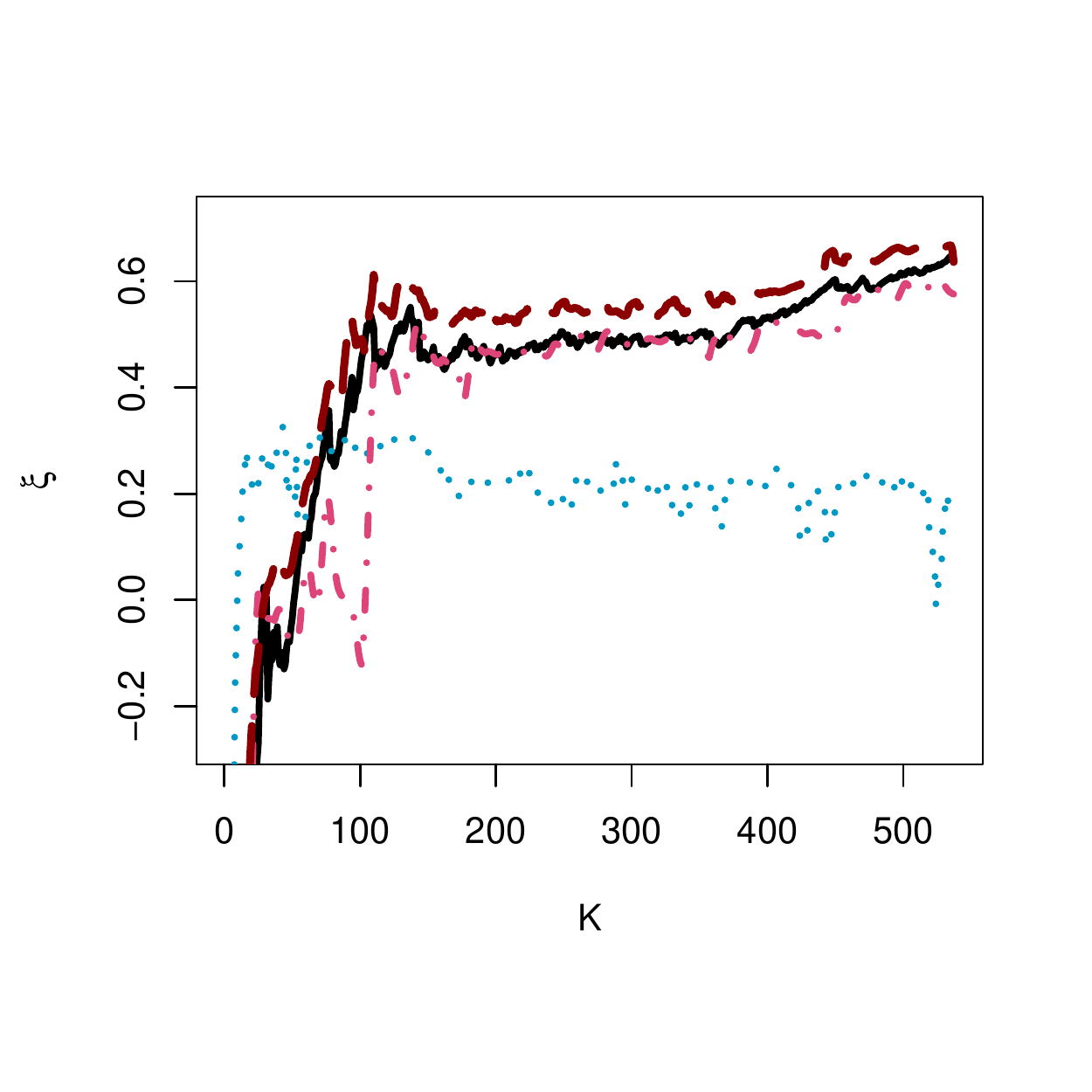} 
\includegraphics[width=0.45\textwidth]{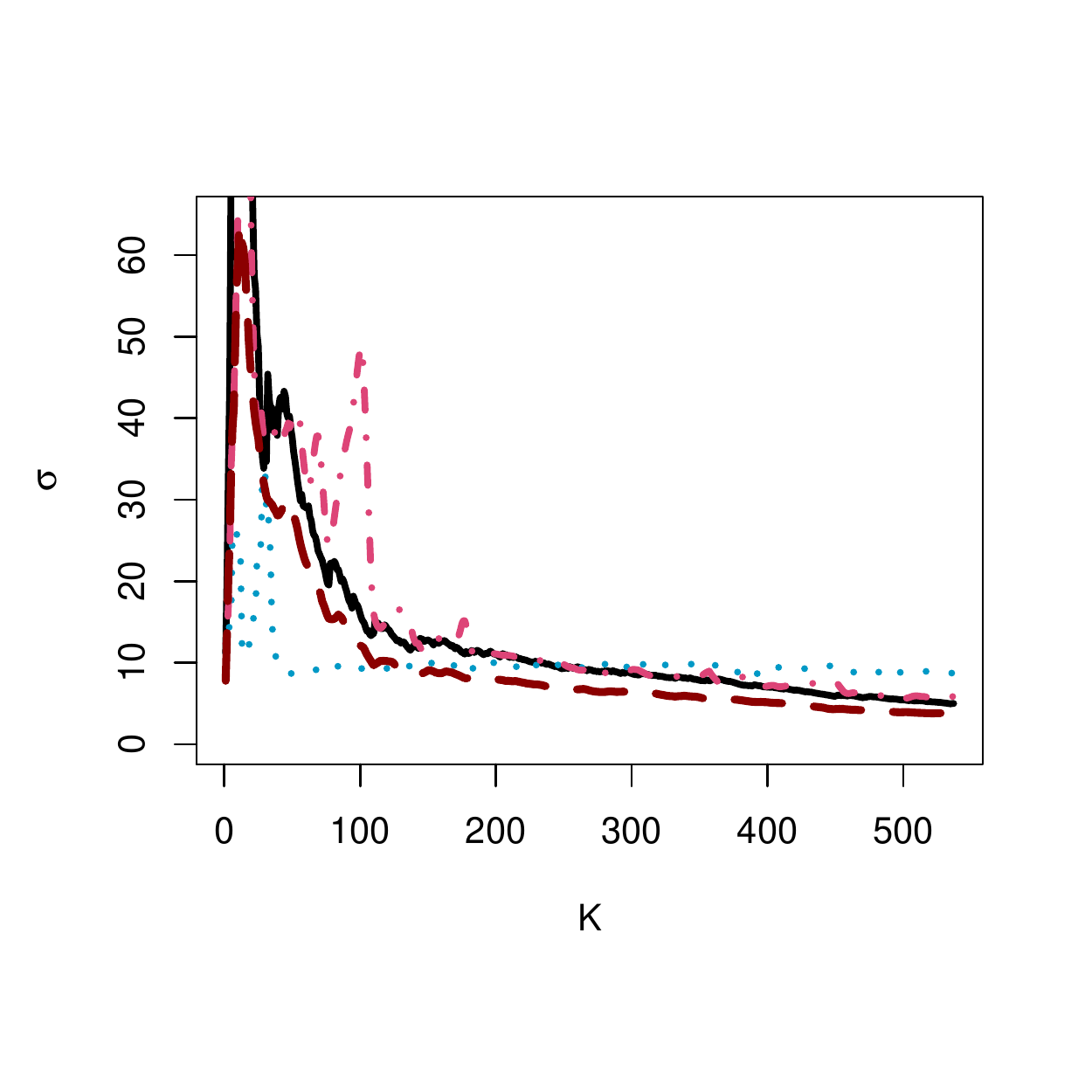} \\
\includegraphics[width=0.45\textwidth]{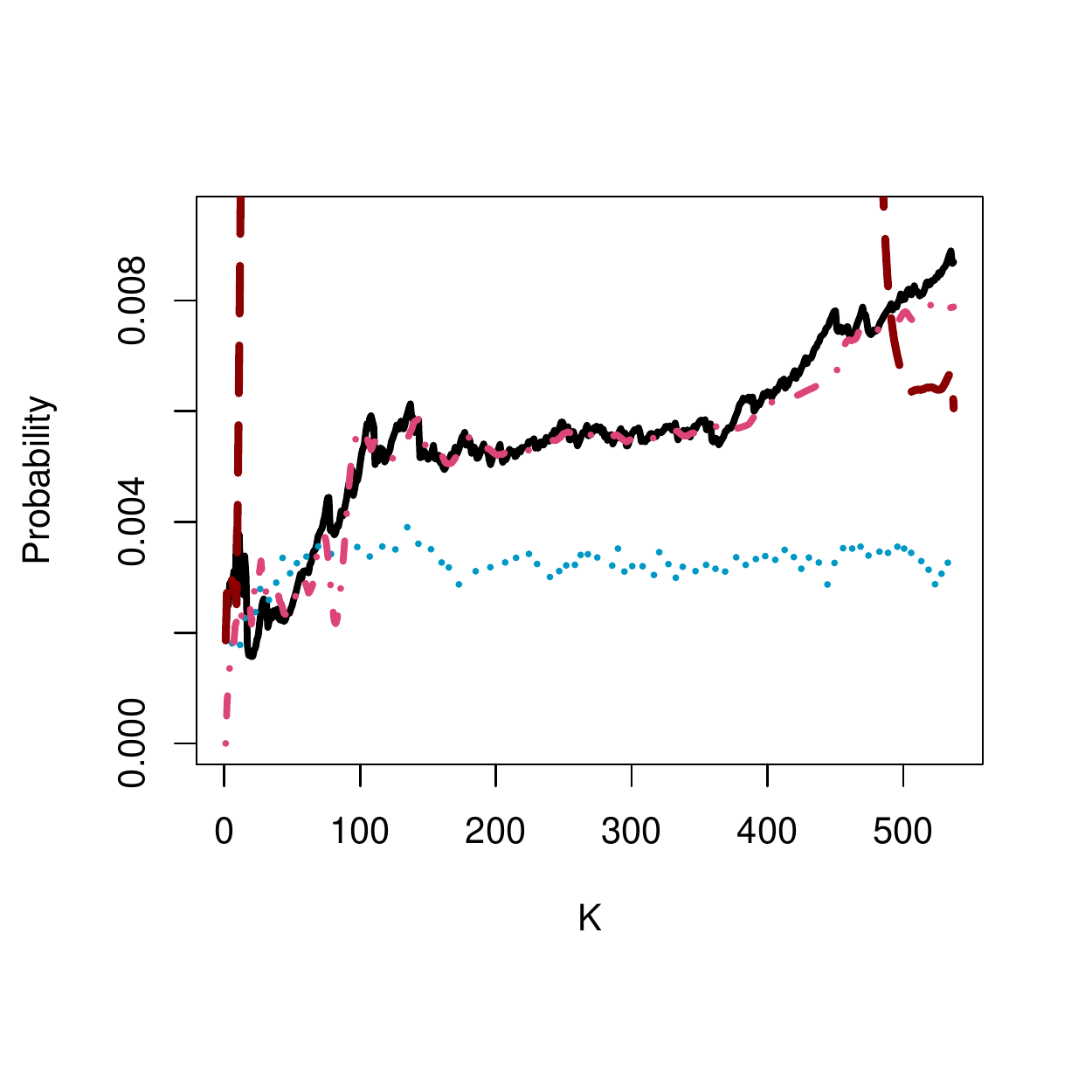}
\includegraphics[width=0.45\textwidth]{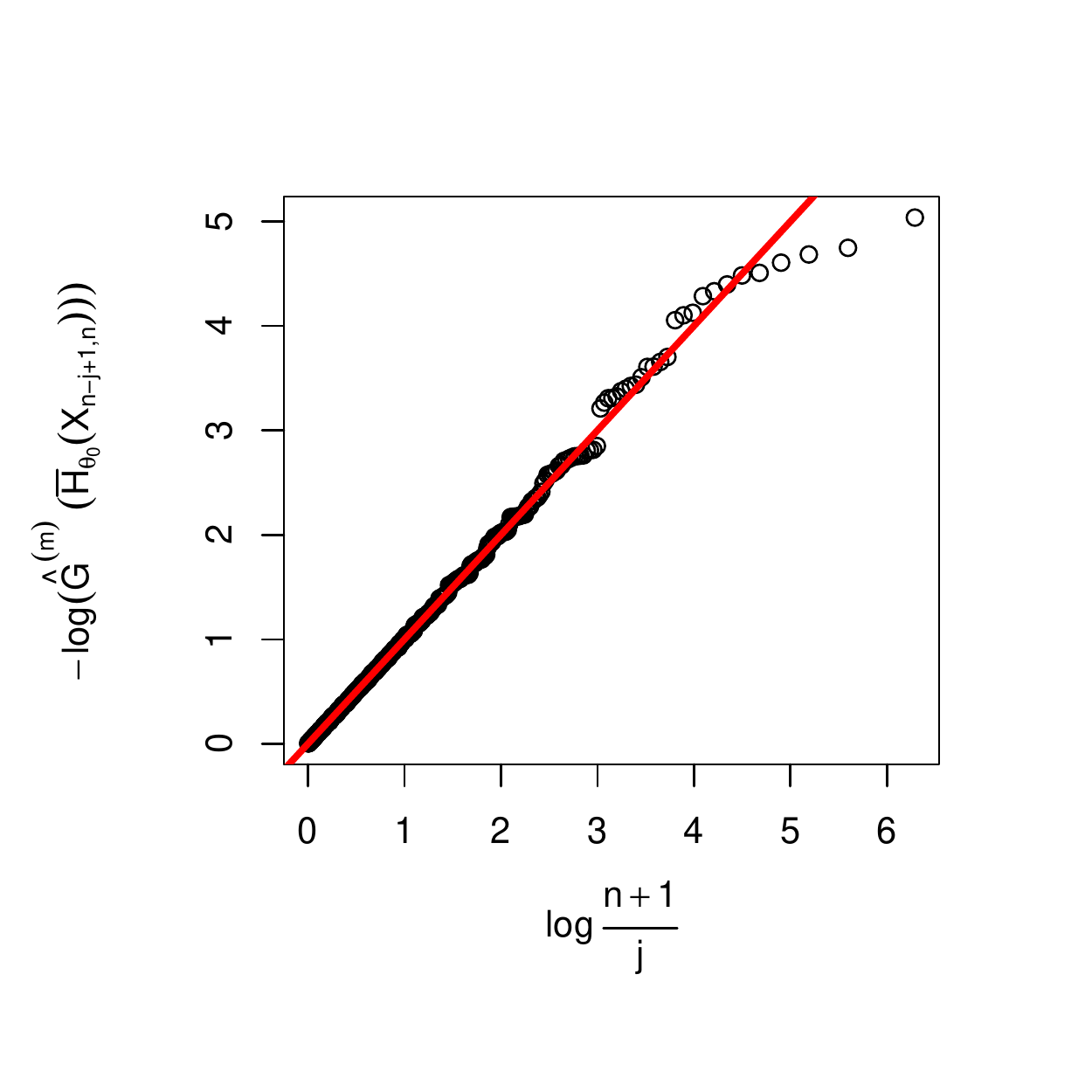}
 
 \caption{Winter rain data at Mont-Aigoual: estimation of $\xi$ and $\sigma$ (top) and tail probability  (bottom left) using minimum variance principle: GPD-ML (full line), $T\bar{p}$ (dotted), $Ep$ (dashed) and $E\bar{p}$ (dash-dotted). Goodness-of-fit plot (bottom right). }
\end{figure}

\end{document}